\documentclass[preprint,10pt,1p,sort&compress]{elsarticle}
\usepackage{amsmath,amssymb}
\usepackage[T1]{fontenc}
\usepackage{graphicx,bm,anyfontsize,xspace,hyperref,xcolor,booktabs}

\newcommand{\clkgnobm}{\ensuremath{C^{\kappa \text{g}}_{\ell}}\xspace}
\newcommand{\clkgth}{$C^{\kappa \text{g}}_{\ell,\text{th}}$\xspace}
\newcommand{\clkgobs}{$C^{\kappa \text{g}}_{\ell,\text{obs}}$\xspace}
\newcommand{\base}{\textit{base}\xspace}

\newcommand{\pggnobm}{\ensuremath{P^{\text{gg}}(k)}\xspace}
\newcommand{\pggth}{$P^{\text{gg}}_{\text{th}}(k)$\xspace}
\newcommand{\pggobs}{$P^{\text{gg}}_{\text{obs}}(k)$\xspace}
\newcommand{\pmmnobm}{$P^{\text{mm}}(k)$\xspace}
\newcommand{\pmgnobm}{$P^{\text{mg}}(k)$\xspace}
\newcommand{\bao}{\textit{BAO}$_{z_1}$\xspace}
\newcommand{\act}{\textit{ACT}\xspace}
\newcommand{\rk}{$r(k,z)$\xspace}

\newcommand{\mnu}{$\Sigma m_\nu$\xspace}
\newcommand{\ttt}{\textit{TT}}
\newcommand{\te}{\textit{TE}}
\newcommand{\ee}{\textit{EE}}
\newcommand{\planckobs}{\textit{Planck}\xspace}
\newcommand{\baoplanck}{\textit{BAO}$_{\text{cons}}$\xspace}
\newcommand{\abias}{\ensuremath{b_{\text{lin}}}\xspace}
\newcommand{\cbias}{\ensuremath{b_{k^2\text{cross}}}\xspace}
\newcommand{\dbias}{\ensuremath{b_{k^2\text{auto}}}\xspace}
\newcommand{\pshot}{\ensuremath{P_{\text{shot}}}\xspace}
\newcommand{\chisq}{\ensuremath{\chi^2}\xspace}
\newcommand{\chisqbest}{\ensuremath{\chi^2_\text{bestfit}}\xspace}
\newcommand{\lcdm}{$\Lambda$CDM\xspace}
\newcommand{\ccross}{$C_{\text{cross}}$\xspace}

\newcommand{\kmax}{\ensuremath{k_\text{max}}\xspace}
\newcommand{\pgghat}{\ensuremath{\hat{P}^{\text{gg}}(k_i)}\xspace}

\newcommand{\clkghat}{\ensuremath{{\hat{C}}^{\kappa \text{g}}_{\ell_j}}\xspace}
\newcommand{\hmpc}{\ensuremath{h\text{Mpc}^{-1}}\xspace}
\newcommand{\veff}{\ensuremath{\nu_\text{eff}}\xspace}

\hypersetup{colorlinks   = true,
            urlcolor     = blue,
            citecolor    = blue,
            linkcolor    = blue,
            menucolor    = blue,
            anchorcolor  = blue,
            filecolor    = blue
            pdfauthor    = author}

\journal{Journal of High Energy Astrophysics}

\begin{document}

\begin{frontmatter}

\title{Updated neutrino mass constraints from galaxy clustering and CMB lensing-galaxy cross-correlation measurements}

\author[okc]{Isabelle Tanseri}
\ead{isabelle.tanseri@gmail.com}

\author[lmu,origins,okc]{Steffen Hagstotz}
\ead{steffen.hagstotz@physik.uni-muenchen.de}

\author[kicc]{Sunny Vagnozzi\corref{cor1}}
\ead{sunny.vagnozzi@ast.cam.ac.uk}\cortext[cor1]{Corresponding author}

\author[mtu]{Elena Giusarma}
\ead{egiusarm@mtu.edu}

\author[okc,texas,nordita]{Katherine Freese}
\ead{ktfreese@utexas.edu}

\address[okc]{The Oskar Klein Centre for Cosmoparticle Physics, Department of Physics, Stockholm University, Roslagstullsbacken 21A, SE-106 91 Stockholm, Sweden}
\address[lmu]{Universit\"{a}ts-Sternwarte, Fakult\"{a}t f\"{u}r Physik, Ludwig-Maximilians Universit\"{a}t M\"{u}nchen, Scheinerstra\ss e 1, D-81679 M\"{u}nchen, Germany}
\address[origins]{Excellence Cluster ORIGINS, Boltzmannstra\ss e 2, D-85748 Garching, Germany}
\address[kicc]{Kavli Institute for Cosmology, University of Cambridge, Cambridge CB3 0HA, UK}
\address[mtu]{Department of Physics, Michigan Technological University, Fisher Hall 118, 1400 Townsend Drive, Houghton, MI 49931, USA}
\address[texas]{Theory Group, Department of Physics, The University of Texas at Austin, 2515 Speedway, C1600, Austin, TX 78712-0264, USA}
\address[nordita]{Nordita, KTH Royal Institute of Technology and Stockholm University, Roslagstullsbacken 23, SE-106 91 Stockholm, Sweden}

\begin{abstract}
\noindent We revisit cosmological constraints on the sum of the neutrino masses $\Sigma m_\nu$ from a combination of full-shape BOSS galaxy clustering [$P(k)$] data and measurements of the cross-correlation between \textit{Planck} Cosmic Microwave Background (CMB) lensing convergence and BOSS galaxy overdensity maps [$C^{\kappa \text{g}}_{\ell}$], using a simple but theoretically motivated model for the scale-dependent galaxy bias in auto- and cross-correlation measurements. We improve upon earlier related work in several respects, particularly through a more accurate treatment of the correlation and covariance between $P(k)$ and $C^{\kappa \text{g}}_{\ell}$ measurements. When combining these measurements with \textit{Planck} CMB data, we find a 95\% confidence level upper limit of $\Sigma m_\nu<0.14\,{\rm eV}$, while slightly weaker limits are obtained when including small-scale \textit{ACTPol} CMB data, in agreement with our expectations. We confirm earlier findings that (once combined with CMB data) the full-shape information content is comparable to the geometrical information content in the reconstructed BAO peaks given the precision of current galaxy clustering data, discuss the physical significance of our inferred bias and shot noise parameters, and perform a number of robustness tests on our underlying model. While the inclusion of $C^{\kappa \text{g}}_{\ell}$ measurements does not currently appear to lead to substantial improvements in the resulting $\Sigma m_{\nu}$ constraints, we expect the converse to be true for near-future galaxy clustering measurements, whose shape information content will eventually supersede the geometrical one.
\end{abstract}


\end{frontmatter}

\section{Introduction}
\label{sec:intro}

Neutrinos, while being among the most abundant particle species in the Universe, remain also one of the most elusive~\cite{Mohapatra:2005wg}. The observation of solar and atmospheric neutrino oscillations indicates that at least two out of three neutrino mass eigenstates are massive~\cite{Fukuda_1998,SNO:2002tuh,Gonzalez_Garcia_2008,Gonzalez_Garcia_2012,de_Salas_2018}, a fact which remains the only direct evidence for new physics beyond the Standard Model of Particle Physics.~\footnote{See e.g.~\cite{deSalas:2017kay,Capozzi:2017ipn,Esteban:2018azc,Hagstotz:2020ukm,Esteban:2020cvm,Capozzi:2021fjo,Archidiacono:2022ich} for recent global fits to active and sterile neutrino parameters (which include both cosmological and non-cosmological probes), and discussions thereof.} It should therefore not come as a surprise that the value of the sum of the neutrino masses \mnu is an extremely important experimental target. Oscillation experiments are insensitive to the absolute neutrino mass scale, and therefore to \mnu, which instead can be constrained by others types of probes: the kinematics of $\beta$-decay~\cite{Otten_2008, aker2021direct}, neutrino-less double-$\beta$ decay searches~\cite{DellOro:2016tmg,Dolinski_2019} and, last but not least, cosmological observations~\cite{LESGOURGUES_2006,lattanzi2017status,Sakr:2022ans}. Moreover, oscillation experiments are currently insensitive to the sign of the largest (atmospheric) mass-squared splitting, $\vert \Delta m_{31}^2 \vert$, leaving two possibilities open for the so-called neutrino mass ordering (or hierarchy): the \textit{normal ordering} with $\Delta m_{31}^2>0$ and $m_1<m_2<m_3$, and the \textit{inverted ordering} with $\Delta m_{31}^2<0$ and $m_3<m_1<m_2$, where $m_1$, $m_2$, and $m_3$ are the masses of the three neutrino mass eigenstates, and the mass-squared splittings are defined as $\Delta m_{ij}^2 \equiv m_i^2-m_j^2$.

As of today, cosmological probes provide the tightest constraints on \mnu, although such constraints are inevitably associated to a certain degree of model-dependence (see e.g.~\cite{lattanzi2017status} for an up-to-date review). In particular, measurements of anisotropies in the thermal radiation from recombination, the Cosmic Microwave Background (CMB)~\cite{Bennett_2003,Planck_2015_cosmoparams, Planck_2018_cosmoparams}, in combination with measurements of Baryon Acoustic Oscillations (BAO) in galaxy clustering data~\cite{Cole_2005,Eisenstein_2005,Cuesta_2016}, have been able to provide extremely strong bounds on \mnu. Currently, one of the tightest upper bounds on \mnu is \mnu$<0.12\,{\rm eV}$ at 95\% confidence level (C.L.), inferred from a combination of CMB data from the \textit{Planck} 2018 legacy data release and BAO data from the MGS, 6dFGS, and BOSS DR12 galaxy surveys~\cite{Planck_2018_cosmoparams}.~\footnote{See for instance~\cite{Cuesta:2015iho,Wang:2016tsz,Lorenz:2017fgo,Wang:2017htc,Chen:2017ayg,Zhao:2017jma,Nunes:2017xon,Vagnozzi:2018jhn,Guo:2018gyo,RoyChoudhury:2018gay,RoyChoudhury:2018vnm,Bonilla:2018nau,Lorenz:2018fzb,Bolliet:2019zuz,Vagnozzi:2019ezj,Yang:2019uog,Palanque-Delabrouille:2019iyz,Nunes:2020hzy,Yang:2020tax,Zhang:2020mox,Li:2020gtk,Yang:2020ope,Giare:2020vzo,RoyChoudhury:2020dmd,Brinckmann:2020bcn,DiValentino:2021zxy,DiValentino:2021hoh,Anchordoqui:2021gji,Feng:2021ipq,DiValentino:2021rjj,Renzi:2021xii,Jin:2022tdf,Kumar:2022vee,Reeves:2022aoi,DiValentino:2022njd} for examples of other recent works investigating constraints on neutrino masses and properties within various cosmological scenarios. See~\cite{Wetterich:2007kr,WaliHossain:2014usl,Geng:2015haa,Chacko:2020hmh,Lorenz:2021alz} for examples of mass-varying or decaying neutrino scenarios. See also~\cite{Green:2021xzn,Alvey:2021xmq} for discussions on the implications of a detection or non-detection of the cosmological neutrino background.} Cosmology is in principle also able to constrain the mass ordering, and recent cosmological observations have been argued to slightly favor the normal mass ordering (see e.g.~\cite{Huang:2015wrx,Hannestad:2016fog,Xu:2016ddc,Gerbino:2016ehw,Yang:2017amu,Simpson:2017qvj,Schwetz:2017fey,Long:2017dru,Gariazzo:2018pei,Heavens:2018adv,Mahony:2019fyb,RoyChoudhury:2019hls,Hergt:2021qlh,Jimenez:2022dkn,Gariazzo:2022ahe}).

Neutrinos decouple from the primordial plasma at temperatures of ${\cal O}({\rm MeV})$, thus while highly relativistic, therefore behaving as radiation early on, including at the time of recombination (given current constraints on \mnu which exclude the possibility that neutrinos were heavy enough to already behave as matter then). After they decouple, neutrinos start free-streaming with high thermal velocities. At late times, at least two out of three neutrino mass eigenstates become non-relativistic, and contribute to the matter component of the Universe. This aspect, in combination with their free-streaming nature, leads to massive neutrinos behaving as a hot dark matter component and suppressing power on small-scales. This small-scale power suppression is in principle one of the tell-tale cosmological signatures of massive neutrinos~\cite{Bond_1980,Eisenstein:1997jh,Hu_1998}, and can be searched for instance through measurements of the clustering of tracers of the large-scale structure (LSS), such as galaxies and quasars.

The effect of massive neutrinos on the CMB is instead more subtle. In general, discussions as to what the effects on the CMB of changing a given cosmological parameter are require specifying what other quantities are being kept fixed while the parameter in question is varied, and \mnu is no exception. In the CMB, the effect of non-zero \mnu is best discussed while fixing both the acoustic scale $\theta_s$~\footnote{The acoustic scale is given by the ratio between the comoving sound horizon at recombination and the angular diameter distance to recombination.} and the redshift of matter-radiation equality $z_{\rm eq}$. This choice \textit{a)} ensures that the position and height of the first acoustic peak in the CMB, both tightly constrained by observations~\cite{Planck_2018_cosmoparams}, are left unchanged while varying \mnu, and \textit{b)} helps disentangling ``direct'' neutrino perturbation effects from ``background'' effects which may instead be re-absorbed by suitably shifting other cosmological parameters. Increasing \mnu while keeping $\theta_s$ and $z_{\rm eq}$ fixed leads to a small reduction of power on large angular scales (due to a reduced late integrated Sachs-Wolfe effect, but swamped by cosmic variance) alongside tiny shifts in the damping scale.~\footnote{See e.g.\ Fig.~4.10 in~\cite{Vagnozzi:2019utt} and the associated discussion for an example of this exercise. Note, in addition, that fixing $\theta_s$ and $z_{\rm eq}$ ensures that massive neutrinos leave no visible imprint on the early integrated Sachs-Wolfe effect, whose amplitude is very tightly constrained by CMB observations (see for example~\cite{Hou:2011ec,Cabass:2015xfa,Kable:2020hcw,Vagnozzi:2021gjh,Ruiz-Granda:2022bcn}).} The most important effect on the CMB is instead related to the reduction of the lensing potential, a direct consequence of the neutrino-induced small-scale structure suppression. As the effect of lensing is to smooth the higher CMB acoustic peaks, increasing \mnu slightly sharpens these peaks.

The above discussion makes it clear that LSS clustering measurements are a promising probe of neutrino masses, and advances in the field lead us to expect that the strongest bounds on \mnu will soon come from datasets probing the imprint on \mnu on structure growth, rather than on the background evolution~\cite{abazajian2016cmbs4,Oyama_2016,Sprenger_2019}. The neutrino-induced suppressed structure growth is most cleanly imprinted on the small-scale amplitude of the matter power spectrum, \pmmnobm: the latter can be measured indirectly through LSS tracers such as galaxies~\cite{Bernardeau_2002,Desjacques_2018_review}, or through the gravitational lensing of the CMB~\cite{Bernardeau_2002,LEWIS_2006} or of background galaxies~\cite{Hildebrandt_2016,Abbott_2021_DESy3}. Here, we shall mainly focus on galaxies as LSS tracers, and consider full-shape (FS) measurements of the galaxy power spectrum. FS galaxy power spectrum analyses do not come without significant challenges. One important challenge is related to our limited ability to model the underlying matter power spectrum in the mildly non-linear regime. Another related challenge pertains to the fact that galaxies are biased tracers of the underlying matter distribution, and therefore do not faithfully trace the latter. The statistical relation between the galaxy and matter overdensity fields is encapsulated in the galaxy bias parameter(s), which ultimately capture complexities associated to galaxy formation and evolution.

While galaxy power spectrum analyses are undoubtedly challenging, and have often been performed in the context of large collaborations (see e.g.~\cite{Gil-Marin:2014baa,BOSS:2016teh,BOSS:2016chr,BOSS:2016off,BOSS:2016hvq,BOSS:2016psr,Gil-Marin:2018cgo,Gil-Marin:2020bct,Semenaite:2021aen,Neveux:2022tuk}), various works in recent years have attempted to extract information on cosmological parameters from these types of measurements, adopting different theoretical models. Among these we mention the Effective Field Theory of LSS (EFTofLSS), a formalism allowing to predict the clustering of the LSS in the mildly non-linear regime in a robust symmetry-driven way (see e.g.~\cite{Baumann:2010tm,Carrasco:2012cv,Pajer:2013jj,Senatore:2014via,Senatore:2014eva}). Recent advances in the EFTofLSS have allowed for applications on real data from the BOSS survey, with extremely promising results (see e.g.~\cite{DAmico:2019fhj,Ivanov_2020,Colas:2019ret,Ivanov:2019hqk,Philcox:2020vvt,DAmico:2020kxu,Nishimichi:2020tvu,Chudaykin:2020aoj,Ivanov:2020ril,DAmico:2020ods,Philcox:2020xbv,Wadekar:2020hax,Chudaykin:2020ghx,DAmico:2020tty,Ivanov:2021zmi} for examples in these directions).~\footnote{See instead~\cite{Philcox:2021ukg,Philcox:2021hbm,Ivanov:2021kcd,Philcox:2021kcw,Cabass:2022wjy,DAmico:2022gki,Cabass:2022ymb,Philcox:2022frc,Hou:2022wfj,Philcox:2022hkh,DAmico:2022osl} for applications of the EFTofLSS and related models with an eye to analyses of the redshift-space bispectrum as well as higher-order correlators.}

Other works have applied related perturbation theory-based models (see e.g.~\cite{Chen:2020zjt,Chen:2021wdi}) for the redshift-space galaxy power spectrum to real data~\cite{Upadhye:2017hdl,Loureiro:2018qva} or forecasts~\cite{Boyle:2017lzt,Boyle:2018rva,Chudaykin_2019,Boyle_2020,Donald-McCann:2021nxc,Donald-McCann:2022pac}, tested these models on simulations~\cite{Angulo:2015eqa,Eggemeier:2020umu,Eggemeier:2021cam,pezzotta2021testing}, or investigated other ways of extracting (possibly compressed) information from the redshift-space FS galaxy power spectrum and higher order multipoles~\cite{Gualdi:2017iey,Gualdi:2019ybt,Gualdi:2018pyw,Gualdi:2019sfc,Brieden:2021edu,Brieden:2021cfg,Valogiannis:2021chp,Gualdi:2022kwz,Gil-Marin:2022hnv,Brieden:2022ieb,Brieden:2022lsd,Valogiannis:2022xwu}. Finally, various works have used simulations to investigate the imprint of neutrino properties on the clustering of the LSS (see for e.g.~\cite{Carbone:2016nzj,Ruggeri:2017dda,Villaescusa-Navarro:2017mfx,Zennaro:2017qnp,Bird:2018all,2019MNRAS.489.5938Z,2019arXiv191004255G,Massara:2020pli,Euclid:2020rfv,Bose:2021mkz}), as well as related cosmological observables (see e.g.~\cite{Li:2018owg,Coulton:2018ebd,Uhlemann:2019gni,Kamalinejad:2020izi,Boyle:2020bqn,ChengChengSiHao:2021hja,Aviles:2021que,Zhou:2021sgl,MoradinezhadDizgah:2021upg,Kamalinejad:2022yyl,Liu:2022vtr,Lee:2022lbu,Ryu:2022npy}).

The strength and main advantage of several of these well-motivated first-principles theoretical models is that they are able to take into account and separate at the modeling level the effects of various different aspects pertaining to galaxy formation and biasing, with contributions to the redshift-space galaxy power spectrum captured by different operators and/or different galaxy bias parameters. At the same time, this strength comes at the cost of several extra nuisance parameters to be marginalized over, which can be problematic if the level of precision of the data is not sufficient as to be able to constrain them in a meaningful way, effectively leading to saturation of cosmological constraints. One may in principle attempt to exploit theory- or simulations-based relations to enforce prior relations between the various nuisance parameters, or set informative priors on some of them, as is routinely done in order to speed up analyses. However, following this route requires (more or less explicitly) making the assumption that one can indeed reliably model galaxy formation in the mildly non-linear regime, and any incorrect assumption in these assumed relations will propagate (as a modeling systematic) to the inferred cosmological parameters.

One important example in this sense is the EFTofLSS, the most general, symmetry-driven model for the mildly non-linear clustering of biased LSS tracers, which integrates out the complex and poorly-known details of short-scale physics by parametrizing these through a series of counterterms with functional form fixed by symmetry considerations, and amplitudes which are effectively treated as nuisance parameters to be fit to the data (see e.g.~\cite{Cabass:2022avo} for a recent review). The state-of-the-art implementation of the EFTofLSS to model the multipoles of the full-shape redshift-space galaxy power spectrum in the \texttt{CLASS-PT} Boltzmann solver~\cite{Chudaykin:2020aoj} introduces up to 11 nuisance parameters per galaxy sample.~\footnote{These parameters are: linear bias $b_1$, quadratic bias $b_2$, (quadratic) tidal bias $b_{{\cal G}_2}$, third-order tidal bias $b_{\Gamma_3}$, $k^2$ counterterms for the monopole, quadrupole, and hexadecapole $c_0$, $c_2$, and $c_4$, $k^4$ Fingers-of-God counterterm $\tilde{c}_{\nabla^4_\mathbf{z}\delta}$, shot noise parameter $P_{\rm shot}$, and scale-dependent shot noise parameters $a_0$ and $a_2$. The theoretical modeling of higher-order correlators such as the bispectrum requires the introduction of up to an order of magnitude more nuisance parameters.} In practice, however, (physically motivated) Gaussian priors need to be imposed on most of these nuisance parameters in order to aid the convergence of analyses: see e.g. Eqs.~(6.4,6.5) in~\cite{Chudaykin:2020aoj}, Eqs.~(D.1,D.2) in~\cite{Ivanov:2021kcd}, and Eqs.~(11,12) in~\cite{Philcox:2021kcw}. This effectively suggests that current galaxy clustering data may not yet be sufficiently precise to meaningfully constrain a large number of nuisance parameters. This highlights an interesting trade-off between model complexity and data precision, already appreciated in earlier works (see also~\cite{Modi:2017wds}).

An alternative possibility, explored by some of us in recent years (see e.g.~\cite{Giusarma:2016phn,Vagnozzi:2017ovm,Giusarma:2018jei}), is instead to adopt a minimal, phenomenological, yet still physically motivated theoretical (bias) model, at least as long as the (limited) precision of current data allows. This should be sufficiently precise for the purposes of current data, but not overly complex to the point that the associated nuisance parameters cannot be meaningfully constrained by data. Of course, as future data becomes more precise, such a model should be refined with the introduction of additional ingredients and nuisance parameters, and at a certain point adopting well-motivated first-principles approaches such as the EFTofLSS becomes inevitable.

For the underlying matter power spectrum, we adopt a linear model with non-linear corrections on top following the \texttt{HALOFIT} prescription~\cite{Bird_2012}, while also including effects due to survey geometry, linear redshift-space distortions, the Alcock-Paczynski effect, and a phenomenological modeling of non-linear redshift-space distortions (Fingers-of-God). As done by some of us in~\cite{Giusarma:2018jei}, we go beyond the simple large-scale linear bias model by considering the leading scale-dependent correction, scaling as $k^2$. This scale-dependent bias model is simple but strongly motivated from various independent theoretical approaches, including but not limited to peaks theory~\cite{Desjacques_2010,Schmidt_2013}, the excursion-set approach~\cite{Musso_2012}, and standard perturbation theory~\cite{Senatore_2015,Desjacques_2018_review}. We do not make prior assumptions on the relations between the bias parameters, but leave these to be constrained by the data. Similar simplified models, in some cases including an extra $k^4$ bias term, have been used in other recent works (see e.g.~\cite{Raccanelli:2017kht,Chiang:2018laa,Valcin:2019fxe}).

Moreover, in addition to galaxy clustering data, we use measurements of the cross-correlation between the CMB lensing convergence and galaxy overdensity fields (CMB lensing-galaxy cross-correlation in short). Joint analyses of galaxy clustering and CMB lensing-galaxy cross-correlation measurements can help constrain the linear galaxy bias parameter (as it enters differently in the two measurements), which in turn is beneficial for improving constraints on \mnu, as demonstrated in~\cite{Giusarma:2018jei}. We also note that cross-correlations between CMB lensing and tracers of the LSS (including galaxies, but also other tracers such as galaxy clusters, quasars, filaments, and galaxy groups) have recently been considered in a huge number of works, including but not limited to~\cite{Sherwin:2012mr,ACT:2014ivk,Baxter:2014frs,ACT:2015eyl,Bianchini:2015yly,DES:2017fyz,He:2017owu,Raghunathan:2017qai,Han:2018izq,DES:2018miq,Singh:2018kmr,Marques:2019aug,Krolewski:2019yrv,Alonso:2020jcy,ACT:2020izl,Hang:2020gwn,Kitanidis:2020xno,Lin:2020sbb,Krolewski:2021yqy,Dong:2021nmq,Sun:2021rhp,White:2021yvw,Chen:2022itx,Kusiak:2022xkt,DES:2022ign,Chen:2022jzq}.

Our aim in the present work is to revisit and improve the analysis performed by some of us in~\cite{Giusarma:2018jei}, which found that \clkgnobm had a small but not insignificant impact on the bound on the neutrino mass, and highlighted the importance of moving beyond the simplified scale-independent bias model adopted previously. We go beyond the earlier work of~\cite{Giusarma:2018jei} in various respect, which include but are not limited to the following:

\begin{itemize}
\item accounting for the reduction of power in the matter-galaxy cross-power spectrum caused by the decorrelation between galaxy and matter fields (see Sec.~\ref{subsec:observables});
\item accounting for the non-negligible cross-covariance between galaxy-galaxy and galaxy-matter power spectra, and therefore between galaxy clustering and CMB lensing-galaxy cross-correlation measurements, which had previously been neglected, although we find a posteriori that the effect of including the cross-covariance is very small (see Sec.~\ref{sec:model-fitting}).
\item updating CMB data from \textit{Planck} 2015 to the \textit{Planck} 2018 legacy data release~\cite{Planck_2018_cosmoparams};
\item investigating and testing in more detail the robustness of the underlying theoretical model and whether the resulting bounds of \mnu are competitive with the bounds gained obtained from a combination of CMB and BAO (non-FS) measurements.
\end{itemize}    
With these improvements, we find that the addition of \clkgnobm has a negligible impact with current data, while it will become important in future datasets. The reason why it does not appear to be important in current datasets (contrary to the findings in~\cite{Giusarma:2018jei}) is the use of the updated \textit{Planck} observations (including importantly small-scale polarization data), which by themselves significantly reduce the error on \mnu (implicitly putting much stronger requirements on other datasets, or equivalently reducing the benefits of including additional datasets), and of the relatively low signal-to-noise of current \clkgnobm measurements. Nonetheless, we expect that the inclusion of \clkgnobm (as well as the adoption of a flexible, simple, but well-motivated galaxy bias model) will be very important when considering upcoming galaxy clustering data, e.g.\ from Euclid~\cite{Amendola:2016saw} or DESI~\cite{2016arXiv161100036D}. We also stress that the scale-dependent bias we are considering here is relevant on small scales (large $k$), and is totally distinct from the spurious scale-dependence studied by some of us in~\cite{Vagnozzi:2018pwo}, appearing on large scales (small $k$) in the presence of massive neutrinos if the bias itself is not correctly defined (see also~\cite{Castorina:2013wga,Raccanelli:2017kht,Fidler:2018dcy,Valcin:2019fxe}).

The rest of this work is then structured as follows. In Sec.~\ref{sec:theory} we review signatures of \mnu in galaxy clustering measurements, and issues pertaining to modeling the scale-dependent galaxy bias in auto- and cross-correlation measurements. In Sec.~\ref{sec:model-datasets} we discuss the adopted observational datasets, theoretical modeling thereof, and analysis methodology. In Sec.~\ref{sec:results} we discuss the resulting constraints on \mnu and investigate the robustness of the underlying theoretical model. Finally, in Sec.~\ref{sec:conclusions} we provide concluding remarks. 

\section{Massive neutrinos and large-scale structure: theory and modeling}
\label{sec:theory}

In this section, we first review the physical imprints of massive neutrinos on the clustering of the large-scale structure (LSS). We follow this up by a description of the scale-dependent galaxy bias model we adopt, before discussing in more detail our modeling of the theoretical (redshift-space) galaxy power spectrum.

\subsection{Signatures of massive neutrinos in LSS data}
\label{subsec:signatures-of-mnu-in-data}

Neutrinos decouple from the primordial plasma at a temperature of ${\cal O}({\rm MeV})$, when they were highly relativistic. While ultra-relativistic, neutrinos are unable to cluster on scales smaller than their free-streaming wavenumber, $k \gg k_{\rm fs}$, as their large thermal velocities prevent them from falling within gravitational potential wells. This leads to a well-known small-scale suppression of structure growth, which is one of the most distinctive signatures of neutrino masses in cosmological observations~\cite{Bond_1980}.

In our work, as discussed in more detail later in Sec.~\ref{sec:model-datasets}, we shall adopt the so-called degenerate approximation wherein the total neutrino mass \mnu is equally distributed among the three mass eigenstates. This approximation is extremely robust given the sensitivity of current cosmological observations, which are only sensitive to the neutrino mass sum rather than the masses of the individual eigenstates (see e.g.~\cite{Gerbino:2016sgw,Archidiacono:2016lnv,Archidiacono:2020dvx}). At the time when neutrinos are ultra-relativistic, the free-streaming wavenumber $k_{\rm fs}$ is roughly equal to the inverse of the Hubble horizon scale. This implies that during this regime neutrinos do not contribute to clustering on any physical scale. However, at late times massive neutrinos transition to becoming non-relativistic. At this point, $k_{\rm fs}$ starts growing slower than the horizon scale. Therefore, on scales $k \ll k_{\rm fs}$, the massive neutrino eigenstates behave as a cold dark matter component, and are able to cluster. On the other hand, for smaller scales, $k \gg k_{\rm fs}$, the massive neutrino eigenstates still cannot cluster, and structure formation is suppressed. As a result, matter perturbation modes entering the horizon after the non-relativistic transition evolve without ever experiencing free-streaming due to the massive neutrino eigenstates. These effects lead to a characteristic step-like suppression in power on scales below the free-streaming wavenumber of neutrinos at their non-relativistic transition ($k_{\rm nr}$), which is given by (see e.g.~\cite{LESGOURGUES_2006}):
\begin{align}
k_{\rm nr} = 0.0178\Omega_\text{m}^{1/2} \left (\frac{\Sigma m_\nu}{\rm eV} \right ) ^{1/2}\,h{\rm Mpc}^{-1}\,,
\end{align}
where $\Omega_\text{m}$ is the present matter density parameter. Besides suppressing power on small scales, massive neutrinos also slow down the growth of matter perturbations $\delta(a)$, with $a$ the scale factor. In the presence of massive neutrinos, matter perturbations grow at a rate $\propto a^{1-3f_{\nu}/5}$ rather than $\propto a^1$, where $f_{\nu} \equiv \Omega_{\nu}/\Omega_{\rm m}$ is the fractional neutrino contribution to the matter density, and similarly $\Omega_{\nu}$ is the density parameter of (massive) neutrinos.

In the linear regime and at redshift $z=0$, the relative suppression of the matter power spectrum $P^{\rm mm}(k)$ due to the presence of massive neutrinos has a characteristic step-like/``kink'' feature, and saturates at~\cite{Hu_1998}:
\begin{align}
\frac{P^\text{mm}(k)^{f_\nu}}{P^\text{mm}(k)^{f_\nu=0}}\approx 1-8f_\nu\;\;\;\;\;\;\;\;\;\;(k\gg k_{\rm nr}, z=0)\,.
\label{eq:pmm-suppression}
\end{align}
In the non-linear regime, the suppression actually saturates at $\simeq -10f_{\nu}$, a result which has been independently confirmed by means of N-body simulations (see e.g.~\cite{Brandbyge:2008rv,Brandbyge:2008js,Brandbyge:2009ce,Viel:2010bn,Ali-Haimoud:2012fzp,Castorina:2015bma,Banerjee:2016zaa,Liu:2017now,Banerjee:2018bxy,Partmann:2020qzb}), as well as higher-order perturbative calculations (see e.g.~\cite{Wong:2008ws,Lesgourgues:2009am,Blas:2014hya,Fuhrer:2014zka,Hannestad:2020rzl}). This step-like suppression is a special feature of massive neutrinos not easily mimicked by other cosmological parameters or systematic effects.~\footnote{Note that light but massive relics (such as certain eV-scale relics) which become non-relativistic during radiation domination (unlike neutrinos which become non-relativistic during matter domination) lead to a similar feature, which however saturates at $-14f_X$, with $f_X$ the fractional light relic contribution to the matter density (see for instance~\cite{Viel:2005qj,DePorzio:2020wcz,Xu:2021rwg}).} Therefore, a possible way of constraining \mnu from LSS clustering data is to measure the full-shape power spectrum of LSS tracers (such as galaxies) for wavenumbers around $k_{\rm nr}$.

However, besides its amplitude and shape, LSS full-shape power spectrum measurements also contain precious geometrical information which helps pinning down \mnu. In particular, the position of the Baryon Acoustic Oscillation (BAO) wiggles in $k$-space is directly related to the ratio $r_s/D_V$, with $r_s$ the sound horizon at baryon drag, and $D_V$ the volume-averaged distance to the effective redshift of the sample of LSS tracers whose power spectrum is being measured (see e.g. recent discussions on this point, and more generally on the information content of full-shape power spectrum measurements, in~\cite{Ivanov_2020,Chudaykin:2020ghx,Vagnozzi:2020rcz}). This geometrical information is crucial in breaking the ``geometrical degeneracy''~\cite{Bond:1997wr,Zaldarriaga:1997ch,Efstathiou:1998xx}, which in the context of a spatially flat Universe involves the matter density parameter $\Omega_{\rm m}$ and the Hubble constant $H_0$, various combinations of which lead to a virtually identical CMB power spectrum.~\footnote{For more recent detailed discussions on the geometrical degeneracy and implications for parameter estimation, as well as different ways of breaking it with additional late-time datasets, see for instance~\cite{Efstathiou:2020wem,Vagnozzi:2020dfn,Dhawan:2021mel,Khadka:2020vlh,Khadka:2020hvb,Khadka:2020tlm,Khadka:2021vqa,Khadka:2021ukv,Cao:2021ldv,Cao:2021cix,Cao:2022wlg}.} The geometrical information contained in the full-shape power spectrum is helpful in breaking the geometrical degeneracy as it helps pinning down both $H_0$ and $\Omega_{\rm m}$ (by excluding extremely low/high values of $H_0$ or $\Omega_{\rm m}$ respectively, which would otherwise be tolerated by CMB data alone), and thereby improving constraints \mnu which, we recall, contributes to $\Omega_{\rm m}$ at late times: see for instance~\cite{Ivanov_2020,Vagnozzi:2020rcz} for related discussions on these aspects.

\subsection{Galaxy bias}
\label{subsec:observables}

Although the LSS neutrino mass signature is cleanly imprinted in the matter power spectrum $P^{\rm mm}(k)$, this is not a directly observable quantity, as we cannot directly observe the clustering of the matter field, but only that of its luminous tracers, such as galaxies. Galaxies (and other tracers) are biased tracers of the underlying matter density field: their clustering properties are related, but not identical, to those of the matter field. On large scales, the relation between matter overdensity $\delta_{m}$ and galaxy overdensity $\delta_{g}$ is fully deterministic and can be captured via a linear relation~\cite{Kaiser_1984}:
\begin{align}
\delta_g = b\delta_m\,,
\label{eq:bias}
\end{align}
where the proportionality factor $b$ is a constant referred to as (linear) galaxy bias. We stress that Eq.~(\ref{eq:bias}) is valid only on sufficiently large, linear scales, where gravitational interactions are dominant. The exact value of the galaxy bias varies depending on the properties of the galaxy sample of interest and is therefore generally treated as a nuisance parameter which is subsequently marginalized over: see~\cite{Desjacques_2018_review} for a recent very complete review on galaxy bias. We stress that, in the presence of massive neutrinos, the linear galaxy bias is scale-independent only if the bias is defined with respect to the dark matter-plus-baryons field rather than the total matter field (the latter including massive neutrinos). Nonetheless, this distinction is not important given the sensitivity of current galaxy clustering data, and will only become relevant with upcoming data (e.g.\ from Euclid or DESI), as discussed for instance in~\cite{Castorina:2013wga,Raccanelli:2017kht,Vagnozzi:2018pwo,Fidler:2018dcy,Valcin:2019fxe}.

The validity of the linear bias model breaks down as we enter mildly non-linear scales ($k \gtrsim {\cal O}(0.1)$~\hmpc at $z=0$), where complications associated to galaxy formation become increasingly relevant over gravitational interactions. As a result, the biasing relation between the galaxy and matter fields, while still deterministic, is expected to become scale-dependent (see e.g.~\cite{Sheth:1999mn,Seljak:2000jg,Schulz:2005kj} for early seminal works in this direction). This can be understood in terms of an expansion of the galaxy overdensity field in higher-order spatial derivatives of the matter overdensity field $\nabla^{(n)} \delta_\text{m}$, which in Fourier space is simplified to an expansion in factors of $k^{n}$:
\begin{align}
\delta_\text{g}=\Big(b_{\rm lin}+b_{k^2}k^2\Big)\delta_\text{m} +{\cal O}(k^4)\,.
\label{eq:scale-dep-bias-exp}
\end{align}
Note that odd powers of $k$ are excluded on the basis of statistical isotropy and the equivalence principle~\cite{Matsubara:2011ck,Desjacques_2018_review}.

We note that the linear-plus-$k^2$ galaxy bias parametrization in the mildly non-linear regime we will adopt is minimal, yet highly motivated from different independent theoretical frameworks, including but not limited to peaks theory~\cite{Desjacques_2010,Schmidt_2013}, the excursion-set approach~\cite{Musso_2012}, and standard perturbation theory~\cite{Senatore_2015,Desjacques_2018_review}. Within the EFTofLSS, the $k^2$ correction can be viewed as the leading order counterterm to the redshift-space monopole, with the associated coefficient being a generalization of the real-space dark matter sound speed (more concretely, see for instance Eqs.~(2.7,2.15,2.23) of~\cite{Chudaykin:2020aoj}).

Importantly, we also note that other works have adopted similar simplified models, without the inclusion of a $k^4$ term, which could further improve the fit (see e.g.~\cite{Raccanelli:2017kht,Chiang:2018laa,Valcin:2019fxe}). A more physically motivated model has also been proposed by~\cite{Saito:2014qha}, and includes non-local bias terms up to third-order in the density field (see e.g. Eq.~(3.3) of~\cite{Valcin:2019fxe}). Recent work by~\cite{Desjacques_2018_1loop} instead computed the complete expression for the redshift-space galaxy power spectrum up to 1-loop order, which includes 28 independent loop integrals and 5 additional free parameters, and the same has been done within the context of the EFTofLSS. In general, on sufficiently small scales, various bias contributions enter in such a way that the relation between the redshift-space galaxy power spectrum and the underlying linear matter power spectrum is no longer a direct (albeit scale-dependent) proportionality such as in Eq.~(\ref{eq:scale-dep-bias-exp}), and introduce several additional nuisance parameters. In this work, our aim is instead to adopt a minimal yet physically motivated bias model going beyond linear bias which, while simplified, is still sufficiently useful given the precision of current data, provided the analysis is limited to sufficiently large scales.

We note that the amplitude of the large-scale full-shape galaxy power spectrum scales as $b_{\rm lin}^2$, and more precisely depends on the combination $b_{\rm lin}^2\sigma_8^2$, where $\sigma_8$ is the present day linear theory amplitude of matter fluctuations averaged in spheres of radius $8\,h^{-1}{\rm Mpc}$. It is therefore clear that jointly fitting another observable which carries a different functional dependence on $b_{\rm lin}$ (and $\sigma_8$) can significantly help improving cosmological parameter constraints obtained from galaxy clustering measurements (and possibly help break the $b_{\rm lin}-\sigma_8$ degeneracy). To this end, we shall include an observable which is sensitive to the matter-galaxy cross-spectrum $P^{\text{mg}}(k,z)$: as $P^{\text{mg}}(k,z)$ only correlates one power of the galaxy density field, its large-scale amplitude scales as $b_{\rm lin}^1\sigma_8^2$. As we will discuss later in this Section, we use measurements of the angular cross-spectrum between the CMB lensing convergence and the galaxy overdensity field $C_{\ell}^{\kappa g}$, see Eq.~(\ref{eq:clkgth}), connected to the matter-galaxy cross-spectrum.

We adopt the following linear-plus-$k^2$ scale-dependent galaxy bias model for measurements of the galaxy-galaxy (auto) and galaxy-matter (cross) power spectra (where, for notational simplicity, all redshift dependencies will be omitted henceforth):
\begin{alignat}{1}
P^{\text{gg}}(k) & = b_{\rm auto}(k)^2P^{\text{mm}}(k) \approx \Big(b_{\rm lin}+b_{k^2\text{auto}}k^2\Big)^2 P^{\text{mm}}(k)\,,
\label{pggideal}\\
P^{\text{mg}}(k) & = b_{\rm cross}(k) P^{\text{mm}}(k) \approx \Big(b_{\rm lin}+b_{k^2\text{cross}}k^2\Big)\: P^{\text{mm}}(k)\,,
\label{pmgideal}
\end{alignat}
where we have assumed that the scale-dependent bias parameters in the auto-power ($b_{k^2\text{auto}}$) and cross-power ($b_{k^2\text{cross}}$) spectra are not necessarily identical. This assumption is supported by theoretical predictions from~\cite{Desjacques_2018_1loop}, and evidence from N-body simulations carried out in~\cite{Okumura:2012xh} and~\cite{Modi_2020}. The theoretical predictions from~\cite{Desjacques_2018_1loop} foresee that potential velocity bias contributions -- e.g. those arising from galaxy formation effects or baryonic pressure perturbations -- would affect the mapping between redshift- and rest-frame. As such, contributions from a velocity bias would not enter into the cross-power spectra, but only into the auto-power spectra (scaled as $\propto k^2$).~\footnote{An expression of the cross-power spectra using the framework of~\cite{Desjacques_2018_1loop} is missing therein but can be retrieved from e.g.~\cite{Boyle_2020}.}

As for the evidence from simulations,~\cite{Okumura:2012xh} and~\cite{Modi_2020} clearly show that, as $k$ is increased, $db_{\rm cross}(k)/dk>0$ and $db_{\rm auto}(k)/dk<0$. Physically speaking this different behavior can be explained as follows. The small-scale matter-galaxy cross-correlation function in real space traces the density profile of host halos~\cite{Hayashi:2007uk}, and therefore increases on small scales, which in Fourier space translates into $b_{\rm cross}$ increasing with increasing $k$. On the other hand, halos are extended objects, which cannot overlap in the initial Lagrangian space~\cite{Sheth:1998xe}. This halo exclusion principle places strong constraints on the small-scale behavior of the real-space galaxy 2-point correlation function, which has to approach $\xi(r) \to -1$ on sufficiently small scales. In Fourier space, this requirement translates into $b_{\rm auto}$ decreasing with increasing $k$. We refer the reader to~\cite{Giusarma:2018jei} for further discussions on these two different behaviors.

\begin{figure}
\includegraphics[width=\columnwidth]{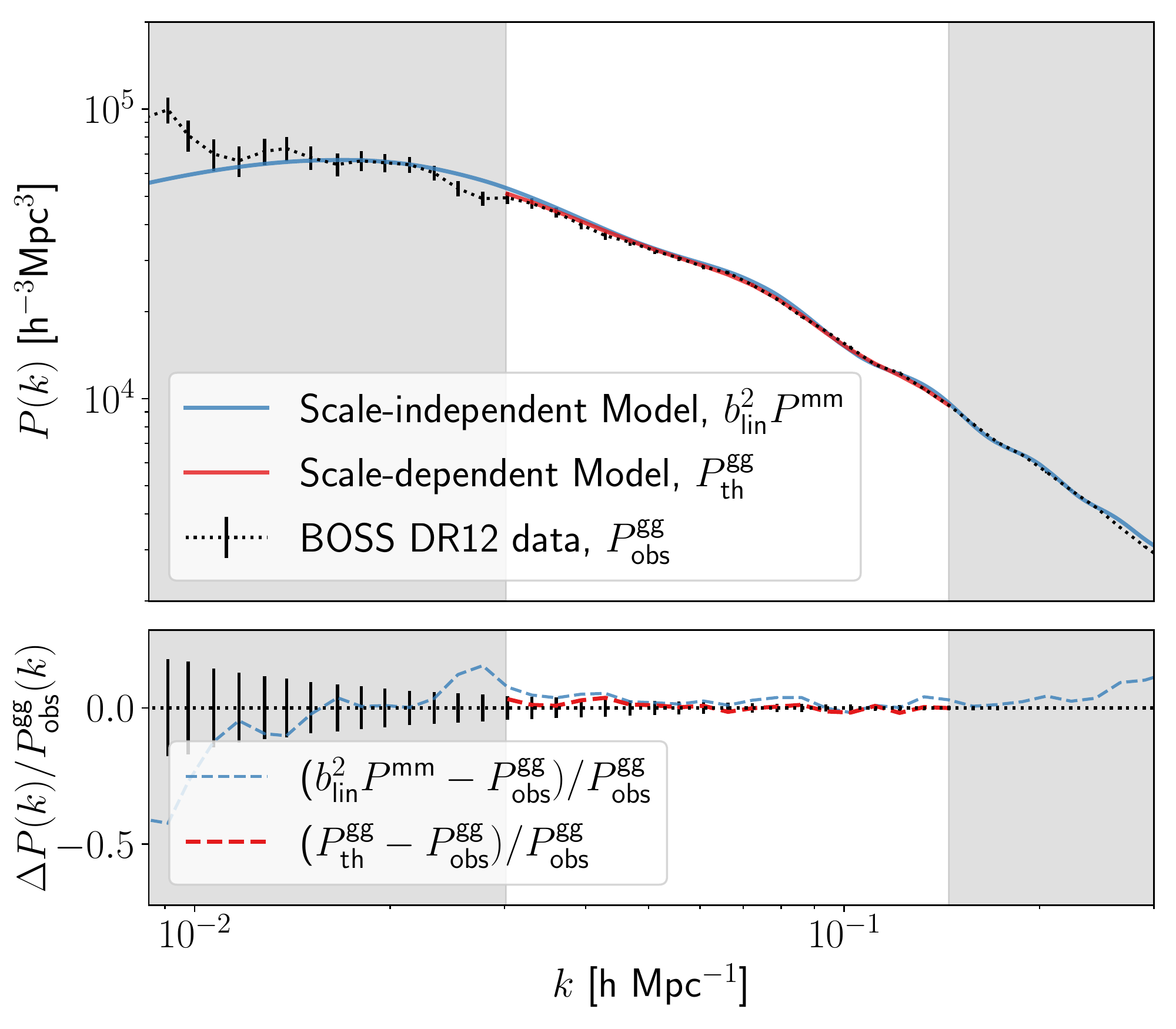}
\caption{\textit{Upper panel}: measured monopole of the BOSS DR12 CMASS power spectrum (\textit{black}), alongside the best-fit theoretical prediction from the model used in this work, with a scale-independent (\textit{blue}) or scale-dependent (\textit{red}) bias. The \textit{white} $k$-band range represents the wavenumber range to which we limit our fit ($0.03<k/(h{\rm Mpc}^{-1})<0.13$ at $z=0.57$): we exclude the remaining range of wavenumbers (\textit{grey}) due to observational systematics (large scales, small $k$) or to avoid non-linearities (small scales, large $k$), as clarified in Sec.~\ref{subsec:datasets}. \textit{Lower panel}: residuals with respect to a fit adopting a scale-independent (\textit{blue}) or scale-dependent (\textit{red}) bias. The residuals clearly show the improvement in fit which follows from considering a scale-dependent bias. The effect of linear redshift-space distortions (Kaiser effect) is included, together with a phenomenological model for non-linear redshift-space-distortions (Fingers-of-God) which is modeled by following Eq.~(\ref{eq:FoG}).}
\label{fig:pkdr12}
\end{figure}

We have so far ignored stochastic contributions to the relation between $\delta_\text{g}$ and $\delta_\text{m}$. We refer to stochastic contributions as being those which are independent of the matter density field~\cite{Bernardeau_2002}. The largest stochastic contribution relevant to our work arises from the fact that the tracers we are using to sample the underlying matter density field, namely galaxies, are discrete rather than continuous. In the simplest scenario, this leads to the appearance of a Poisson noise term, $1/\bar{n}$, where $\bar{n}$ is the mean number density of galaxies in our sample. The second stochastic contribution that we consider emerges from the fact that processes associated to galaxy formation eventually lead to decorrelations between galaxy and matter density fields. This results in the presence of small-scale fluctuations which are decorrelated from (and thus largely independent of) the large-scale fluctuations \cite{pezzotta2021testing}. This noise and the Poisson noise enter the matter power spectrum as additive terms (to leading order), and therefore are mutually degenerate. We therefore include an effective stochastic parameter, labeled \pshot, in the auto-power spectrum, as follows:
\begin{alignat}{1}
P^{\text{gg}}(k) &\approx \Big(b_{\rm lin}+b_{k^2\text{auto}}k^2\Big)^2 P^{\text{mm}}(k) + \pshot\,.
\label{pgg+pshot}
\end{alignat}
In our later discussion, we normalize \pshot to the fiducial Poisson noise of our galaxy sample $1/\bar{n}$, where the average galaxy number density for the sample we are considering is $\bar{n} \simeq 3 \times 10^{-4}\,h^3{\rm Mpc}^{-3}$~\cite{reid2015sdssiii}. Therefore, if \pshot deviates from unity, it should be interpreted as the presence of non-Poissonian noise. We have only considered scale-independent components, although in principle the stochastic contributions may be expanded similarly to the deterministic components of the galaxy bias in Eq.~(\ref{eq:scale-dep-bias-exp}). We are assuming that scale-dependent stochastic contributions are negligible on the scales of interest (i.e. $k<0.13$\hmpc), an assumption which is supported by earlier findings (see for instance~\cite{Desjacques_2010,Baldauf_2013,Modi:2016dah,Eggemeier:2021cam,pezzotta2021testing}). As a consistency check, we verified that this scale-dependent model improves the residual fit to the observed galaxy power spectrum as compared to a scale-independent model (only involving the linear galaxy bias), as we see in Fig.~\ref{fig:pkdr12}.

Note that \pshot leaves the matter-galaxy cross-power spectrum (directly) unchanged. The underlying reason is that any stochastic contributions are completely independent of the matter field by definition, and thus can only show up in the auto-power spectrum. Nevertheless, the matter-galaxy cross-power spectrum can still be indirectly affected by stochastic contributions, in a way which can instead be captured by the cross-correlation coefficient:
\begin{align}
r(k,z) &= \frac{P^{\text{mg}}(k,z)}{\sqrt{P^{\text{gg}}(k,z)P^{\text{mm}}(k,z)}}\,,
\label{eq:cross-corr}
\end{align}
which quantifies the loss of information caused by scatter in the $\delta_\text{g}-\delta_\text{m}$ relation~\cite{Matsubara_1999}. This scatter may originate from other effects besides those associated to the stochastic components, such as differing values of the scale-dependent bias parameters $b_{k^2\text{auto}}$ and $b_{k^2\text{cross}}$, as well as redshift-space distortions (RSD), which are discussed later on in this section.

To include the cross-correlation coefficient, we rescale the matter-galaxy cross-power spectrum as follows:
\begin{align}
P^{\text{mg}}(k,z) &\rightarrow r(k,z) P^{\text{mg}}(k,z)\,,
\end{align}
By including \rk, the cross-power spectrum is damped even on linear scales (where $b_{k^2}k^2\approx0$) as a result of \rk not converging to unity but to a value that is proportional to the relative size between the linear galaxy auto-power spectrum and the stochastic component $[{\approx1- \pshot/(2\abias^2P^\text{mm}(k))}]$. On smaller scales, the cross-power spectrum experiences a larger amount of damping as the contribution of the scale-dependent galaxy bias parameter grows. We show the behaviour of \rk in Fig.~\ref{fig:rk} of Appendix~\ref{app:cross-corr}, which displays all the aforementioned effects. 

Having now defined our theoretical galaxy bias model in auto- and cross-correlation measurements, we discuss our modeling of the observed redshift-space galaxy power spectrum and the angular cross-spectrum between the CMB lensing convergence and the galaxy overdensity field. The galaxy auto-power spectrum is observed from three-dimensional galaxy clustering data, and therefore in so-called redshift space, since distances along the third dimension are computed from the observed redshift assuming a fiducial cosmology. It is thus sensitive to peculiar velocities along the line-of-sight. On linear scales, peculiar velocities are dominated by the coherent motion of galaxies falling into the gravitational wells of overdense regions, an effect commonly referred to as the ``Kaiser effect'', or linear RSD. The Kaiser effect induces a dependency on the line-of-sight angle for the otherwise isotropic galaxy power spectrum, the strength of this line-of-sight angle dependency being connected to the infall rate of the galaxies. To model this effect, we assume that on large scales the coherent motion of galaxies is described by linear perturbation theory, which makes the distortion proportional to the linear growth rate of structures~\cite{Kaiser_1987}. In the following, we shall only be interested in the monopole of the full-shape power spectrum, i.e. the angle-averaged (spherically averaged) power spectrum. This average results in the dependency on the line-of-sight angle being integrated out, so that the effect of linear RSD is captured by a growth rate-dependent global enhancement of the power spectrum. The linear RSD-corrected galaxy power spectrum monopole is given by:
\begin{align}
P^{\rm gg}_{\rm th}(k) &= b_{\rm auto}^2(k)\Big(1+\frac{2}{3}\beta(k) + \frac{1}{5}\beta^2(k)\Big)P^{\rm cb}(k) + P_{\rm shot}\,,
\label{eq:pgg+pshot+rsd}
\end{align}
where $b_{\rm auto}^2(k)=b_{\rm lin}+b_{k^2{\rm auto}}k^2$ and $\beta=f/b_{\rm auto}(k)$, and $f$ is the linear growth rate of structure, which we approximate as $\Omega_{\rm m}(z)^{0.545}$, an approximation which is valid to very high accuracy under the assumption of general relativity and a cosmological constant as the dark energy component (see for instance~\cite{Lahav:1991wc,Linder_2007,Nesseris_2008}). Finally, $P^{\rm cb}$ is the cold dark matter-plus-baryons power spectrum. It is this quantity which appears in Eq.~(\ref{eq:pgg+pshot+rsd}) rather than $P^{\rm mm}$ as this choice has been shown to result in the linear galaxy bias being scale-independent and universal (independent of \mnu) on large scales, reflecting the fact that neutrinos do not participate in clustering on the scales relevant for galaxy formation~\cite{Castorina:2013wga,Raccanelli:2017kht,Vagnozzi:2018pwo,Fidler:2018dcy,Valcin:2019fxe}.~\footnote{Of course, on sufficiently large scales $P^{\rm cb}=P^\text{mm}$, as neutrinos behave as a cold dark matter component. Note that we do not model the residual neutrino-induced scale-dependent bias, also referred to as growth-induced scale-dependent bias, as the effect is too small to make a difference given the precision of current data~\cite{LoVerde:2014pxa,LoVerde:2014rxa,LoVerde:2016ahu,Munoz:2018ajr,Xu:2020fyg}.} At any rate, given the precision of current observational data, the distinction between $P^{\rm cb}$ and $P^{\rm mm}$ in Eq.~(\ref{eq:pgg+pshot+rsd}) is irrelevant, as shown in~\cite{Raccanelli:2017kht,Vagnozzi:2018pwo}, but this difference will become important for upcoming LSS surveys. We append the subscript ``${\rm th}$'' to \pggnobm to distinguish the theoretical and observed galaxy auto-power spectra.

On top of the large-scale infall described by the Kaiser effect, and captured by the $\beta$-dependent terms in Eq.~(\ref{eq:pgg+pshot+rsd}), one of the key non-linear RSD contributions arises from random peculiar velocities of galaxies which further distort small-scale clustering information in redshift space, an effect known as Fingers-of-God (FoG). The simplest modeling of FoG exponentially suppresses Eq.~(\ref{eq:pgg+pshot+rsd}) on small scales (see e.g.~\cite{Jackson:1971sky,Bull:2014rha}):~\footnote{For recent works modeling non-linear redshift-space distortions within the EFTofLSS (and potentially doing away with the need to model FoG), see for instance~\cite{Ivanov:2021fbu} and~\cite{2021arXiv211000016D}.}
\begin{align}
P^{\rm gg}_{\rm th}(k)-P_{\rm shot} \to \left ( P^{\rm gg}_{\rm th}(k)-P_{\rm shot} \right ) \exp \left [ (-k\sigma_{\rm FoG})^2 \right ] \,,
\label{eq:FoG}
\end{align}
where $\sigma_{\rm FoG}$ is the typical scale above which the power spectrum is suppressed (or equivalently, the suppression occurs above a typical wavenumber $k_{\rm FoG} \sim \sigma_{\rm FoG}^{-1}$). For the BOSS CMASS galaxy sample we are interested in and given the scale cuts we will apply (all of which will be discussed in Sec.~\ref{sec:model-datasets}), we expect the FoG contribution to be dominated by the virialized motions of satellite galaxies inside host halos and to be small, as shown for instance in~\cite{Baldauf_2013}.~\footnote{As shown in Fig.~5 of~\cite{Okumura:2015fga}, the contributions from central-central and central-satellite galaxy pairs to the multipole moments of the redshift-space galaxy power spectrum are negligible for $k \lesssim 0.2\,h{\rm Mpc}^{-1}$. In addition, BOSS CMASS galaxies are expected to have a low satellite fraction ($\lesssim 10\%$). These considerations lead us to expect that the FoG contribution to the BOSS CMASS galaxy power spectrum should be small.} Nonetheless, for completeness we test this expectation in our analysis, finding that it is met (see Sec.~\ref{subsubsec:fog}).

For the cross-power spectrum, we use measurements of the cross-correlation between \textit{Planck} 2015 CMB lensing (convergence) maps and BOSS DR12 galaxy overdensity maps, \clkgnobm \cite{Pullen_2016}. The CMB lensing convergence field is related to the integrated effect of the intervening matter between the last-scattering surface and us~\cite{LEWIS_2006}. The cross-correlation between CMB lensing convergence and galaxy overdensity reads~\cite{Bianchini_2015}:
\begin{align}
C^{\kappa g}_{\ell,\text{th}} = \int dz \frac{H(z)}{\chi^2(z)}W^{\kappa}(z)f^g(z)P^{\text{mg}}\bigg(k=\frac{\ell+ 1/2}{\chi(z)},z\bigg)\,,
\label{eq:clkgth}
\end{align}
where $H(z)$ is the Hubble parameter, $\chi$ is the comoving scale, and we apply the Limber approximation~\cite{LoVerde_2008}. $W^{\kappa}(z)$ is the lensing convergence kernel:
\begin{align}
W^{\kappa}(z) = \frac{3\Omega_{\text{m}}}{2}\frac{H_0^2}{H(z)}(1+z)\chi(z)\frac{\chi(z_{\text{CMB}})-\chi(z)}{\chi(z_{\text{CMB}})}\,,
\label{eq:wkappa}
\end{align}
where $H_0$ is the Hubble parameter today, and $z_{\text{CMB}}$ is the redshift of recombination. Lastly, in Eq.~(\ref{eq:clkgth}), $f^{g}(z)$ is the normalized redshift distribution of the galaxy overdensity maps:
\begin{align}
f^\text{g}(z) = \frac{dN/dz}{\int dz' dN/dz'}\,,
\label{eq:fg}
\end{align}
where $dN/dz$ is the redshift distribution of the galaxy sample. We neglect the effect of lensing magnification given that this effect is dependent on redshift and our redshift bin is fairly small~\cite{Raccanelli:2013gja,MoradinezhadDizgah:2016pqy}. Finally, we do not include relativistic effects, as these are only relevant on very large scales, beyond those probed here (see e.g.~\cite{Yoo:2010ni,Maartens:2012rh,Raccanelli:2015vla,Grimm:2020ays,Castorina:2021xzs}). Finally, to model non-linear corrections to the underlying matter power spectrum in the presence of massive neutrinos (which nonetheless are very small compared to current observational uncertainties on the scales we are interested in, see Fig.~1 of~\cite{Vagnozzi:2017ovm}) we make use of the \texttt{HALOFIT} prescription discussed in~\cite{Bird_2012}.

\section{Datasets and methodology}
\label{sec:model-datasets}

We consider a 7-parameter model, which extends the 6-parameter \lcdm model by allowing the sum of the neutrino masses \mnu to vary. The 7 free parameters we consider are then: the physical baryon and cold dark matter densities $\omega_\text{b} h^2$ and $\omega_\text{cdm} h^2$, the acoustic scale $\theta_s$, the optical depth to reionization $\tau$, the amplitude and tilt of the primordial scalar power spectrum $A_\text{s}$ and $n_\text{s}$, and finally the sum of the neutrino masses \mnu. Concerning the neutrino mass spectrum, we adopt the so-called degenerate approximation, wherein the total neutrino mass \mnu is equally distributed among the three mass eigenstates, each carrying an individual mass $m_{\nu_i} = \Sigma m_{\nu}/3$. Various works have argued that this approximation is extremely robust given the sensitivity of current cosmological observations, which are only sensitive to the neutrino mass sum rather than the masses of the individual eigenstates (see e.g.~\cite{Gerbino:2016sgw,Archidiacono:2016lnv,Archidiacono:2020dvx}). Prospects for distinguishing the normal and inverted orderings based on physical effects associated to the individual mass eigenstates (as opposed to overall parameter space volume effects) do not appear promising in near-future cosmological data (see~\cite{Archidiacono:2020dvx}).~\footnote{See however~\cite{Lesgourgues:2004ps,DeBernardis:2009di,Jimenez:2010ev,Hall:2012kg,Jimenez:2016ckl} for other works reaching slightly different conclusions.}

\subsection{Datasets}
\label{subsec:datasets}

We now discuss the datasets adopted, starting from galaxy clustering and CMB lensing-galaxy cross-correlation data:
\begin{itemize}
\item Angle-averaged (monopole moment) full-shape power spectrum of the BOSS DR12 CMASS galaxies, measured at an effective redshift $z_{\rm eff}=0.57$, as measured in~\cite{Gil-Marin:2015sqa}. We only use measurements within the wavenumber range $0.03 < k [\hmpc]< 0.13$. The choice of large-scale cutoff $k_{\min}=0.03\,\hmpc$ is dictated by the the fact that larger scales have significantly lower signal-to-noise ratio and are dominated by observational systematics (e.g.\ related to stellar density, seeing requirements, missing close-pairs, fiber collisions, and redshift failures, see~\cite{BOSS:2012coo}). The choice of small-scale cutoff $k_{\max}=0.13\,\hmpc$ is instead limited by the ability to reliably model non-linear effects (discussed further below). This dataset is referred to as $\boldsymbol{P^{{\rm gg}}_{{\rm obs}}(k)}$. For simplicity and especially for consistency and ease of comparison to our earlier related work~\cite{Giusarma:2016phn,Vagnozzi:2017ovm,Giusarma:2018jei}, here we have chosen not to include measurements of the quadrupole moment of the BOSS DR12 CMASS full-shape power spectrum.~\footnote{However, we note that the peculiar velocity information contained within the quadrupole through redshift-space distortions would help tightening parameter constraints compared to the monopole-only ones, as this information helps breaking the \abias-$\sigma_8$ degeneracy. Nonetheless, the inclusion of CMB lensing-galaxy cross-correlation measurements also helps breaking this degeneracy.}.
\item Measurements of the cross-correlation between CMB lensing convergence maps from the \textit{Planck} 2015 data release and galaxy overdensity maps from the BOSS DR12 CMASS sample~\cite{Pullen_2016}. This dataset is referred to as $\boldsymbol{C^{\kappa {\rm g}}_{\ell,{\rm obs}}}$. These is some degree of overlap between the \clkgnobm and \pggnobm measurements as the part of sky covered by respective galaxy samples overlap with the \textit{Planck} lensing maps. In~\cite{Giusarma:2018jei} this overlap had not been taken into account and the two measurements had been treated as independent. Here, we go beyond this simplification, and self-consistently take into account the overlap between the \pggobs and \clkgobs measurements by including the cross-covariance between the two (see Sec.~\ref{sec:model-fitting} for further discussions), although we find a posteriori that the effect of neglecting the cross-covariance is small given the precision of current CMB lensing and full-shape galaxy clustering data.
\end{itemize}

As discussed earlier, \pggnobm measurements are particularly useful when complemented with CMB data, as they help breaking the geometrical degeneracy, through the geometrical information contained in the reconstructed BAO peak(s): see for instance~\cite{Ivanov_2020,Vagnozzi:2020rcz}. In particular, \pggnobm can help excluding low/high values of $H_0$ or $\Omega_{\rm m}$ respectively, which would otherwise be tolerated by CMB data alone. This in turn improves constraints \mnu, which contributes to $\Omega_{\rm m}$ at late times. We therefore complement the above datasets by the latest CMB measurements, along with BAO distance and expansion rate measurements. CMB data is particularly helpful in constraining the 6 $\Lambda$CDM parameters. BAO data, on the other hand, provide a late-time standard ruler calibrating the matter density parameter $\Omega_\text{m}$, and absolute scale of the expansion rate $H_0$. More specifically, we consider the following datasets:
\begin{itemize}
\item Measurements of the CMB temperature (\ttt), E-mode polarization (\ee), and temperature-polarization cross-correlation (\te) anisotropy spectra from the \textit{Planck} 2018 data release~\cite{planck2018likelihoods}. We include the full \ttt~($2 \leq \ell \leq 2508$) and \ee~($2 \leq \ell \leq 1996$) ranges, as well as the high-$\ell$ \te~($30 \leq \ell \leq 1996)$ range. For the high-$\ell$ ($\ell \geq 30$) \ttt, \te, and \ee\, measurements, we adopt the \texttt{Plik} likelihood~\cite{planck2018likelihoods}. We refer to this dataset as \textit{\textbf{Planck}}.
\item Small-scale CMB \ttt,\te,\ee~anisotropy measurements from the Atacama Cosmology Telescope Polarimeter (ACTPol) Data Release 4 (DR4)~\cite{Aiola_2020,Choi_2020}. We use measurements in the multipole range $\ell_{min} \leq \ell \leq 4000$. In particular, we truncate at the large-scale cut-off $\ell_{min}=1800$ for \ttt and $\ell_{min}=350$ for \te,\ee, as suggested in \cite{Aiola_2020} in order to ensure that the errors arising from neglecting the cross-covariance between ACTPol and \textit{Planck} datasets are negligible and in any case at most $5\%$. We refer to this dataset as \textit{\textbf{ACTPol}}.~\footnote{The \texttt{actpollite\_dr4}-software is available at \url{https://lambda.gsfc.nasa.gov/product/act/act_dr4_likelihood_get.cfm}.}
\item Baryon Acoustic Oscillation (BAO) measurements from the: SDSS Main Galaxy Sample (MGS, $z_{\rm eff}=0.15$) \cite{Ross_2015}; Six-degree-Field Galaxy Survey (6dFGS, $z_{\rm eff}=0.106$) \cite{Beutler_2011}; and lastly, the post-reconstructed (consensus) results constructed from the BOSS DR12 galaxy samples ($z_1=0.38, z_2=0.51, z_3=0.61$) \cite{Alam_2017}. Note that our galaxy probes -- \pggobs and \clkgobs -- are also constructed from the BOSS DR12 galaxy samples, however they are only dependent on the CMASS sample that is situated at an effective redshift $z_\text{eff}=0.57$, which overlaps with the two upper $z$-bins of the BAO BOSS DR12 consensus dataset. Hence, we exclude the redshift bins $z_2$ and $z_3$ from the BAO BOSS DR12 consensus dataset whenever it is used simultaneously with our \pggobs and \clkgobs measurements. We denote this reduced BAO dataset, together with MGS and 6dFGS, \textit{\textbf{\textit{BAO}$_{z_1}$}}. Whenever we exclude the full-shape galaxy power spectrum measurements and thus also include the redshift bins $z_2$ and $z_3$, we refer to this as \textit{\textbf{\textit{BAO}$_{\text{cons}}$}}. Therefore, \textit{BAO}$_{z_1}$ includes BAO measurements from the MGS, 6dFGS, and BOSS DR12 $z_1$ galaxy samples, whereas \textit{BAO}$_{\text{cons}}$ includes BAO measurements from the MGS, 6dFGS, and complete BOSS DR12 ($z_1$, $z_2$, and $z_3$) galaxy samples.
\end{itemize}
For conciseness, hereafter we shall refer to the combination \planckobs + \bao + \pggnobm  as \textit{\textbf{base}}: this combination defines our reference baseline dataset against which we will compare all our results later on.

We make a few final amendments to our theoretical galaxy power spectrum, \pggth, in order to correct for survey-specific effects impacting the observed galaxy power spectrum. Firstly, the finite survey geometry introduces mode-coupling between otherwise independent $k$-modes. In practice, we model this effect through the window function $W(k_i\,,k_j)$, which we convolve with the theoretical galaxy power spectrum as follows:
\begin{align}
P^{\text{gg}}_{\text{th}}(k_i,z_\text{eff}) \rightarrow \sum_{k_j} W(k_i, k_j) \frac{P^{\text{gg}}_{\text{th}}\Big(k=\frac{k_j}{a_{\text{scl}}},z_\text{eff}\Big)}{a_{\text{scl}}^3}\,,
\end{align}
where $z_{\rm eff}=0.57$ is the effective redshift of the BOSS DR12 CMASS sample ($z_{\text{eff}}$=0.57), and the parameter $a_{\text{scl}}$ is defined in Eq.~(\ref{eq:ascl}) below. In addition, we also model the Alcock-Paczynski (AP) effect, a well-known effect resulting from the need to assume a fiducial cosmology in order to convert redshifts into comoving coordinates to estimate the power spectrum~\cite{Alcock:1979mp}, where the assumption of a wrong fiducial cosmology will lead to geometrical distortions in the observed clustering pattern. To model the AP effect, we follow~\cite{SDSS:2006lmn,Ferramacho:2008ap,Reid:2009xm,Parkinson_2012}, and adopt the scaling factor $a_{\rm scl}$:
\begin{align}
a_{\rm scl} = \frac{D_A(z_{\text{eff}})^2/H(z_{\text{eff}})}{D^{\text{fid}}_A(z_{\text{eff}})^2/H^{\text{fid}}(z_{\text{eff}})}\,,
\label{eq:ascl}
\end{align}
where $D_A$ is the angular diameter distance, and the superscript ``fid'' denotes quantities evaluated in the fiducial cosmology assumed by the BOSS collaboration to compute \pggobs. The AP effect is implemented by evaluating the theoretical power spectrum at re-scaled wavenumbers $\hat{k} = k(a_{\rm scl})^{-1/3}$, and re-scaling the power spectrum by a factor of $a_{\rm scl}$. We note, however, that the effect on parameter estimation of not including the AP effect is negligible given the precision of current galaxy clustering data.

\subsection{Parameter estimation}
\label{sec:model-fitting}

Our \pggnobm and \clkgnobm datasets probe large, linear scales, where density perturbations are approximately Gaussian. Hence, they are approximately described by normal random variables. This enables us to express the joint \pggnobm-\clkgnobm, $\cal{L}$, as a multivariate normal probability density function:
\begin{eqnarray}
\ln {\cal L} \sim (\bm{t}(\bm{\theta}) - \bm{d})^\text{T} {\cal C}^{-1} (\bm{t}(\bm{\theta}) - \bm{d})\,,
\end{eqnarray}
where $\bm{t}(\bm{\theta})$ is the theoretical prediction for the observational datavector $\bm{d}$ given a set of model parameters $\bm{\theta}$, and $C_{ij}$ is the covariance matrix quantifying the amount of covariance between two elements $d_i$ and $d_j$. In our case, in order to properly model the fact that our \pggnobm and \clkgnobm measurements are not independent (as they are obtained from datasets which overlap on the sky), we are considering a \textit{joint} \pggnobm-\clkgnobm likelihood, which means that the datavector $\bm{d}$ holds the measurements of both \pggobs and \clkgobs, and similarly for $\bm{t}$ with the corresponding theoretical predictions:
\begin{align}
\arraycolsep0.0pt\def\arraystretch{1.8}
\begin{array}{cccccc}
\begin{array}{r}
\bm{d}=\Big[ \\
\bm{t}(\bm{\theta})=\Big[
\end{array}
&
\begin{array}{cccccc}
P^{\text{gg}}_{\text{obs}}(k_1), & \dots, & P^{\text{gg}}_{\text{obs}}(k_n), & C^{\kappa \text{g}}_{\ell_1,\text{obs}}, & \dots, & C^{\kappa \text{g}}_{\ell_m,\text{obs}}\\
P^{\text{gg}}_{\text{th}}(k_1), & \dots, & P^{\text{gg}}_{\text{th}}(k_n), & C^{\kappa \text{g}}_{\ell_1,\text{th}}, & \dots, & C^{\kappa \text{g}}_{\ell_m,\text{th}}\
\end{array}
&
\begin{array}{l}
\Big] \\ \Big]
\end{array}
\end{array}\,,
\end{align}
where we are considering $n$ bins with \pggobs measurements in the wavenumber range $k_1 \leq k \leq k_n$ and $m$ bins with \clkgobs measurements within the multipole range $\ell_1 \leq \ell \leq \ell_m$.

The fact that \pggobs and \clkgobs are not statistically independent is reflected in the full covariance matrix. More specifically, it is useful to think of the covariance matrix ${\cal C}$ as a $(n+m)\times(n+m)$ block matrix, partitioned into 2 row groups and 2 column groups:
\begin{align}
{\cal C} = 
\begin{bmatrix}
\text{Cov}\big[\hat{P}^{\text{gg}}(k),\hat{P}^{\text{gg}}(k')\big] && \text{Cov}\big[\hat{P}^{\text{gg}}(k),{\hat{C}}^{\kappa \text{g}}_{\ell}\big]\\ \text{Cov}\big[\hat{P}^{\text{gg}}(k),{\hat{C}}^{\kappa \text{g}}_{\ell}\big]^\text{T} && \text{Cov}\big[{\hat{C}}^{\kappa \text{g}}_{\ell},{\hat{C}}^{\kappa \text{g}}_{\ell'}\big]
\end{bmatrix}\,,
\label{eq:fullcovariancematrix}
\end{align}
where $\text{Cov}\big[\hat{P}^\text{gg}(k),\hat{P}^\text{gg}(k')\big]$ and $\text{Cov}\big[{\hat{C}}^{\kappa \text{g}}_{\ell},{\hat{C}}^{\kappa \text{g}}_{\ell'}\big]$ are the covariance matrices of the individual $\hat{P}^{\text{gg}}(k)$ and ${\hat{C}}^{\kappa \text{g}}_{\ell}$ measurements, themselves estimators of the observational datasets \pggobs and \clkgobs respectively. Then, the statistical correlation between \pggobs and \clkgobs is captured by the off-diagonal block $\text{Cov}\big[\hat{P}^{\text{gg}}(k),{\hat{C}}^{\kappa \text{g}}_{\ell}\big]$ and its transpose: we shall refer to this block as the cross-covariance between $\hat{P}^{\text{gg}}(k)$ and ${\hat{C}}^{\kappa \text{g}}_{\ell}$, occasionally denoting it by \ccross.

The covariance matrix for \pggobs has been measured by the BOSS collaboration using dedicated mocks~\cite{Cuesta_2016}. On the other hand, we have assumed a Gaussian likelihood for \clkgobs, with covariance matrix estimated by jackknife resampling of 37 equal-weight regions of the CMASS survey area. We refer the reader to~\cite{Pullen_2016} for further details. In order to write down the full joint \pggobs-\clkgobs likelihood, we therefore require an estimate for \ccross. We write down an analytical estimator for \ccross based on two assumptions: Gaussian density perturbations, implying that the cross-covariance is independent of the matter trispectra; and flat-sky approximation for \clkgobs, valid as long as we are not observing perturbations on ultra-large scales. The full derivation of \ccross is reported in Appendix~\ref{app:cross-cov}, and we simply cite the result of this calculation below:
\begin{align}
\text{Cov}&\Big[\hat{P}^\text{gg}(k_i,z_\text{eff}),\hat{C}^{\kappa \text{g}}_{\ell_j}\Big] = \int dz \frac{H(z)}{\chi^2(z)}W^{\kappa}(z)f^{g}(z) \label{eq:cross-cov-maintext}\\
& \times \frac{V_\text{f}}{V_\text{s}(k_i)}\frac{D^{2}_+(z)}{D^2_+(z_\text{eff})}2P^\text{mg}(k_i,z_\text{eff})P^\text{gg}_*(k_i,z_\text{eff})\Big|_{\big|k_i - \frac{\ell_j}{\chi(z)}\big|\leq \frac{\delta k_i}{2}}\,.\nonumber
\end{align}
In the above, we have denoted the shot noise-less galaxy power spectrum by ${P^\text{gg}_*=P^\text{gg}_\text{th}-\pshot}$ and the linear growth function by $D_+$. Furthermore, ${V_\text{s}(k_i)}$ is the volume of a spherical shell centred upon $k_i$ and $\delta k_i$ is the size of the bin associated to $k_i$: $\delta k_i = (k_{i+1}-k_{i-1})/2$. Finally, ${V_\text{f}=(2\pi)^3/V_\text{surv}}$ is the volume of the fundamental cell that depends on the galaxy survey volume $V_\text{surv}$. Note that this result ignores non-Gaussian corrections to the covariance, an approach which was also adopted in the earlier related work of~\cite{Schmittfull:2017ffw}, see also~\cite{Philcox:2019xzt,Philcox:2019hdi}.

As can clearly be seen in Eq.~(\ref{eq:cross-cov-maintext}), the cross-covariance is by definition a function of the scale-dependent bias parameters and the \lcdm parameters. Therefore, it would in principle require a new evaluation for each sample in our Markov Chain Monte Carlo (MCMC) analysis. However, as often done in these contexts with current data, we fix all parameters to given fiducial values, evaluate the covariance matrix for these sets of parameters, and assume that the covariance matrix is then fixed and does not vary with parameters.~\footnote{Note that with future more precise galaxy clustering data, this simplification may no longer be adequate (see for instance~\cite{Kodwani:2018uaf,Harnois-Deraps:2019rsd}).} We fix the bias parameters to the following values: ${\abias = 2}$, ${ \cbias=\dbias=0 [h^{-2}\text{Mpc}^2]}$. Note that while $\abias$ is dimensionless, $\cbias$ and $\dbias$ carry dimensions of $h^{-2}\text{Mpc}^{2}$, reflecting the fact that the quantities $\cbias k^2$ and $\dbias k^2$ need to be dimensionless (as they carry the same units as $\abias$). Moreover, we normalize the stochastic shot noise component \pshot in units of the fiducial Poisson shot noise $1/\bar{n}$, where $\bar{n}$ is the average number density of the galaxy survey in question, which for the BOSS DR12 CMASS sample is $\bar{n} \simeq 3 \times 10^{-4}\,h^3{\rm Mpc}^{-3}$~\cite{reid2015sdssiii}. Therefore, our fiducial Poisson shot noise is $1/\bar{n} \simeq 0.33 \times 10^4\,h^{-3}{\rm Mpc}^3$, and we implicitly normalize \pshot in these units (e.g.\ $\pshot=1$ really means $\pshot=0.33 \times 10^4\,h^{-3}{\rm Mpc}^3$). Finally, we fix the \lcdm parameters to their respective best-fit values as inferred from the \planckobs 2018 TT, TE, EE (across the full $\ell$-range) legacy measurements alone~\cite{Planck_2018_cosmoparams}, i.e.\ $\Omega_b h^2=0.0224$, $\Omega_c h^2=0.120$, $\theta_s=0.0104$, $\tau=0.054$, $\ln(10^{10}A_s)=3.045$, and $n_s=0.966$.

We compute theoretical predictions for the cosmological observables we consider through the Boltzmann solver \texttt{CAMB}~\cite{Lewis:1999bs}. To sample the joint posterior distribution for the cosmological and nuisance parameters (including the scale-dependent bias parameters), we employ MCMC methods, with samples generated through a suitably modified version of the cosmological MCMC sampler \texttt{CosmoMC}~\cite{Lewis_2002_cosmomc}. The convergence of the generated chains is evaluated by computing the Gelman-Rubin parameter $R-1$~\cite{Gelman:1992zz}, a measure of the ratio between the intra-chain and inter-chain variances, adopting $R-1<0.01$ as stopping criterion. We set uniform priors on all cosmological parameters. We allow \mnu to be as small as $0\,{\rm eV}$, ignoring prior information from oscillation experiments, which set a lower limit of $0.06\,{\rm eV}$ (see e.g.~\cite{Vagnozzi:2018jhn} for further discussions on advantages associated to using this prior). Table~\ref{tab:priors} summarizes the priors on the galaxy bias and shot noise parameters. Finally,to compute  parameter constraints and produce plots of the respective posterior distributions, we make use of the \texttt{GetDist} Python analysis package~\cite{Lewis_2019_getdist}.

\begin{table}
\centering
\caption{Ranges for the (flat) priors on the galaxy bias and shot noise parameters. We normalize the stochastic shot noise component \pshot in units of the fiducial Poisson shot noise $1/\bar{n}$, where $\bar{n}$ is the average number density of the galaxy survey in question, which for the BOSS DR12 CMASS sample is $\bar{n} \simeq 3 \times 10^{-4}\,h^3{\rm Mpc}^{-3}$~\cite{reid2015sdssiii}, so our fiducial Poisson shot noise ($P_{\rm shot}=1$) actually corresponds to $1/\bar{n} \simeq 0.33 \times 10^4\,h^{-3}{\rm Mpc}^3$.}
\label{tab:priors}
\begin{tabular}{cccc}
\toprule
\abias& \cbias & \dbias& \pshot\\[0.3ex]
& [$h^{-2}$Mpc$^{2}$] & [$h^{-2}$Mpc$^{2}$] & [$1/\bar{n}$]\\
\midrule \relax
$[0.0,5.0]$ & $[-70.0,30.0]$ & $[-70.0,30.0]$ & $[0.03,5.00]$\\
\bottomrule
\end{tabular}
\end{table}

\section{Results \& Discussion}
\label{sec:results}

The obtained marginalized constraints on \mnu and the galaxy bias parameters are summarized in Tab.~\ref{tab:results}. We always report 68\%~C.L. intervals except for cases where only an upper/lower limit is available (as with \mnu), in which case we quote a 95\%~C.L. upper/lower limit. We begin by discussing the 95\%~C.L. upper limits on \mnu. Posterior distributions for \mnu obtained from different dataset combinations are presented in Fig.~\ref{fig:mnu}, whereas the corresponding 95 \% C.L. upper limits are given in the last column of Tab.~\ref{tab:results}.

\begin{table*}
\centering
\scalebox{0.7}{
\begin{tabular}{|c||ccccc|}       
\hline\hline
Datasets & \abias & \cbias & \dbias & \pshot & \mnu (95\%~C.L.) \\
 & [1] & [$h^{-2}{\rm Mpc}^2$] & [$h^{-2}{\rm Mpc}^2$] & [$1/\bar{n}$] & [${\rm eV}$] \\\hline
\base $\equiv$ \planckobs + \bao + \pggnobm & $1.97 \pm 0.05$ & -- & $-27.2 \pm 10.0$ & $1.39 \pm 0.44$ & $<0.14$ \\
\base + \clkgnobm & $1.97 \pm 0.05$ & $9.1 \pm 3.1$ & $-29.7 \pm 10.1$ & $1.51 \pm 0.44$ & $<0.14$ \\
\base + \clkgnobm (including $C_{\rm cross}$) & $1.97 \pm 0.05$ & $8.8 \pm 3.0$ & $-28.7 \pm 10.0$ & $1.47 \pm 0.43$ & $<0.14$ \\
\base + \clkgnobm (\pshot$=1$) & $2.00 \pm 0.05$ & $5.1 \pm 0.7$ & $-16.1 \pm 1.0$ & -- & $<0.14$ \\ \hline
\base + \clkgnobm ($k_{\max} = 0.145\,h{\rm Mpc}^{-1}$) & $1.98 \pm 0.05$ & $7.0 \pm 2.4$ & $-22.2 \pm 7.8$ & $1.18 \pm 0.36$ & $<0.14$ \\
\base + \clkgnobm ($k_{\max} = 0.160\,h{\rm Mpc}^{-1}$) & $2.01 \pm 0.05$ & $4.8 \pm 1.8$ & $-15.8 \pm 5.8$ & $0.86 \pm 0.29$ & $<0.14$ \\
\base + \clkgnobm ($k_{\max} = 0.188\,h{\rm Mpc}^{-1}$) & $2.00 \pm 0.05$ & $5.3 \pm 1.9$ & $-17.0 \pm 6.5$ & $0.92 \pm 0.31$ & $<0.14$ \\
\base + \clkgnobm ($k_{\max} = 0.205\,h{\rm Mpc}^{-1}$) & $2.01 \pm 0.05$ & $4.3 \pm 1.6$ & $-13.5 \pm 4.7$ & $0.75 \pm 0.24$ & $<0.13$ \\ \hline
\base + \act & $1.98 \pm 0.06$ & -- & $-27.2 \pm 10.0$ & $1.39 \pm 0.45$ & $<0.17$ \\
\hline \hline     
\end{tabular}}
\caption{Constraints on the scale-dependent bias parameters and \mnu. For the scale-dependent bias parameters we report 68\%~C.L. intervals, whereas for \mnu we report the 95\%~C.L. upper limit. If the value of $\kmax$ is not mentioned, it is set to $0.13\hmpc$. We normalize the stochastic shot noise component \pshot in units of the fiducial Poisson shot noise $1/\bar{n}$, where $\bar{n}$ is the average number density of the galaxy survey in question, which for the BOSS DR12 CMASS sample is $\bar{n} \simeq 3 \times 10^{-4}\,h^3{\rm Mpc}^{-3}$~\cite{reid2015sdssiii}, so our fiducial Poisson shot noise ($P_{\rm shot}=1$) actually corresponds to $1/\bar{n} \simeq 0.33 \times 10^4\,h^{-3}{\rm Mpc}^3$. For the dataset combination including $C_\text{cross}$, the cross-covariance between \pggobs and \clkgobs has been properly included, although we find a posteriori that the impact thereof is negligible.}
\label{tab:results}
\end{table*}

\subsection{Baseline constraints on neutrino masses}

We begin by discussing the constraints on \mnu we obtain from our \base dataset combination, which we recall is given by the combination \planckobs + \bao + \pggobs. By comparing the 95\%~C.L. upper limit obtained from the \base combination against the same limit obtained from \planckobs alone, we see that the inclusion of LSS data has significantly improved constraints on \mnu, bringing the upper limit from $0.26\,{\rm eV}$ to $0.14\,{\rm eV}$. This is not unexpected, as the inclusion of LSS data helps easing the geometrical degeneracy affecting $H_0$ and $\Omega_\text{m}$, by cutting out the part of parameter space associated to low/high values of $H_0$/$\Omega_\text{m}$ respectively, which would otherwise be tolerated by CMB data alone. The tighter constraints on $\Omega_\text{m}$ naturally results in tighter constraints on \mnu. We have checked that neutrino masses below $0.14\,{\rm eV}$ would result in an induced suppression of power in the galaxy power spectrum which is of the order or less than the suppression obtained by propagating the uncertainty on $b_\text{auto}^2(k)$.

We find that the upper limit on \mnu obtained from the \base dataset combination is comparable to the one we would obtain if we were to use purely geometrical information from the reconstructed BAO peak(s) using the \planckobs+\baoplanck dataset combination, i.e.\ removing the full-shape \pggnobm measurement and replacing it by the BOSS DR12 $z_2$ and $z_3$ BAO measurements. In the latter case, we find an upper limit of $0.12\,{\rm eV}$, consistent with the bound reported by the Planck collaboration from the same dataset combination~\cite{Planck_2018_cosmoparams}. We recall once more that \textit{BAO}$_{z_1}$ includes BAO measurements from the MGS, 6dFGS, and BOSS DR12 $z_1$ galaxy samples, whereas \textit{BAO}$_{\text{cons}}$ includes BAO measurements from the MGS, 6dFGS, and complete BOSS DR12 ($z_1$, $z_2$, and $z_3$) galaxy samples. We also recall that the reason why we only use the BOSS DR12 $z_1$ BAO measurements (in addition to the MGS and 6dFGS BAO measurements, which are always included) when including the \pggnobm dataset is that the $z_2$ and $z_3$ samples partially overlap with the BOSS DR12 CMASS sample. Therefore, the power spectrum of the BOSS DR12 CMASS sample (i.e.\ the \pggnobm dataset) cannot be used simultaneously with the BOSS DR12 $z_2$ and $z_3$ BAO measurements to avoid double-counting data.

These findings suggests that current BOSS full-shape information and purely geometrical information from the reconstructed BAO peak(s)~\footnote{We use the plural for ``peaks'' as the peak in the real-space correlation function translates to a series of (damped) peaks in the power spectrum.} carry comparable constraining power once combined with \planckobs CMB data. This somewhat surprising conclusion agrees with the same conclusion reached in~\cite{Ivanov_2020,Ivanov:2019hqk}, where it was argued that this fact is merely a coincidence given the current volume and redshift coverage of the BOSS survey as well as the efficiency of current BAO reconstruction algorithms~\cite{Eisenstein_2007,Padmanabhan_2012,Cuesta_2016}. With future spectroscopic galaxy surveys covering a much larger volume and redshift range, together with expected substantial improvements in the efficiency of BAO reconstruction algorithms (see e.g.~\cite{Levy:2020emr,vonHausegger:2021luu}), this trend is expected to be reversed, with the full-shape information eventually superseding the purely geometrical information (see~\cite{2016arXiv161100036D,Brinckmann:2018owf,Chudaykin_2019}).~\footnote{See also~\cite{Hamann:2010pw,Vagnozzi:2017ovm} for similar conclusions reached using earlier data. Moreover, these same works argued that this result may be reversed in extensions to $\Lambda$CDM where shape information can play a crucial role. A recent explicit example of this has been provided in~\cite{Nunes:2022bhn}.}

Thus, for what concerns \mnu bounds, we conclude that \pggnobm-only shape information is approximately as informative as geometrical information from reconstructed BAO peak(s). Alongside the reasons outlined in~\cite{Ivanov:2019hqk} and discussed above, another possibility previously raised in~\cite{Vagnozzi:2017ovm} and~\cite{Giusarma:2018jei} is that this may be at least partially due to the introduction of extra nuisance parameters when analyzing full-shape \pggnobm data, such as the scale-dependent bias and shot noise parameters. To more thoroughly harness the shape information, it is therefore desirable to add other measurements which help nailing down or at least breaking degeneracies related to the bias parameters. To this end, we include measurements of the CMB lensing-galaxy angular cross-power spectrum \clkgnobm: this dataset is sensitive to $\abias\sigma_8^2$, while \pggnobm is sensitive to $\abias^2\sigma_8^2$: therefore, the \base+\clkgnobm combination can help disentangle \abias and $\sigma_8$. Moreover, \clkgnobm suffers from a different set of observational systematics compared to \pggnobm, as discussed for instance in~\cite{Singh_2016,Doux:2017tsv,Schmittfull:2017ffw,Bermejo-Climent:2021jxf,Ballardini:2021frp}.

\subsubsection{Impact of Fingers-of-God}
\label{subsubsec:fog}

Earlier in Sec.~\ref{subsec:observables}, we argued that the impact of FoG is expected to be negligible given our galaxy sample and scale cuts. We test this expectation explicitly, by including our FoG modeling given in Eq.~(\ref{eq:FoG}). More specifically, we include an extra parameter $\sigma_{\rm FoG}$, for which we set a prior linear in the range $[1;100]\,h^{-1}\,{\rm Mpc}$. Considering the \textit{base} dataset combination, we then test whether the inclusion of $\sigma_{\rm FoG}$ significantly improves the fit and/or alters the inferred values of the other parameters.

We find that including FoG does not lead to meaningful changes in the inferred cosmological or bias parameters. To within the precision at which we report constraints, the upper limit on \mnu is unchanged, and so are the inferred values of all the bias parameters. The only exception is \dbias, which shifts very slightly to less negative values, to compensate the extra FoG-induced suppression. As expected, we only infer upper limit on $\sigma_{\rm FoG}$, with $\sigma_{\rm FoG}<3.2\,h^{-1}{\rm Mpc}$ at 68\%~C.L. and $<4.7\,h^{-1}\,{\rm Mpc}$ at 95\%~C.L.: these limits can be roughly translated to lower limits on the wavenumber $k_{\rm FoG}$ at which FoG become non-negligible, $k_{\rm FoG} \gtrsim 0.33\,h{\rm Mpc}^{-1}$ (68\%~C.L.) and $\gtrsim 0.21\,h{\rm Mpc}^{-1}$ (95\%~C.L.), limits within which our scale cuts are safely inside. We thus conclude that for the purposes of our analysis FoG can be safely neglected, although we stress that all our subsequent results include FoG modeling.

\subsection{Including the CMB lensing-galaxy cross-correlation} 
\label{subsec:cmblensing}

We now complement the previously discussed \base dataset combination with measurements of \clkgnobm. Doing so, we find that the upper limit on \mnu is essentially unchanged. In order to investigate whether this is due to a ``poor'' fit to the data or to the data uncertainties we perform a goodness-of-fit analysis that is detailed in Appendix~\ref{app:goodness-of-fit}. We find that there is an underfit between the data and the model, as the significance is determined to $p$-value$\approx$0.01. This may either imply that the dataset errors are too optimistic or that our model is insufficient to represent the data.

Another possibility is that there is some tension between the \clkgnobm and \pggnobm measurements. In fact, as already pointed out in~\cite{Giusarma:2018jei}, \clkgnobm measurements (including the one we adopted) systematically appear show a lack of power on large angular scales~\cite{DES:2015eqk,Kuntz:2015wza,Pullen_2016,saraf2021crosscorrelation}, which can be interpreted as a preference for a lower value of the linear galaxy bias compared to that inferred from galaxy clustering.~\footnote{See for instance~\cite{Bianchini_2015,Bianchini:2015yly,Bianchini:2018mwv} for other measurements of CMB lensing-galaxy cross-correlations which do not find this deficit of power.} The most plausible explanations for this lack of power attribute it to systematics in CMB lensing mapp (see for instance~\cite{Hirata:2004rp,Smith:2007rg,Das:2011ak,vanEngelen:2013rla,Liu:2015xfa,Kuntz:2015wza,Ferraro:2017fac}), such as thermal Sunyaev-Zel'dovich contamination, for which a novel cleaning procedure was recently proposed in~\cite{Madhavacheril:2018bxi} and applied to ACT data in~\cite{Darwish_2020}. Overall, we therefore find that including shape information from \clkgnobm has not improved our constraints on \mnu. It is however expected that this conclusion should change with expected improvements in the quality of future CMB lensing maps and overlapping galaxy redshift surveys, where CMB lensing-galaxy cross-correlations will be a major science driver (see for instance~\cite{Schmittfull:2017ffw,SimonsObservatory:2018koc,SimonsObservatory:2019qwx,Fang:2021ici,Yu:2021vce}).

The previous work of~\cite{Giusarma:2018jei} found that \clkgnobm had a small but not insignificant impact on the bound on the neutrino mass, while here we find that the impact of \clkgnobm is essentially negligible. This can be attributed to the use of the updated \planckobs dataset (from \planckobs 2015 to \planckobs 2018), as this dataset by itself leads to a large reduction of the uncertainties on the neutrino mass. Implicitly, this puts much stronger requirements on other datasets for them to make an impact. The main improvement in going from \planckobs 2015 to 2018 is that for the latter we have also included small-scale polarization data: the use of high-$\ell$ polarization data in \planckobs 2015 was earlier cautioned against due to possible residual systematics in the dataset, which is no longer the case for the \planckobs legacy data release.

\begin{figure}
\includegraphics[width=\columnwidth]{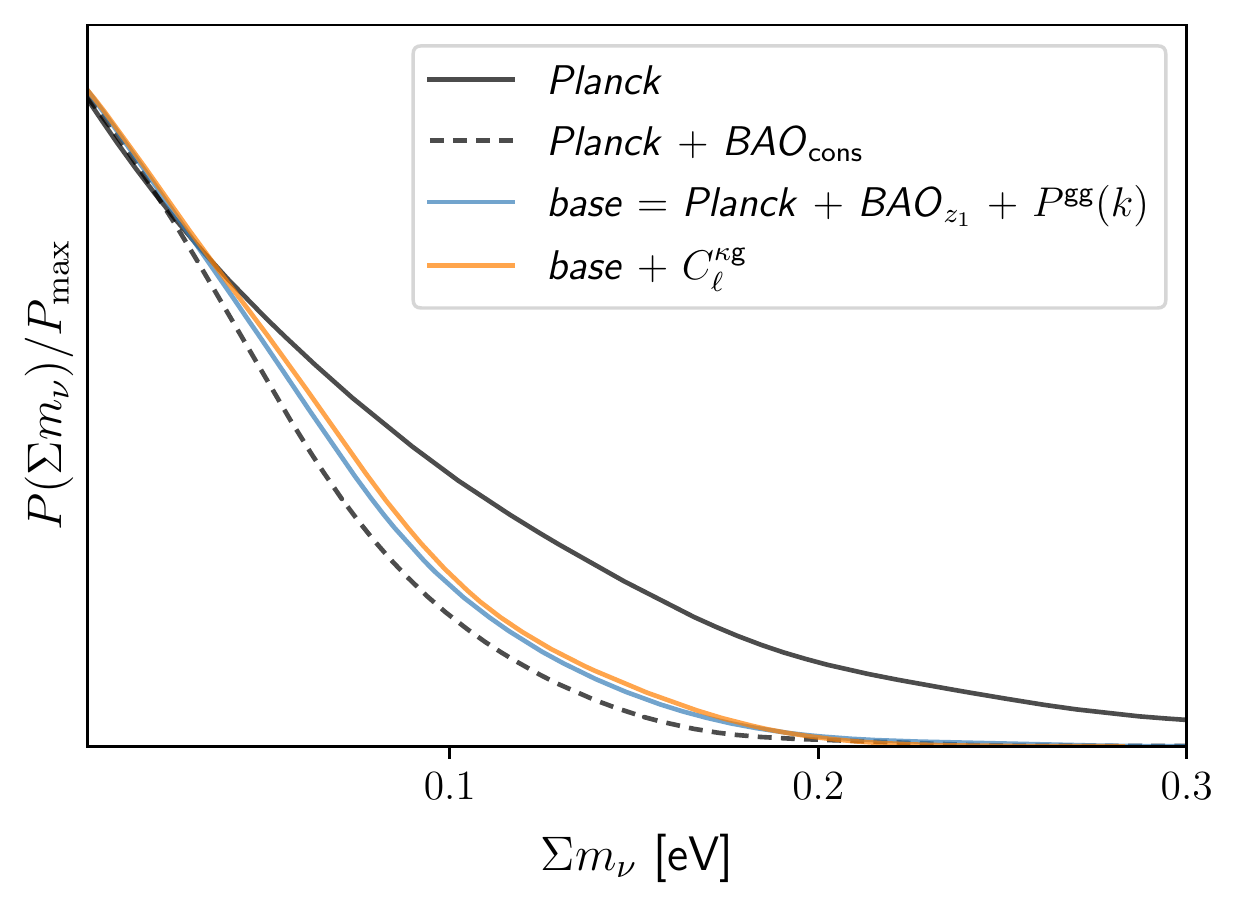}
\caption{1D marginalized posterior distributions for \mnu obtained from various dataset combinations discussed in the main text. The \base dataset combination significantly improves the bound on \mnu compared to \planckobs data alone (from $0.26\,{\rm eV}$ to $0.14\,{\rm eV}$). The bound resulting from the \base dataset combination is also comparable to the \planckobs+\baoplanck bound ($0.12\,{\rm eV}$). The $y$ axis is in arbitrary units, as we are plotting normalizable probability distributions.}
\label{fig:mnu}
\end{figure}
 
\subsection{Including ACTPol data}
\label{subsec:actpol}

Finally, we further include the latest \act small-scale CMB temperature and polarization anisotropy measurements~\cite{Aiola_2020}. Unlike \planckobs, \act does not display a preference for extra lensing (as captured by the lensing amplitude $A_{\rm lens}>1$ in \planckobs). Therefore, we expect the \planckobs+\act dataset combination to prefer slightly higher values of \mnu, or in any case for the \mnu constraints resulting from such a dataset combination to be slightly degraded compared to the same dataset not including \act. The reason is that increasing \mnu decreases the amplitude of lensing in the CMB, in the direction required by \act. This expectation is confirmed by our analysis, as reported in the last rows of Tab.~\ref{tab:results}, where we find that the 95\%~C.L. upper limit of $<0.14\,{\rm eV}$ from the \base dataset combination is degraded to $0.17\,{\rm eV}$ within the \base+\act dataset combination.

\begin{figure}
\includegraphics[width=\columnwidth]{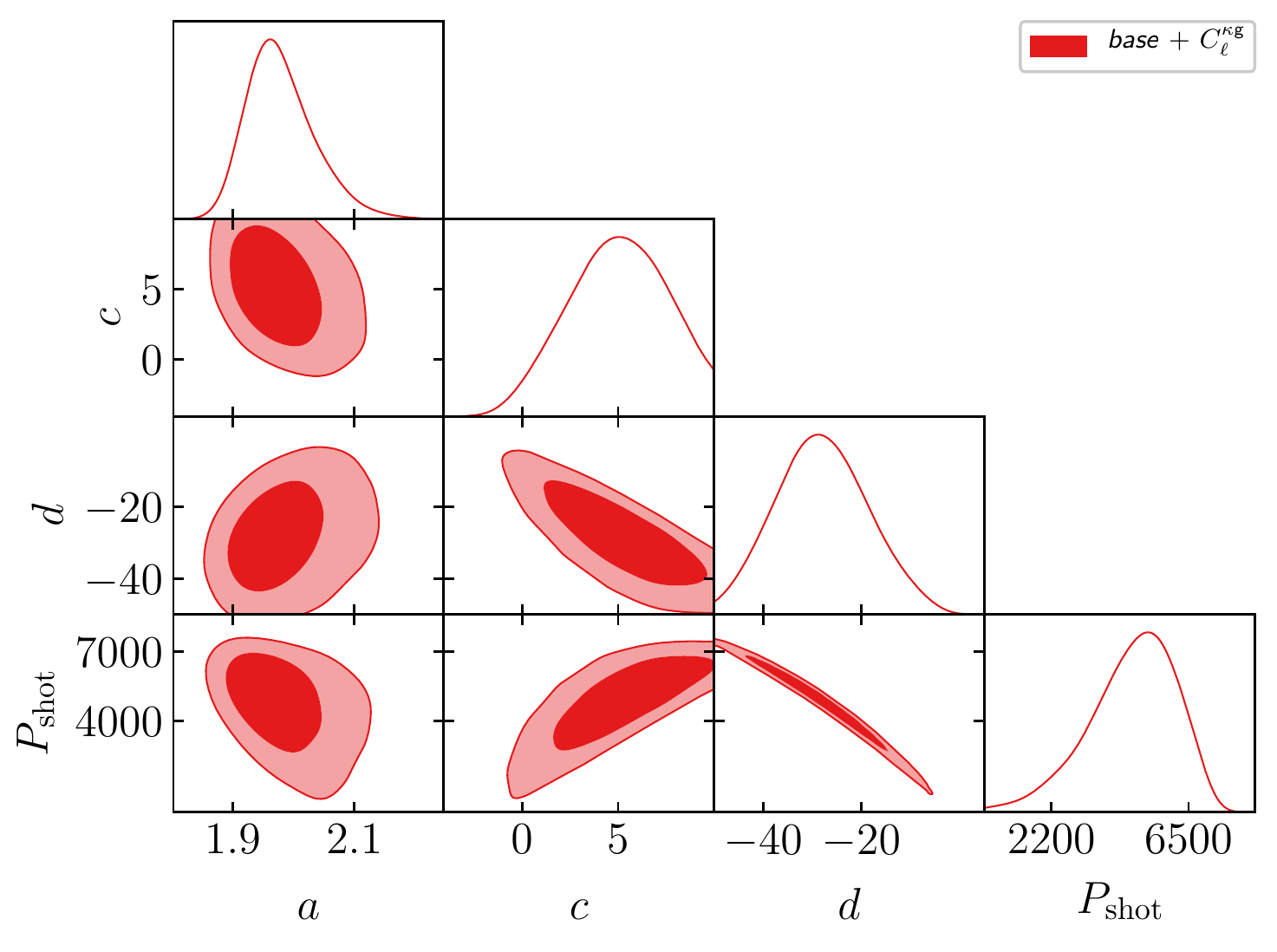}
\caption{Triangular plot showing 2D joint and 1D marginalized posterior probability distributions for the galaxy bias and shot noise parameters resulting from the \base+\clkgnobm dataset combination.}
\label{fig:degeneracies}
\end{figure}

The extent to which the bound degrades is not very drastic, since the \bao and \pggnobm datasets (included in the \base dataset) are still the main drivers for the improvement in the constraints on \mnu compared to the CMB-only constraints, through the improved determination of $\Omega_\text{m}$. Finally, we further include the \clkgnobm dataset, thus considering the \base+\act+\clkgnobm dataset combination, finding no significant shift in the upper bound on \mnu, which remains compatible with $0.17\,{\rm eV}$. Marginalized posterior distributions for \mnu obtained from the dataset combinations discussed so far are shown in Fig.~\ref{fig:mnu}, whereas the corner plot in Fig.~\ref{fig:degeneracies} shows 2D joint and 1D marginalized posteriors for \mnu and the scale-dependent bias parameters obtained from the \base+\clkgnobm dataset combination.

Our interpretation of these results is that the difference between the bounds on \mnu when including \act vs \planckobs is partially a reflection of the mild $\approx 2.5\sigma$ tension existing between these CMB measurements, which has been well documented in the literature both by the \act collaboration~\cite{Aiola_2020} and by~\cite{Handley:2020hdp}. Ultimately, part of this tension can be brought down to the fact that \planckobs primary CMB measurements appear at face value to prefer extra lensing in the small-scale temperature data (as the higher acoustic peaks are more smoothed than expected), as indicated by $A_{\rm lens}>1$. This naturally disfavors heavier neutrinos. as these would suppress structure and reduce the lensing signal, whereas \act sees no preference for extra lensing, and therefore can accommodate heavier neutrinos. This explains why unsurprisingly the upper limit on \mnu degrades when including \act data.

We note that, while is some disagreement between \act and \planckobs as to the height of the first acoustic peak, this is not relevant to our discussion, as we are only using small-scale (high-$\ell$) data from \act. Our conclusion is that until the reason underlying the preference for $A_{\rm lens}>1$ in \planckobs data is well understood, within a \lcdm+\mnu model the CMB side of the data is in principle still able to tolerate neutrino mass limits $\gtrsim 40\%$ weaker than those obtained when making use of \planckobs data.~\footnote{See also the recent work of~\cite{DiValentino:2021imh}, where a stronger version of this point was made, and the related results of~\cite{Sharma:2022ifr,Chudaykin:2022rnl} obtained from other datasets.}
 
\subsection{Galaxy bias parameters and shot noise: detection significance and degeneracies}
\label{subsec:detection}

The small-scale galaxy bias parameters (\dbias and \cbias), as well as the shot noise parameter \pshot, are all detected at moderate significance: $2.9\sigma$, $2.9\sigma$, and $3.4\sigma$ respectively, see Tab.~\ref{tab:results} for the inferred mean values and uncertainties. We note that \dbias and \cbias are anti-correlated, which further justifies our choice of treating \dbias and \cbias as separate parameters modeling the galaxy bias behavior in the the galaxy-galaxy auto-spectrum and galaxy-matter cross-spectrum, respectively. 

Another important degeneracy we find is that between the shot noise parameter \pshot and the scale-dependent term of the auto-spectrum bias (\dbias). In fact, we find that fixing \pshot decreases the uncertainty in \dbias by an order of magnitude (see the fourth row in Tab.~\ref{tab:results}). This increases the detection significance from $2.9\sigma$ to $16.5\sigma$, confirming that it is of vital importance to include (and have precise measurement of) the shot noise term in order to better constrain the scale-dependent bias, and viceversa. Moreover, the negative correlation between the two parameters, already noted earlier in~\cite{Giusarma:2018jei}, is not surprising. Decreasing \pshot decreases power on all scales, but the effect is particularly noticeable on small scales, as power is naturally larger on larger scales, for wavenumbers beyond the matter-radiation equality turn-around in \pggnobm. This can be compensated by increasing \dbias, as it enhances clustering and hence power on small scales.

We also comments on the signs and values of the bias parameters \cbias and \dbias. In general, the bias parameters are considered nuisance parameters which are marginalized over. Nevertheless, the inferred values of the bias parameters can to some extent provide information on the scales at which physical processes connected to the galaxy bias parameters themselves start to play an important role. In the case of the phenomenological \cbias and \dbias parameters, these are loosely speaking associated to complexities inherent to galaxy formation. In particular, galaxy formation gives rise to non-local interactions associated to a characteristic scale, $k_{\rm gf}$, which we can loosely speaking identify as $\approx 1/\sqrt{\vert \cbias \vert}$ or $\approx 1/\sqrt{\vert \dbias \vert}$ (for details, see Sec.~2.6 of~\cite{Desjacques_2018_review}). Given our values of \cbias and \dbias, the estimated scale is approximately $k_{\rm gf} \sim 0.3\,h{\rm Mpc}^{-1}$, which agrees with our expectation concerning the scale at which effects due to galaxy formation start becoming important.

Moreover, the signs of the inferred constraints on \cbias and \dbias also agree with our expectation that $\cbias>0$ and $\dbias<0$. We recall, as discussed in Sec.~\ref{subsec:observables}, that the expectation that $\cbias>0$ comes from the fact that the small-scale matter-galaxy cross-correlation function in real space traces the density profile of host halos, whereas we expect $\dbias<0$ on the basis of the halo exclusion principle. We stress that we had not provided any information on the expected signs of \cbias and \dbias at the level of priors. Therefore, the fact that the inferred signs of these two parameters are consistent with theoretical expectations is completely data-driven. Finally, we note that similar observations had been made earlier in~\cite{Giusarma:2018jei}.
 
\subsection{Robustness tests}
\label{subsec:mnukmax}

We perform a set of robustness tests on our model, investigating the impact of including versus excluding the cross-covariance as well as the cross-correlation coefficient, and examining potential dependencies of \mnu and the bias model parameters on the \kmax cut-off of the galaxy power spectrum.

\subsubsection{Cross-covariance}

One of the aims of this work was to investigate the impact of including versus excluding the usually neglected cross-covariance between \pggnobm and \clkgnobm, see Eq.~(\ref{eq:cross-cov-maintext}). We used a Gaussian analytical approximation to estimate of the effects of including the cross-covariance. We ran the combination \base + \clkgnobm with and without cross-covariance, with results given in Tab.~\ref{tab:results}. We found no significant impact of including the cross-covariance. We interpret this as indicating that the effect of the cross-covariance is negligible at least with the current datasets, which justifies a posteriori the approximation adopted in~\cite{Giusarma:2018jei}. With future datasets, however, the cross-covariance might become an important contributor, in which case the non-Gaussian contributions could also be considered, e.g. those related to mode-couplings or dependent on the binning-scheme \cite{Scoccimarro_1999, Takada_2013,Mohammed_2016}.

\subsubsection{Cross-correlation coefficient}

As another robustness test, we examined the difference caused by excluding versus including the cross-correlation coefficient, \rk. The inclusion of the cross-correlation coefficient is detailed in the paragraphs following Eq.~(\ref{eq:cross-corr}). The exclusion of the cross-correlation coefficient was achieved by setting $r(k,z)=1$. Note that in doing so one is in a sense assuming that $\cbias=\dbias$, as the cross-correlation coefficient accounts for decorrelation effects caused by $\cbias\neq\dbias$\footnote{The cross-correlation coefficient also accounts for decorrelations caused by stochastic contributions, as well as redshift-space distortions. However, compared to \cbias and \dbias these are relatively small and can therefore be ignored in the following discussion.}. However, we know that $\cbias=\dbias$ does not agree with predictions from simulations nor from theory, as discussed in Sec. ~\ref{subsec:observables}. Indeed, we find that the data strongly prefers to treat \cbias and \dbias as separate parameters.

To test if including the cross-correlation coefficient is preferred by the data, we treated \cbias and \dbias as separate parameters while still setting \rk to unity. We found that effectively including \rk in this way gave significantly tighter constraints on \cbias, as its detection significance increased approximately by a factor of 2. We did not find any significant change to the other bias parameters (\abias, \dbias, \pshot). The increased significance of our \cbias inference can be understood from the fact that the term in \pmgnobm associated to \cbias is increased by a factor of $2\abias$: without \rk, \cbias enters into \pmgnobm as a factor of $\cbias k^2$, whereas with \rk it enters as a factor of $2\abias\cbias k^2$. Despite achieving a tighter constraint on \cbias, we did not find strong evidence that a better fit was achieved when including the effects of \rk. In fact, the \chisq for \clkgnobm remained relatively unchanged. However, we stress that there are no drawbacks of adding \rk into the model on mildly non-linear scales: \rk only depends on existing bias parameters and thus does not add any degrees of freedom that could lead to an overfitting of the data.

\subsubsection{Small-scale wavenumber cut}

We now investigate to what extent the inferred bounds on \mnu are affected by the choice \kmax while remaining within the mildly non-linear regime. The full BOSS DR12 galaxy power spectrum monopole measurements cover wavenumbers within the range $0.002\lesssim k/(\hmpc) \lesssim 0.317$~\cite{Gil-Marin:2015sqa}. Previously, we chose to restrict our analysis to the wavenumber range between $k_{\min}=0.030\,$\hmpc and $k_{\max}=0.13\,$\hmpc. The choice of large-scale cut-off $k_{\min}$ was dictated by the fact that \pggnobm measurements on larger scales become increasingly contaminated by observational systematics related to stellar density, seeing requirements, missing close-pairs, fiber collisions, and redshift failures, as discussed in detail in~\cite{BOSS:2012coo}. While these effects are modelled through systematic weights at the map level, we have conservatively chosen to exclude scales where these systematics play an important role, following earlier analyses~\cite{Giusarma:2016phn,Vagnozzi:2017ovm,Giusarma:2018jei}. Similarly, the choice of small-scale cut-off $k_{\max}$ is dictated by the fact that non-linearities and complexities associated to galaxy formation start playing a very important role~\cite{Cooray_2004}, and are not adequately captured by our simplified theoretical (bias) model.

With these caveats in mind, we investigate how the inferred constraints on \mnu and other parameters (including the scale-dependent bias parameters) change if a higher $k_{\max}$ is adopted. To do so, we adopt the \base+\clkgnobm dataset and increase $k_{\max}$ starting from our baseline $k_{\max}$ of $0.13\,h{\rm Mpc}^{-1}$ up to $0.205\,h{\rm Mpc}^{-1}$. The results of this test are reported in Tab.~\ref{tab:results}, from the fifth to the eighth row. We find that there are no significant changes neither in the inferred limits on \mnu (see the last column of the Table), nor in the inferred values of the 6 \lcdm parameters (not reported in the Table).

One possible interpretation of these results is related to our earlier observation, and similar earlier findings in~\cite{Vagnozzi:2017ovm,Ivanov_2020,Ivanov:2019hqk}, that current BOSS full-shape information and purely geometrical information coming from the reconstructed BAO peak(s) carry comparable constraining power once combined with CMB data, particularly given that the latter carries significant statistical power. If this is the case, most of the improvement gained by adding \pggnobm measurements to CMB data comes from breaking the geometrical degeneracy and better constraining $\Omega_{\rm m}$ (and thus indirectly \mnu). The first two BAO peaks in \pggnobm lie between $0.05\,h{\rm Mpc}^{-1}$ and $0.1\,h{\rm Mpc}^{-1}$, and between $0.1\,h{\rm Mpc}^{-1}$ and $0.15\,h{\rm Mpc}^{-1}$ respectively. Once the first BAO peak(s) in \pggnobm have been measured, the geometrical information in \pggnobm has mostly been ``exhausted'' (also given that FoG damp peaks at higher $k$), and there is not much gain in moving to smaller scales.

Another possible interpretation of these results is that the bias nuisance parameters (almost) completely absorb the additional shape information when moving into the more non-linear regime. If so this would imply that our simplified theoretical model is able to reliably obtain the shape information from \pggnobm while extensively covering the mildly non-linear regime ($\leq0.205\,$\hmpc), \textit{at least at the current level of precision of BOSS full-shape data}. Nonetheless, while this is an instructive test, we caution against over-interpreting its results since our theoretical \pggnobm model cannot safely be extended down to the scales to which we pushed the test. The fact that nuisance parameters absorb at least part of the small-scale information, even when moving across different cosmological models, has been noted also in the context of the EFTofLSS (see e.g.~\cite{Ivanov:2020ril,DAmico:2020ods,Nunes:2022bhn}), consistently with our findings.

We have discussed above that the inferred constraints on \mnu are hardly affected by the choice of \kmax beyond $0.13\,h{\rm Mpc}^{-1}$, as most of the geometrical information has been ``exhausted'' by then, and the galaxy bias parameters absorb the additional shape information which would be gained from moving within the more non-linear regime. Here we perform a similar analysis focused on the galaxy bias and shot noise parameters. The inferred values thereof for different choices of \kmax are given in Tab.~\ref{tab:results}. We find that as \kmax increases, the detection significance for all three parameters (\cbias, \dbias, and \pshot) remains roughly constant. What changes are the inferred central values of the parameters, which decrease by $\approx$0.6-$\sigma$ for each added $k$-bin. These changes may be understood in terms of different bias contributions (e.g.\ tidal bias and other higher-order bias terms) entering and playing a role at different scales (see for instance~\cite{Desjacques_2018_review,Boyle_2020}). On the other hand, we are instead capturing these contributions through an ``effective'' $k^2$ scale-dependent term, which is the leading-order correction to a constant linear bias. Note that the constancy of the relative uncertainties implies that the errors propagated from \pggnobm into the galaxy bias parameters is not improved by including additional scales. Thus, additional shape information beyond $k_{\max}=0.13\,h{\rm Mpc}^{-1}$ does not appear to improve the precision at which the galaxy bias parameters and \mnu are inferred, at least when considering current data, and within our simplified model.

It is also interesting to look at deviations in the inferred value of the shot noise parameter \pshot from the fiducial Poissonian shot noise of the BOSS DR12 CMASS galaxy sample (i.e. deviations from \pshot=1, given our choice of normalization) while varying \kmax. We find that \pshot goes from being super-Poissonian~\footnote{A super-Poissonian shot noise contribution may be understood as an enhancement due to a high fraction of satellite galaxies (see for instance~\cite{Baldauf_2013}.} for $k_{\max}=0.13\,h{\rm Mpc}^{-1}$ to sub-Poissonian for $k_{\max}=0.12\,h{\rm Mpc}^{-1}$. We have assumed that any higher-order stochastic contributions to \pggnobm can be neglected. However, the differences in \pshot for variations in \kmax might suggest potential benefits in including next-to-leading-order scale-dependent contributions, i.e. $k^4$ terms besides the $k^2$ term we considered, as a $k^4$ scale-dependent term is consistent with the bias upturn observed on small scales from both theoretical considerations and simulations (see for instance~\cite{Assassi:2014fva,Biagetti:2014pha} for considerations of this sort). For instance~\cite{Raccanelli:2017kht,Chiang:2018laa,Valcin:2019fxe} recently adopted a phenomenological underlying theoretical model which is very close to ours, while including both $k^2$ and $k^4$ terms, and finding a very good agreement with simulations.

In the EFTofLSS context, recent work has found improvements with the inclusion of additional parameters capturing the scale-dependence of the shot noise (referred to as $a_0$ and $a_2$ in a large number of recent papers, see e.g.~\cite{Ivanov:2021kcd,Philcox:2021kcw}), originally not included in the fit. These phenomenological parameters may capture effects such as scale-dependent stochasticity, halo exclusion, and so on. It may be beneficial to include similar parameters in our model as well.

\section{Conclusions}
\label{sec:conclusions}

In this work, we have revisited cosmological neutrino mass constraints from current full-shape galaxy power spectrum data (BOSS DR12 CMASS), in combination with measurements of the cross-correlation between CMB lensing convergence and galaxy overdensity maps. We adopt an underlying model which is minimal yet theoretically motivated, particularly in light of the precision of current data. We improve on the earlier work carried out by some of us in~\cite{Giusarma:2018jei} in several respects, most notably through a more careful treatment of the correlations and covariance between galaxy clustering and CMB lensing-galaxy cross-correlation measurements, for which we construct a tractable model, and by performing a number of additional robustness tests.

When combining galaxy clustering data with current CMB data from \planckobs, we find a 95\%~C.L. upper limit on the sum of the neutrino masses \mnu of $0.14\,{\rm eV}$, compatible with the bound of $0.12\,{\rm eV}$ one would obtain when replacing the full-shape information with purely geometrical information from the reconstructed BAO peak(s). This conclusion, already reached independently with a similar theoretical model in~\cite{Vagnozzi:2017ovm} and in the EFTofLSS-based analysis in~\cite{Ivanov:2019hqk} and related works, indicates that full-shape and purely geometrical information carry the same level of constraining power \textit{given the level of precision, volume, and reconstruction efficiency of current BOSS data}. This interpretation is confirmed by our robustness tests which show that including clustering information from smaller scales does not improve our parameter constraints, suggesting that beyond a certain wavenumber all the geometrical information has been ``exhausted''.

We find that the inclusion of CMB lensing-galaxy cross-correlation measurements does not have a significant impact on our results, which slightly disagrees with the earlier findings of~\cite{Giusarma:2018jei}. This is partially due to the use of the updated \planckobs dataset (and in particular to the use of small-scale polarization data), as this dataset by itself leads to tight constraints on \mnu: this implicitly sets much stronger requirements on other datasets or, equivalently, reduces the benefits of including additional datasets. In addition, the fact that including CMB lensing-galaxy cross-correlation measurements appears to not have a significant impact is a direct consequence of the relatively low signal-to-noise level of the current measurements. Furthermore, we have explored the role of CMB data by including small-scale temperature and polarization data from ACT. We have found that including the latter degrades the previous constraints by $\approx 40\%$. This is related to the fact that unlike \planckobs data, \act data does not appear to show any indication for extra lensing (as captured by the phenomenological $A_{\rm lens}$ parameter).

We expect that the full-shape information content of near-future galaxy clustering measurements at much higher signal-to-noise (for instance from Euclid or DESI) will supersede the geometrical one. In turn, this will significantly increase the importance of CMB lensing-galaxy cross-correlation measurements, which appear to not (yet) play a significant role in current data. Therefore, improvements in the precision and robustness of neutrino mass constraints from future galaxy surveys will require a more robust theoretical modeling, ultimately requiring the introduction of several extra nuisance parameters beyond the ones considered here (including possibly scale-dependent stochastic terms, as suggested by our robustness tests on the inferred shot noise parameter, and similar results in the context of the EFTofLSS finding improvements with the addition of scale-dependent shot noise terms). This will require a study weighing the systematic biases introduced by including too few nuisance parameters against the parameter degeneracies introduced by including many parameters: in other words, whether the information gain from the decrease in observational uncertainties overcomes the increased complexity of the required theoretical model, or if there is a sweet spot compromising between the two, an issue which we plan to return to.

Finally, it is worth noting that cross-correlations between future CMB lensing~\cite{abazajian2016cmbs4,SimonsObservatory:2018koc,SimonsObservatory:2019qwx} and galaxy clustering~\cite{LSST:2008ijt,2016arXiv161100036D} measurements will be detected at much higher statistical significance, particularly in light of the expected substantial overlap in sky fraction between future surveys. This will considerably increase the information content gain from the proposed joint analysis of galaxy clustering and CMB lensing-galaxy cross-correlations~\cite{Fang:2021ici}. Therefore, future work along these lines is very timely and warranted.

\section*{Acknowledgements}
\noindent We thank Rishi Babu and Shirley Ho for collaboration in the initial stages of the project. We acknowledge the use of computing facilities at NERSC and at the Texas Advanced Computing Center. S.V. thanks Misha Ivanov, Shubham Kejriwal, and Oliver Philcox for useful discussions. I.T., S.H., and K.F. acknowledge support by the Vetenskapsr\aa det (Swedish Research Council) through contract No. 638-2013-8993 and the Oskar Klein Centre for Cosmoparticle Physics. S.V. is supported by the Isaac Newton Trust and the Kavli Foundation through a Newton-Kavli Fellowship, and by a grant from the Foundation Blanceflor Boncompagni Ludovisi, n\'{e}e Bildt. S.V. acknowledges a College Research Associateship at Homerton College, University of Cambridge. E.G. acknowledges support from the Physics Department at Michigan Technological University. K.F. is grateful for support from the Jeff and Gail Kodosky Chair of Physics at the University of Texas, Austin, and from the U.S. Department of Energy, Office of Science, Office of High Energy Physics program under Award Number DE-SC0022021.

\appendix

\section{Goodness-of-Fit Test}
\label{app:goodness-of-fit}

\begin{figure*}
\centering
\includegraphics[width=0.49\linewidth]{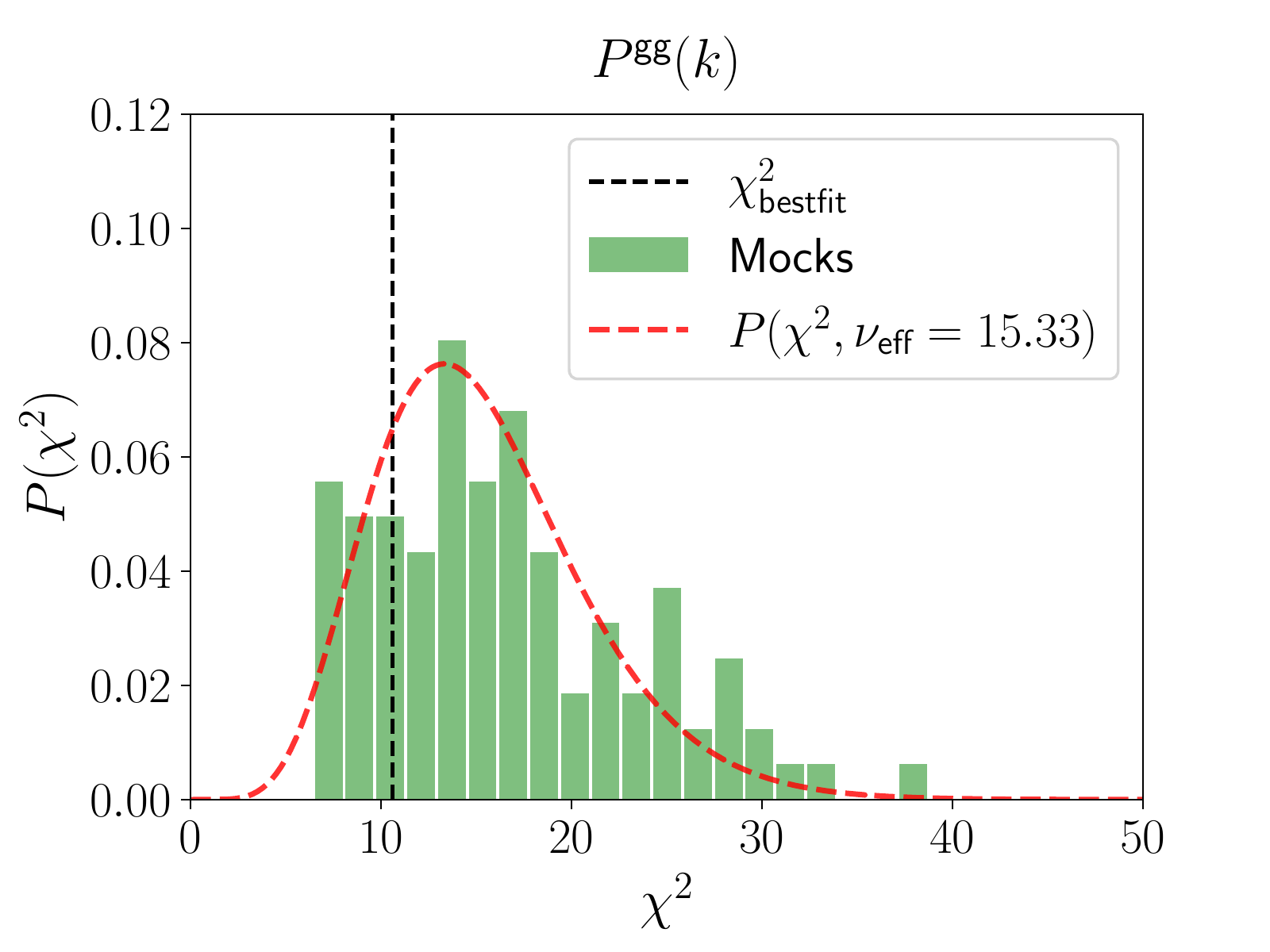}\,
\includegraphics[width=0.49\linewidth]{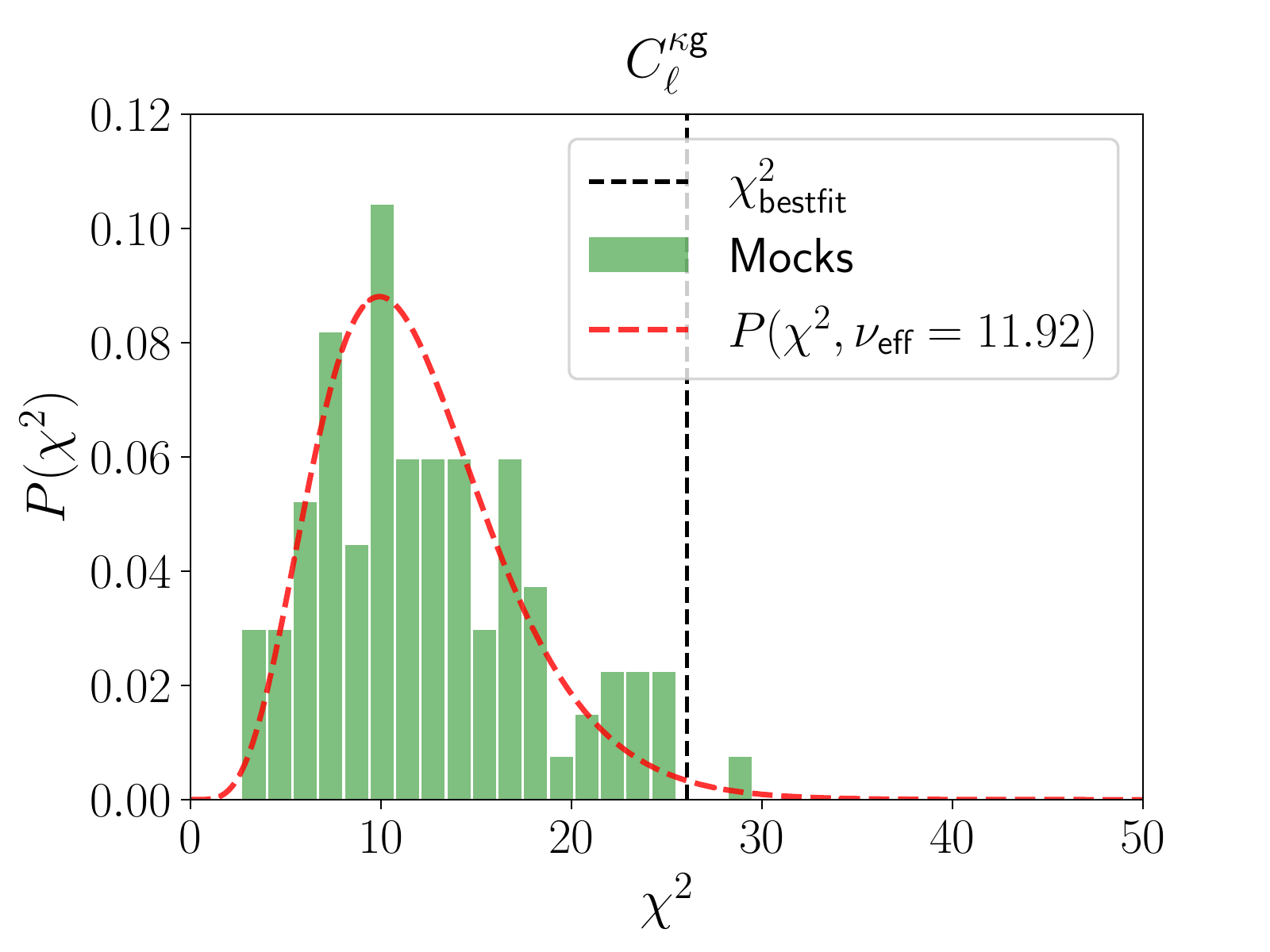}
\caption{\chisq-distributions for \pggnobm (\textit{left}) and \clkgnobm (\textit{right}). In each figure, the vertical dotted line represents the \chisq value of the bestfit. The \textit{green} histogram contains mock datasets that were generated using the method described in detail in the text. The green histogram was least-square fitted to a theoretical \chisq-distribution, illustrated by the dotted \textit{red} curve.}
\label{fig:test}
\end{figure*}

We performed a goodness-of-fit analysis in order to evaluate how closely our theoretical model fits the observed data. This analysis can indicate if there are any under-/overestimation of errors, as well as general discrepancies between theory and data, including under-/overfits. To perform this analysis, we estimated the effective degrees of freedom (\veff) for \clkgnobm and \pggnobm, which are later used to calculate the significance ($p$-value) of our theoretical models.

We wish to estimate \veff because of the general rule that the minimized \chisq-values (\chisqbest) should approximately be equal to \veff. For example, if \chisqbest is much smaller than \veff (i.e. large $p$-value) we may have an overfit to the data (potentially due to degenerate parameters), and/or an overestimation of errors. On the contrary, if \chisqbest is much larger than the \veff (i.e. small $p$-value) we may have an underfit to the data (due to disagreements between model or data), and/or underestimation of errors.

The quantity \veff may be estimated in several different ways: a rudimentary method is to calculate $\veff = D - M$ where $D$ is the number of data points and $M$ is the number of model parameters, both of which we know a priori. However, we can only use this estimate with confidence if we know how many of the model parameters are actually effective. For example, we have seen that some of the model parameters, such as \pshot, are tightly constrained by \pggth but not by \clkgth. Therefore, if we were to assume that all model parameters are important for \clkgth, we would be subsequently underestimating \veff for \clkgnobm.

To have a more robust estimate of the effective number of degrees of freedom for \clkgnobm and \pggnobm, we sampled their \chisq-distributions from mock data, generated according to the following steps:

\begin{enumerate}
\item Assume that \pggnobm (\clkgnobm) can be described by multivariate Gaussian distributions. Let the variance be represented by the covariance matrices we specified in Sec.~\ref{sec:model-datasets}, and the means by the theoretical bestfits of \pggth (\clkgth) that we obtained in Sec.~\ref{tab:results} for the \base + \clkgnobm dataset combination.
\item From these distributions, we draw a number of $n$ random samples, and thereby generate a number of $n$ mock datasets. In our case, we set $n=100$.
\item For each generated mock dataset, calculate the minimized \chisq-value.
\item With this $\chi^2$-distribution, estimate \veff.
\end{enumerate}

The results from the above steps are presented in Fig.~\ref{fig:test}: the left panel contains the results for \pggnobm, whereas the right panel shows the equivalent for \clkgnobm.

We start by commenting on the results for \pggnobm. Based on the obtained significance value ($p$=0.8), we find that \pggnobm is a good fit to the data. We also find a good agreement between the \veff shown in the Figure with the \veff that we would estimated by using the equation mentioned earlier: $\veff = D - M = 19 - 4 = 15$, where the value of $M$ has been taken directly from the number of bias model parameters in our model: \abias, \cbias, \dbias, and \pshot, as these were indicated to be effective for \pggnobm in Tab.~\ref{tab:results}.

For \clkgnobm, we find that the significance is relatively low ($p$=0.01, i.e. within a 3-$\sigma$ confidence level). This low significance indicates the possibility that the errors for \clkgnobm are too optimistic, and/or that there are disagreements between the model and data. As for the latter, there have been consistent findings for discrepancies between theory and modeling of \clkgnobm, possibly related to observational systematics: we discuss this in detail in Sec.~\ref{subsec:cmblensing}. Lastly, we find a good agreement between the \veff from the Figure with the \veff that we would estimated from: $\veff = D - M = 13 - 1 = 12$. Indeed, the results in Table~\ref{tab:results} indicates that it is mostly \cbias which is constrained by \clkgnobm ($M$ = 1), whereas the other bias parameters are constrained by \pggnobm.

\section{Cross-correlation coefficient}
\label{app:cross-corr}

\begin{figure}
\centering
\includegraphics[width=\columnwidth]{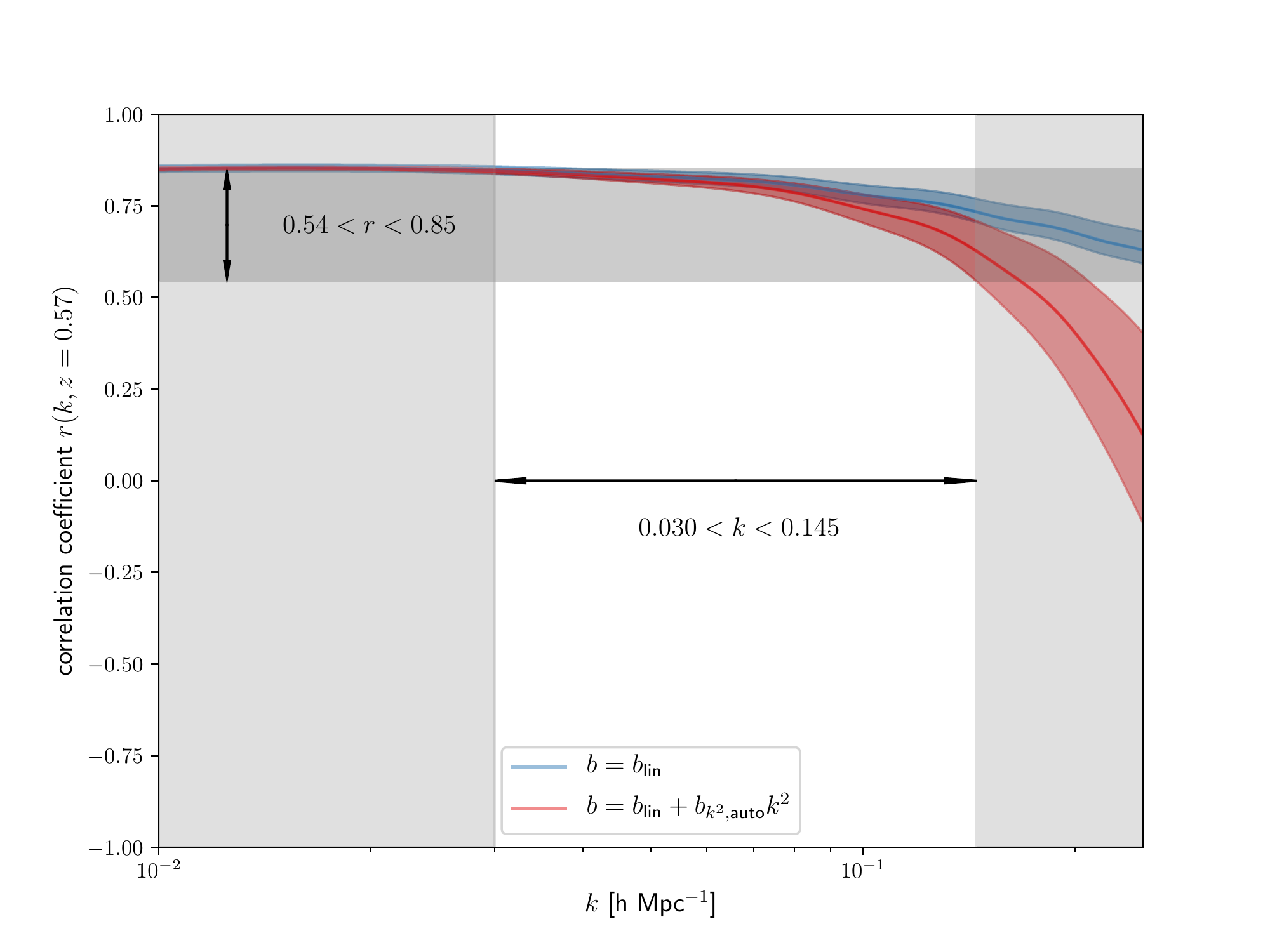}
\caption{Behaviour of the cross-correlation coefficient [see Eq.~(\ref{eq:cross-corr})] for a scale-independent (\textit{blue}) versus scale-dependent (\textit{red}) bias model. As discussed in the main text, the reduction in power is clearly stronger for the scale-dependent model.}
\label{fig:rk}
\end{figure}

As our formulation of the cross-correlation coefficient depends on non-linear quantities, we performed a cross-check that the cross-correlation coefficient behaves as intended. We expect the cross-correlation coefficient to capture the gradual reduction in power while moving to smaller scales, coming from decorrelations between the galaxy and matter field. As we have mentioned in Sec.~\ref{sec:theory}, one such type of decorrelation is the stochastic component, \pshot, which we have demonstrated in Figure~\ref{fig:rk} for a linear (scale-independent) galaxy bias model. Another source for decorrelation is the differing scale-dependent components between the matter and galaxy field. This is also portrayed in Figure~\ref{fig:rk}: in particular, the scale-dependent model predicts a larger suppression than the linear galaxy bias model. Both the scale-dependent component and the stochastic component lead to a gradual reduction in \rk on smaller scales. This demonstrates that the cross-correlation quantity behave as intended within our $k$-range of interest. Lastly, note that the decorrelation effect coming from RSD has been included into both functions in Figure~\ref{fig:rk} and also has been confirmed to contribute in the same way as the two aforementioned effects, although with a smaller contribution.

\section{Cross-covariance calculation}
\label{app:cross-cov}

\begin{figure}
\centering
\includegraphics[width=\columnwidth]{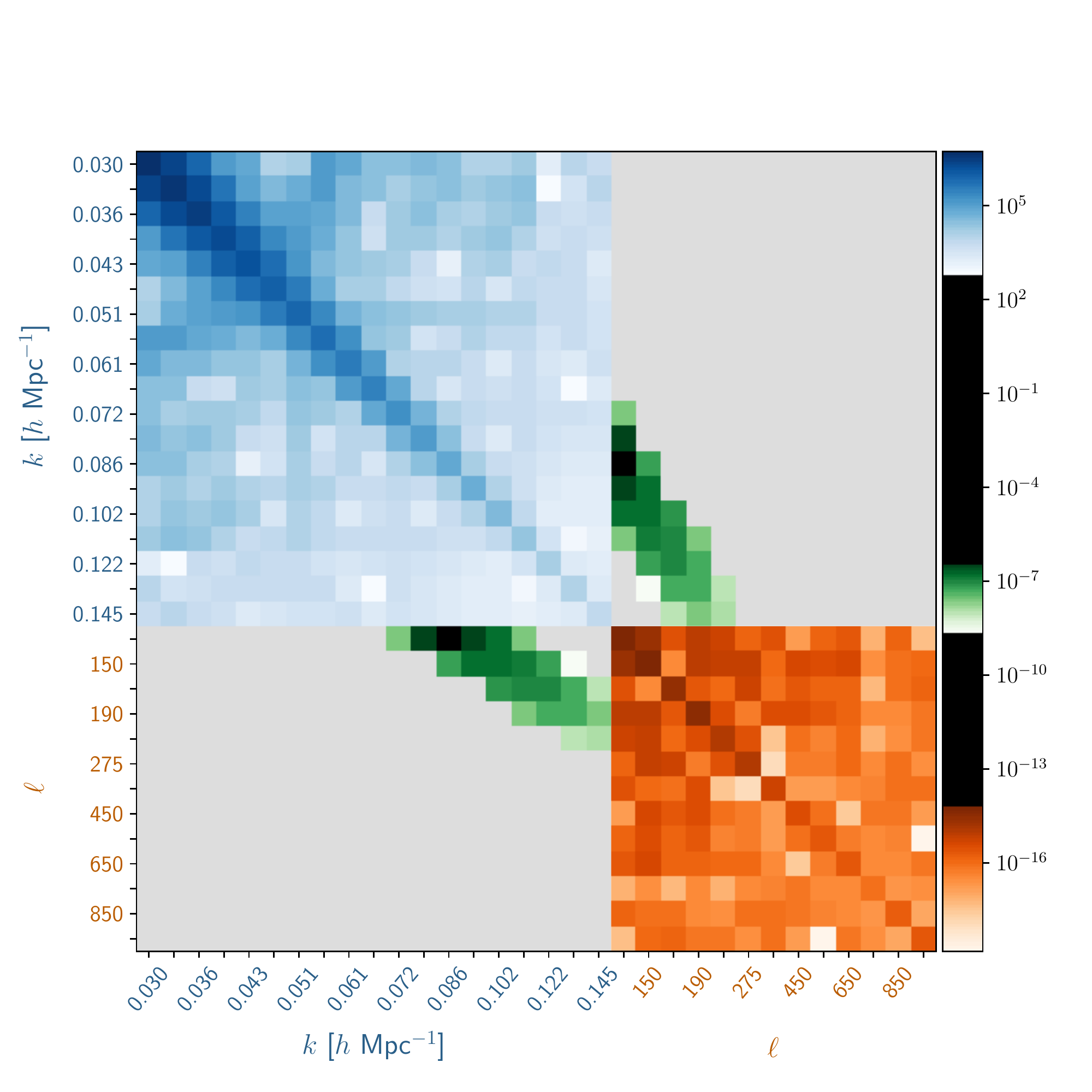}
\caption{Gaussian cross-covariance matrix from Eq.~(\ref{eq:cross-cov-finalapp}) alongside the covariance matrices for the individual probes. The upper left block shows the covariance matrix for \pggobs (\textit{blue}), and the lower right block shows the covariance matrix for \clkgobs (\textit{orange}). For the cross-covariance, the \textit{grey} entries are essentially zero, whereas the non-zero entries are in \textit{green} and capture the Gaussian modes.}
\label{fig:full_cov}
\end{figure}

The cross-covariance we use is based on the formulation of the estimators for 3D and 2D Gaussian fields in \cite{Scoccimarro_1999}. First, we incorporate our galaxy bias model into the aforementioned estimators, and second, we use these estimators to calculate an analytical form of the cross-covariance between the galaxy clustering data, \pggobs, and the CMB lensing convergence-galaxy cross-correlation, \clkgobs. The assumptions that go into this calculation are primarily that non-Gaussian modes can be considered negligible for the purposes of this work (given current observational uncertainties), and that spherical harmonics defining the estimator for \clkgobs can be approximated by 2D Fourier modes. We demonstrate the steps that are required to perform this calculation by first deriving the estimator for the galaxy clustering data, \pgghat, and the estimator for the CMB lensing convergence-galaxy cross-correlation, \clkghat. 

Using the convention in~\cite{Scoccimarro_1999}, we define the unbiased estimator of the matter power spectrum used to compute the estimators for \pggobs and \clkgobs as follows:
\begin{align}
\hat{P}^\text{mm}(k_i,z) &= V_\text{f} \int_{k^\text{bin}_i} \frac{d^3\bm{k}}{V_\text{s}(k_i)} \delta_{\text{m}}(\bm{k},z)\delta_{\text{m}}(-\bm{k},z)\,,
\label{eq:pmmhat}
\end{align}
where $\hat{P}^\text{mm}$ is the bin-averaged matter power spectrum taken over a thin spherical shell, $V_\text{s}$, that radially extends within the interval: $k^\text{bin}_i = [k_i - \delta k_i/2,k_i + \delta k_i/2]$, where $\delta k_i$ is the estimated bin-size associated to $k_i$: $\delta k_i = (k_{i+1}-k_{i-1})/2$. Lastly, $V_\text{f}=(2\pi)^3/V_\text{surv}$ is the volume of the fundamental cell, which takes into account the finite volume of the survey, $V_\text{surv}$.

The bin-averaged galaxy power spectrum depends on the bin-averaged matter power spectrum in Eq.~(\ref{eq:pmmhat}) analogously to how the theoretical galaxy power spectrum depends on the theoretical matter power spectrum in Eq.~(\ref{eq:pgg+pshot+rsd}):
\small
\begin{eqnarray}
\hat{P}^\text{gg}(k_i,z) &=& V_\text{f} \int_{k^\text{bin}_i} \frac{d^3\bm{k}}{V_\text{s}(k_i)} \delta_{\text{g}}(\bm{k},z)\delta_{\text{g}}(-\bm{k},z)\\
&=& b_{\text{auto}}^2(k_i)\Big[1+\frac{2}{3}\beta(k_i,z)+\frac{1}{5}\beta^2(k_i,z)\Big]\hat{P}^\text{mm}(k_i,z) + P_\text{shot}\nonumber\,,
\label{eq:pgghat}
\end{eqnarray}
\normalsize
which works well as long as the scale-dependent galaxy bias, $b_\text{auto}(k)$, as well as the redshift-space distortion corrections (in this case, the Kaiser effect) are approximately constant within $k^\text{bin}_i$.

Having now defined our estimator for the galaxy power spectrum from the formulation of the matter power spectrum in Eq.~(\ref{eq:pmmhat}), we will perform a similar procedure for \clkgobs. \clkgnobm depends on the projection of the mass overdensities:
\begin{align}
A (\hat{\bm{n}}) = \int dz W^A(z) \delta_\text{m}(\chi(z)\hat{\bm{n}})\,,
\label{eq:fields}
\end{align}
where $A$ is either $\kappa$ or $\text{g}$ and therefore represents the convergence field or the 2D galaxy field, respectively. The fields are weighted along the line-of-sight with the help of their respective window functions. As we have projected mass densities, we could expand in terms of spherical harmonic modes over the unit sphere: $A (\hat{\bm{n}}) = \sum_{\ell,m} A_{\ell m} Y_{\ell m}(\hat{\bm{n}})$. However, the use of spherical harmonic modes would involve integration over spherical Bessel functions for the computation of the cross-covariance. To circumvent this issue, we will follow the approach in \cite{Scoccimarro_1999} and employ the flat-sky approximation. For \clkgobs, this relies on expanding the spherical harmonic modes into Cartesian Fourier modes:
\begin{align}
A (\hat{\bm{n}}) = \int \frac{d^2\bm{\ell}}{(2\pi)^2} e^{i\bm{\ell}\cdot\hat{\bm{n}}}A (\bm{\ell})\,.
\label{eq:2dfouriermodes}
\end{align}
Using this formulation, the cross-covariance between the convergence mass overdensity and the projected galaxy overdensity is:
\begin{align}
\langle\kappa(\bm{\ell})g(\bm{\ell}')\rangle = (2\pi)^2\delta^{(2)}_D (\bm{\ell}-\bm{\ell'}) P^{\kappa \text{g}} (\ell)\,,
\end{align}
where $\delta^{(2)}_D$ is the Dirac delta.

We insert the expressions of $\kappa(\bm{\ell})$ and $g(\bm{\ell})$~--~which can be straightforwardly derived from Eq.~(\ref{eq:fields}) and~(\ref{eq:2dfouriermodes})~--~into the above equation and obtain that the cross-power spectrum evaluates to:
\begin{align}
P^{\kappa \text{g}} (\ell) = \int dz \frac{H(z)}{\chi^2(z)}W^{\kappa}(z)f^{g}(z)P^{\text{mg}}\Big(k=\frac{\ell}{\chi(z)},z\Big)\,,
\label{eq:pkappag}
\end{align} 
which is nothing but the Limber approximation to \clkgth given in Eq.~(\ref{eq:clkgth}). Thus, $P^{\kappa \text{g}}(\ell)$ and our \clkgth are equivalent, implying that the statistical estimator for \clkgobs using $\hat{P}^{\kappa \text{g}} (\ell)$ is sufficient for the evaluation of the cross-covariance.

To calculate $\hat{P}^{\kappa \text{g}} (\ell)$, we only need to replace \pmmnobm (contained within \pmgnobm) with its estimator that we defined in Eq.~(\ref{eq:pmmhat}): we assume that the other $k$-dependent components in \pmgnobm (i.e. \cbias and \rk) are approximately constant within a $k$-bin. This yields:
\begin{align}
\hat{P}^{\kappa \text{g}} (\ell_i) &= \int dz \frac{H(z)}{\chi^2(z)}W^{\kappa}(z)f^{g}(z)\nonumber\\ &\times b_\text{cross}(k_i)r(k_i,z)\hat{P}^{\text{mm}}(k_i,z)\Big|_{k_i=\frac{\ell_i}{\chi(z)}}\,,
\end{align}
where we have made the substitution $\hat{P}^{\text{mg}}\rightarrow b_\text{cross}r\hat{P}^{\text{mm}}$ for clarity.

With the estimators for \pggobs and \clkgobs at hand, we calculate the cross-covariance:
\begin{align}
\label{eq:cross-cov-app}
\text{Cov}&\Big[\hat{P}^\text{gg}(k_i,z_\text{eff}),\hat{P}^{\kappa \text{g}} (\ell_j)\Big] \nonumber \\
&= b_{\text{auto}}^2(k_i)\Big[1+\frac{2}{3}\beta(k_i,z_\text{eff})+\frac{1}{5}\beta^2(k_i,z_\text{eff})\Big]\nonumber\\&\times\int dz \frac{H(z)}{\chi^2(z)}W^{\kappa}(z)f^{g}(z)b_\text{cross}(k_j)r(k_j,z)\nonumber \\
& \times \text{Cov}\Big[\hat{P}^\text{mm}(k_i,z_\text{eff}),\hat{P}^{\text{mm}}(k_j,z)\Big]\Big|_{k_j=\frac{\ell_j}{\chi(z)}}\,.
\end{align}
The expression can easily be checked by identifying the pre-factors belonging to \pggth through comparison with Eq.~(\ref{eq:pgg+pshot+rsd}) and similarly, the integrand belonging to \clkgth by comparison with Eq.~(\ref{eq:clkgth}). Note that \pshot does not directly appear in the expression as it is a constant and is statistically decorrelated from the density perturbations, However, \pshot still affects the cross-covariance through its indirect impact on the cross-correlation coefficient, $r(k_j,z)$.

To continue evaluating Eq.~(\ref{eq:cross-cov-app}), we expand the covariance of the matter power spectrum. By using our definition of the estimator of the matter power spectrum in Eq.~(\ref{eq:pmmhat}), the covariance of the matter power spectrum evaluates to:
\begin{align}
\text{Cov}&\Big[\hat{P}^\text{mm}(k_i,z_\text{eff}),\hat{P}^{\text{mm}}(k_j,z)\Big] =V_\text{f}^2 \int_{k^\text{bin}_i} \frac{d^3\bm{k}}{V_\text{s}(k_i)} \int_{k^\text{bin}_j} \frac{d^3\bm{k'}}{V_\text{s}(k_j)}\nonumber\\
&\times\big[\langle \delta_{\text{m}}(\bm{k},z_\text{eff})\delta_{\text{m}}(\bm{k'},z)\rangle\langle \delta_{\text{m}}(-\bm{k},z_\text{eff})\delta_{\text{m}}(\bm{-k'},z)\rangle\nonumber\\
&-\langle \delta_{\text{m}}(\bm{k},z_\text{eff})\delta_{\text{m}}(\bm{-k'},z)\rangle\langle \delta_{\text{m}}(-\bm{k},z_\text{eff}) \delta_{\text{m}}(\bm{k'},z)\rangle\big]\,,
\label{eq:covpmmfirstformul}
\end{align}
where we have neglected higher-order correlators beyond the power spectra. If the matter perturbations are evaluated at the same redshifts (i.e. $z=z_\text{eff}$), the above equation simply gives the Gaussian contributions to the covariance matrix for the matter power spectrum. However, in our case, the matter perturbations are evaluated at different redshifts as the value of $z$ varies within the integration limits set by the redshift distribution [$f^{g}(z)$, see Eq.~(\ref{eq:fg})]. Given that this redshift distribution spans over a relatively small redshift interval ($0.43\leq z\leq 0.69$), which encompasses $z_\text{eff} (=0.57)$, we normalize the matter perturbations evaluated at $z$ to redshift $z_\text{eff}$ by assuming linear growth of matter perturbations $\delta_{\text{m}}(k,z) = D_+(z)\delta_{\text{m}}(k,0)= [D_+(z)/D_+(z_\text{eff})]\delta_{\text{m}}(k,z_\text{eff})$, yielding:
\begin{align}
\text{Cov}&\Big[\hat{P}^\text{mm}(k_i,z_\text{eff}),\hat{P}^{\text{mm}}(k_j,z)\Big] \nonumber\\
&= \frac{V_\text{f}}{V_\text{s}(k_i)}\frac{D^{2}_+(z)}{D^2_+(z_\text{eff})}2\big[P^\text{mm}(k_i,z_\text{eff})\big]^2\Big|_{|k_i - k_j|\leq \frac{\delta k_i}{2}}\,.
\label{eq:pmmgausscov}
\end{align}
Therefore, we retrieve a Gaussian covariance that has been normalized by the amplitude of the growing mode, $D^+$. The Gaussian covariance has been proven to work very well for modes $k\leq 0.1 \hmpc$. For $k>0.1 \hmpc$, however, the covariance is dependent on the mode-couplings to a larger extent \cite{Scoccimarro_1999, Takada_2013,Mohammed_2016}. If the cross-covariance is significant, ignoring these mode-couplings could impact the bounds on the bias parameters. The Gaussian contribution, however, is likely to be sufficient for our purposes: we aim to investigate if there is an indication that including the cross-covariance is of significant importance, and have found the answer to be no, at least given the precision of current galaxy clustering data. Note that with the increased precision of future datasets as well as the with use of more non-linear models, non-Gaussian contributions to the cross-covariance need to be considered. Therefore, in future work, it will be extremely important to extend the covariance to include non-Gaussian contributions, including e.g. trispectrum contributions and binning-scheme dependent contributions.

By inserting the Gaussian covariance in Eq.~(\ref{eq:pmmgausscov}) into Eq.~(\ref{eq:cross-cov-app}), we arrive at the final expression of the cross-covariance:
\begin{align}
\label{eq:cross-cov-finalapp}
\text{Cov}&\Big[\hat{P}^\text{gg}(k_i,z_\text{eff}),\hat{P}^{\kappa \text{g}} (\ell_j)\Big] = \int dz \frac{H(z)}{\chi^2(z)}W^{\kappa}(z)f^{g}(z) \\
& \times \frac{V_\text{f}}{V_\text{s}(k_i)}\frac{D^{2}_+(z)}{D^2_+(z_\text{eff})}2P^\text{mg}(k_i,z_\text{eff})P^\text{gg}_*(k_i,z_\text{eff})\Big|_{\big|k_i - \frac{\ell_j}{\chi(z)}\big|\leq \frac{\delta k_i}{2}}\,,\nonumber
\end{align}
where we have absorbed the following factors: the auto-galaxy bias and the redshift-space distortions into ${P^\text{gg}_*=P^\text{gg}_\text{th}-\pshot}$ and the cross-galaxy bias and cross-correlation coefficient into ${P^\text{mg}=\cbias rP^\text{mm}}$. For illustrative purposes, we plot the full cross-covariance matrix in Fig.~\ref{fig:full_cov} together with the individual covariance matrices for \pggobs and \clkgobs, and find that the results fully confirm our qualitative expectations.

\footnotesize

\bibliography{Mnu_constraints_JHEAp.bib}

\providecommand{\href}[2]{#2}\begingroup\raggedright\begin{thebibliography}{100}

\bibitem{Mohapatra:2005wg}
R.~N. Mohapatra et~al., \emph{{Theory of neutrinos: A White paper}},
  \href{https://doi.org/10.1088/0034-4885/70/11/R02}{\emph{Rept. Prog. Phys.}
  {\bfseries 70} (2007) 1757--1867},
  [\href{https://arxiv.org/abs/hep-ph/0510213}{{\ttfamily hep-ph/0510213}}].

\bibitem{Fukuda_1998}
{\scshape Super-Kamiokande} collaboration, Y.~Fukuda et~al., \emph{{Evidence
  for oscillation of atmospheric neutrinos}},
  \href{https://doi.org/10.1103/PhysRevLett.81.1562}{\emph{Phys. Rev. Lett.}
  {\bfseries 81} (1998) 1562--1567},
  [\href{https://arxiv.org/abs/hep-ex/9807003}{{\ttfamily hep-ex/9807003}}].

\bibitem{SNO:2002tuh}
{\scshape SNO} collaboration, Q.~R. Ahmad et~al., \emph{{Direct evidence for
  neutrino flavor transformation from neutral current interactions in the
  Sudbury Neutrino Observatory}},
  \href{https://doi.org/10.1103/PhysRevLett.89.011301}{\emph{Phys. Rev. Lett.}
  {\bfseries 89} (2002) 011301},
  [\href{https://arxiv.org/abs/nucl-ex/0204008}{{\ttfamily nucl-ex/0204008}}].

\bibitem{Gonzalez_Garcia_2008}
M.~C. Gonzalez-Garcia and M.~Maltoni, \emph{{Phenomenology with Massive
  Neutrinos}}, \href{https://doi.org/10.1016/j.physrep.2007.12.004}{\emph{Phys.
  Rept.} {\bfseries 460} (2008) 1--129},
  [\href{https://arxiv.org/abs/0704.1800}{{\ttfamily 0704.1800}}].

\bibitem{Gonzalez_Garcia_2012}
M.~C. Gonzalez-Garcia, M.~Maltoni, J.~Salvado and T.~Schwetz, \emph{{Global fit
  to three neutrino mixing: critical look at present precision}},
  \href{https://doi.org/10.1007/JHEP12(2012)123}{\emph{JHEP} {\bfseries 12}
  (2012) 123}, [\href{https://arxiv.org/abs/1209.3023}{{\ttfamily 1209.3023}}].

\bibitem{de_Salas_2018}
P.~F. de~Salas, S.~Gariazzo, O.~Mena, C.~A. Ternes and M.~T\'ortola,
  \emph{{Neutrino Mass Ordering from Oscillations and Beyond: 2018 Status and
  Future Prospects}},
  \href{https://doi.org/10.3389/fspas.2018.00036}{\emph{Front. Astron. Space
  Sci.} {\bfseries 5} (2018) 36},
  [\href{https://arxiv.org/abs/1806.11051}{{\ttfamily 1806.11051}}].

\bibitem{deSalas:2017kay}
P.~F. de~Salas, D.~V. Forero, C.~A. Ternes, M.~Tortola and J.~W.~F. Valle,
  \emph{{Status of neutrino oscillations 2018: 3$\sigma$ hint for normal mass
  ordering and improved CP sensitivity}},
  \href{https://doi.org/10.1016/j.physletb.2018.06.019}{\emph{Phys. Lett. B}
  {\bfseries 782} (2018) 633--640},
  [\href{https://arxiv.org/abs/1708.01186}{{\ttfamily 1708.01186}}].

\bibitem{Capozzi:2017ipn}
F.~Capozzi, E.~Di~Valentino, E.~Lisi, A.~Marrone, A.~Melchiorri and A.~Palazzo,
  \emph{{Global constraints on absolute neutrino masses and their ordering}},
  \href{https://doi.org/10.1103/PhysRevD.95.096014}{\emph{Phys. Rev. D}
  {\bfseries 95} (2017) 096014},
  [\href{https://arxiv.org/abs/2003.08511}{{\ttfamily 2003.08511}}].

\bibitem{Esteban:2018azc}
I.~Esteban, M.~C. Gonzalez-Garcia, A.~Hernandez-Cabezudo, M.~Maltoni and
  T.~Schwetz, \emph{{Global analysis of three-flavour neutrino oscillations:
  synergies and tensions in the determination of $\theta_{23}$, $\delta_{CP}$,
  and the mass ordering}},
  \href{https://doi.org/10.1007/JHEP01(2019)106}{\emph{JHEP} {\bfseries 01}
  (2019) 106}, [\href{https://arxiv.org/abs/1811.05487}{{\ttfamily
  1811.05487}}].

\bibitem{Hagstotz:2020ukm}
S.~Hagstotz, P.~F. de~Salas, S.~Gariazzo, M.~Gerbino, M.~Lattanzi, S.~Vagnozzi
  et~al., \emph{{Bounds on light sterile neutrino mass and mixing from
  cosmology and laboratory searches}},
  \href{https://doi.org/10.1103/PhysRevD.104.123524}{\emph{Phys. Rev. D}
  {\bfseries 104} (2021) 123524},
  [\href{https://arxiv.org/abs/2003.02289}{{\ttfamily 2003.02289}}].

\bibitem{Esteban:2020cvm}
I.~Esteban, M.~C. Gonzalez-Garcia, M.~Maltoni, T.~Schwetz and A.~Zhou,
  \emph{{The fate of hints: updated global analysis of three-flavor neutrino
  oscillations}}, \href{https://doi.org/10.1007/JHEP09(2020)178}{\emph{JHEP}
  {\bfseries 09} (2020) 178},
  [\href{https://arxiv.org/abs/2007.14792}{{\ttfamily 2007.14792}}].

\bibitem{Capozzi:2021fjo}
F.~Capozzi, E.~Di~Valentino, E.~Lisi, A.~Marrone, A.~Melchiorri and A.~Palazzo,
  \emph{{Unfinished fabric of the three neutrino paradigm}},
  \href{https://doi.org/10.1103/PhysRevD.104.083031}{\emph{Phys. Rev. D}
  {\bfseries 104} (2021) 083031},
  [\href{https://arxiv.org/abs/2107.00532}{{\ttfamily 2107.00532}}].

\bibitem{Archidiacono:2022ich}
M.~Archidiacono and S.~Gariazzo, \emph{{Two Sides of the Same Coin: Sterile
  Neutrinos and Dark Radiation, Status and Perspectives}},
  \href{https://doi.org/10.3390/universe8030175}{\emph{Universe} {\bfseries 8}
  (2022) 175}, [\href{https://arxiv.org/abs/2201.10319}{{\ttfamily
  2201.10319}}].

\bibitem{Otten_2008}
E.~W. Otten and C.~Weinheimer, \emph{{Neutrino mass limit from tritium beta
  decay}}, \href{https://doi.org/10.1088/0034-4885/71/8/086201}{\emph{Rept.
  Prog. Phys.} {\bfseries 71} (2008) 086201},
  [\href{https://arxiv.org/abs/0909.2104}{{\ttfamily 0909.2104}}].

\bibitem{aker2021direct}
M.~{Aker}, A.~{Beglarian}, J.~{Behrens}, A.~{Berlev}, U.~{Besserer},
  B.~{Bieringer} et~al., \emph{{First direct neutrino-mass measurement with
  sub-eV sensitivity}}, {\emph{arXiv e-prints} (May, 2021) arXiv:2105.08533},
  [\href{https://arxiv.org/abs/2105.08533}{{\ttfamily 2105.08533}}].

\bibitem{DellOro:2016tmg}
S.~Dell'Oro, S.~Marcocci, M.~Viel and F.~Vissani, \emph{{Neutrinoless double
  beta decay: 2015 review}},
  \href{https://doi.org/10.1155/2016/2162659}{\emph{Adv. High Energy Phys.}
  {\bfseries 2016} (2016) 2162659},
  [\href{https://arxiv.org/abs/1601.07512}{{\ttfamily 1601.07512}}].

\bibitem{Dolinski_2019}
M.~J. Dolinski, A.~W.~P. Poon and W.~Rodejohann, \emph{{Neutrinoless
  Double-Beta Decay: Status and Prospects}},
  \href{https://doi.org/10.1146/annurev-nucl-101918-023407}{\emph{Ann. Rev.
  Nucl. Part. Sci.} {\bfseries 69} (2019) 219--251},
  [\href{https://arxiv.org/abs/1902.04097}{{\ttfamily 1902.04097}}].

\bibitem{LESGOURGUES_2006}
J.~Lesgourgues and S.~Pastor, \emph{{Massive neutrinos and cosmology}},
  \href{https://doi.org/10.1016/j.physrep.2006.04.001}{\emph{Phys. Rept.}
  {\bfseries 429} (2006) 307--379},
  [\href{https://arxiv.org/abs/astro-ph/0603494}{{\ttfamily
  astro-ph/0603494}}].

\bibitem{lattanzi2017status}
M.~Lattanzi and M.~Gerbino, \emph{{Status of neutrino properties and future
  prospects - Cosmological and astrophysical constraints}},
  \href{https://doi.org/10.3389/fphy.2017.00070}{\emph{Front. in Phys.}
  {\bfseries 5} (2018) 70}, [\href{https://arxiv.org/abs/1712.07109}{{\ttfamily
  1712.07109}}].

\bibitem{Sakr:2022ans}
Z.~Sakr, \emph{{A Short Review on the Latest Neutrinos Mass and Number
  Constraints from Cosmological Observables}},
  \href{https://doi.org/10.3390/universe8050284}{\emph{Universe} {\bfseries 8}
  (2022) 284}.

\bibitem{Bennett_2003}
{\scshape WMAP} collaboration, C.~L. Bennett et~al., \emph{{First year
  Wilkinson Microwave Anisotropy Probe (WMAP) observations: Preliminary maps
  and basic results}}, \href{https://doi.org/10.1086/377253}{\emph{Astrophys.
  J. Suppl.} {\bfseries 148} (2003) 1--27},
  [\href{https://arxiv.org/abs/astro-ph/0302207}{{\ttfamily
  astro-ph/0302207}}].

\bibitem{Planck_2015_cosmoparams}
{\scshape Planck} collaboration, P.~A.~R. Ade et~al., \emph{{Planck 2015
  results. XIII. Cosmological parameters}},
  \href{https://doi.org/10.1051/0004-6361/201525830}{\emph{Astron. Astrophys.}
  {\bfseries 594} (2016) A13},
  [\href{https://arxiv.org/abs/1502.01589}{{\ttfamily 1502.01589}}].

\bibitem{Planck_2018_cosmoparams}
{\scshape Planck} collaboration, N.~Aghanim et~al., \emph{{Planck 2018 results.
  VI. Cosmological parameters}},
  \href{https://doi.org/10.1051/0004-6361/201833910}{\emph{Astron. Astrophys.}
  {\bfseries 641} (2020) A6},
  [\href{https://arxiv.org/abs/1807.06209}{{\ttfamily 1807.06209}}].

\bibitem{Cole_2005}
{\scshape 2dFGRS} collaboration, S.~Cole et~al., \emph{{The 2dF Galaxy Redshift
  Survey: Power-spectrum analysis of the final dataset and cosmological
  implications}},
  \href{https://doi.org/10.1111/j.1365-2966.2005.09318.x}{\emph{Mon. Not. Roy.
  Astron. Soc.} {\bfseries 362} (2005) 505--534},
  [\href{https://arxiv.org/abs/astro-ph/0501174}{{\ttfamily
  astro-ph/0501174}}].

\bibitem{Eisenstein_2005}
{\scshape SDSS} collaboration, D.~J. Eisenstein et~al., \emph{{Detection of the
  Baryon Acoustic Peak in the Large-Scale Correlation Function of SDSS Luminous
  Red Galaxies}}, \href{https://doi.org/10.1086/466512}{\emph{Astrophys. J.}
  {\bfseries 633} (2005) 560--574},
  [\href{https://arxiv.org/abs/astro-ph/0501171}{{\ttfamily
  astro-ph/0501171}}].

\bibitem{Cuesta_2016}
A.~J. Cuesta et~al., \emph{{The clustering of galaxies in the SDSS-III Baryon
  Oscillation Spectroscopic Survey: Baryon Acoustic Oscillations in the
  correlation function of LOWZ and CMASS galaxies in Data Release 12}},
  \href{https://doi.org/10.1093/mnras/stw066}{\emph{Mon. Not. Roy. Astron.
  Soc.} {\bfseries 457} (2016) 1770--1785},
  [\href{https://arxiv.org/abs/1509.06371}{{\ttfamily 1509.06371}}].

\bibitem{Cuesta:2015iho}
A.~J. Cuesta, V.~Niro and L.~Verde, \emph{{Neutrino mass limits: robust
  information from the power spectrum of galaxy surveys}},
  \href{https://doi.org/10.1016/j.dark.2016.04.005}{\emph{Phys. Dark Univ.}
  {\bfseries 13} (2016) 77--86},
  [\href{https://arxiv.org/abs/1511.05983}{{\ttfamily 1511.05983}}].

\bibitem{Wang:2016tsz}
S.~Wang, Y.-F. Wang, D.-M. Xia and X.~Zhang, \emph{{Impacts of dark energy on
  weighing neutrinos: mass hierarchies considered}},
  \href{https://doi.org/10.1103/PhysRevD.94.083519}{\emph{Phys. Rev. D}
  {\bfseries 94} (2016) 083519},
  [\href{https://arxiv.org/abs/1608.00672}{{\ttfamily 1608.00672}}].

\bibitem{Lorenz:2017fgo}
C.~S. Lorenz, E.~Calabrese and D.~Alonso, \emph{{Distinguishing between
  Neutrinos and time-varying Dark Energy through Cosmic Time}},
  \href{https://doi.org/10.1103/PhysRevD.96.043510}{\emph{Phys. Rev. D}
  {\bfseries 96} (2017) 043510},
  [\href{https://arxiv.org/abs/1706.00730}{{\ttfamily 1706.00730}}].

\bibitem{Wang:2017htc}
S.~Wang, Y.-F. Wang and D.-M. Xia, \emph{{Constraints on the sum of neutrino
  masses using cosmological data including the latest extended Baryon
  Oscillation Spectroscopic Survey DR14 quasar sample}},
  \href{https://doi.org/10.1088/1674-1137/42/6/065103}{\emph{Chin. Phys. C}
  {\bfseries 42} (2018) 065103},
  [\href{https://arxiv.org/abs/1707.00588}{{\ttfamily 1707.00588}}].

\bibitem{Chen:2017ayg}
L.~Chen, Q.-G. Huang and K.~Wang, \emph{{New cosmological constraints with
  extended-Baryon Oscillation Spectroscopic Survey DR14 quasar sample}},
  \href{https://doi.org/10.1140/epjc/s10052-017-5344-1}{\emph{Eur. Phys. J. C}
  {\bfseries 77} (2017) 762},
  [\href{https://arxiv.org/abs/1707.02742}{{\ttfamily 1707.02742}}].

\bibitem{Zhao:2017jma}
M.-M. Zhao, J.-F. Zhang and X.~Zhang, \emph{{Measuring growth index in a
  universe with massive neutrinos: A revisit of the general relativity test
  with the latest observations}},
  \href{https://doi.org/10.1016/j.physletb.2018.02.042}{\emph{Phys. Lett. B}
  {\bfseries 779} (2018) 473--478},
  [\href{https://arxiv.org/abs/1710.02391}{{\ttfamily 1710.02391}}].

\bibitem{Nunes:2017xon}
R.~C. Nunes and A.~Bonilla, \emph{{Probing the properties of relic neutrinos
  using the cosmic microwave background, the Hubble Space Telescope and galaxy
  clusters}}, \href{https://doi.org/10.1093/mnras/stx2661}{\emph{Mon. Not. Roy.
  Astron. Soc.} {\bfseries 473} (2018) 4404--4409},
  [\href{https://arxiv.org/abs/1710.10264}{{\ttfamily 1710.10264}}].

\bibitem{Vagnozzi:2018jhn}
S.~Vagnozzi, S.~Dhawan, M.~Gerbino, K.~Freese, A.~Goobar and O.~Mena,
  \emph{{Constraints on the sum of the neutrino masses in dynamical dark energy
  models with $w(z) \geq -1$ are tighter than those obtained in $\Lambda$CDM}},
  \href{https://doi.org/10.1103/PhysRevD.98.083501}{\emph{Phys. Rev. D}
  {\bfseries 98} (2018) 083501},
  [\href{https://arxiv.org/abs/1801.08553}{{\ttfamily 1801.08553}}].

\bibitem{Guo:2018gyo}
R.-Y. Guo, J.-F. Zhang and X.~Zhang, \emph{{Exploring neutrino mass and mass
  hierarchy in the scenario of vacuum energy interacting with cold dark
  matte}}, \href{https://doi.org/10.1088/1674-1137/42/9/095103}{\emph{Chin.
  Phys. C} {\bfseries 42} (2018) 095103},
  [\href{https://arxiv.org/abs/1803.06910}{{\ttfamily 1803.06910}}].

\bibitem{RoyChoudhury:2018gay}
S.~Roy~Choudhury and S.~Choubey, \emph{{Updated Bounds on Sum of Neutrino
  Masses in Various Cosmological Scenarios}},
  \href{https://doi.org/10.1088/1475-7516/2018/09/017}{\emph{JCAP} {\bfseries
  09} (2018) 017}, [\href{https://arxiv.org/abs/1806.10832}{{\ttfamily
  1806.10832}}].

\bibitem{RoyChoudhury:2018vnm}
S.~Roy~Choudhury and A.~Naskar, \emph{{Strong Bounds on Sum of Neutrino Masses
  in a 12 Parameter Extended Scenario with Non-Phantom Dynamical Dark Energy
  ($w(z)\geq -1$) with CPL parameterization}},
  \href{https://doi.org/10.1140/epjc/s10052-019-6762-z}{\emph{Eur. Phys. J. C}
  {\bfseries 79} (2019) 262},
  [\href{https://arxiv.org/abs/1807.02860}{{\ttfamily 1807.02860}}].

\bibitem{Bonilla:2018nau}
A.~Bonilla, R.~C. Nunes and E.~M.~C. Abreu, \emph{{Forecast on lepton asymmetry
  from future CMB experiments}},
  \href{https://doi.org/10.1093/mnras/stz524}{\emph{Mon. Not. Roy. Astron.
  Soc.} {\bfseries 485} (2019) 2486--2491},
  [\href{https://arxiv.org/abs/1810.06356}{{\ttfamily 1810.06356}}].

\bibitem{Lorenz:2018fzb}
C.~S. Lorenz, L.~Funcke, E.~Calabrese and S.~Hannestad, \emph{{Time-varying
  neutrino mass from a supercooled phase transition: current cosmological
  constraints and impact on the $\Omega_m$-$\sigma_8$ plane}},
  \href{https://doi.org/10.1103/PhysRevD.99.023501}{\emph{Phys. Rev. D}
  {\bfseries 99} (2019) 023501},
  [\href{https://arxiv.org/abs/1811.01991}{{\ttfamily 1811.01991}}].

\bibitem{Bolliet:2019zuz}
B.~Bolliet, T.~Brinckmann, J.~Chluba and J.~Lesgourgues, \emph{{Including
  massive neutrinos in thermal Sunyaev Zeldovich power spectrum and cluster
  counts analyses}}, \href{https://doi.org/10.1093/mnras/staa1835}{\emph{Mon.
  Not. Roy. Astron. Soc.} {\bfseries 497} (2020) 1332--1347},
  [\href{https://arxiv.org/abs/1906.10359}{{\ttfamily 1906.10359}}].

\bibitem{Vagnozzi:2019ezj}
S.~Vagnozzi, \emph{{New physics in light of the $H_0$ tension: An alternative
  view}}, \href{https://doi.org/10.1103/PhysRevD.102.023518}{\emph{Phys. Rev.
  D} {\bfseries 102} (2020) 023518},
  [\href{https://arxiv.org/abs/1907.07569}{{\ttfamily 1907.07569}}].

\bibitem{Yang:2019uog}
W.~Yang, S.~Pan, R.~C. Nunes and D.~F. Mota, \emph{{Dark calling Dark:
  Interaction in the dark sector in presence of neutrino properties after
  Planck CMB final release}},
  \href{https://doi.org/10.1088/1475-7516/2020/04/008}{\emph{JCAP} {\bfseries
  04} (2020) 008}, [\href{https://arxiv.org/abs/1910.08821}{{\ttfamily
  1910.08821}}].

\bibitem{Palanque-Delabrouille:2019iyz}
N.~Palanque-Delabrouille, C.~Y\`eche, N.~Sch\"oneberg, J.~Lesgourgues,
  M.~Walther, S.~Chabanier et~al., \emph{{Hints, neutrino bounds and WDM
  constraints from SDSS DR14 Lyman-$\alpha$ and Planck full-survey data}},
  \href{https://doi.org/10.1088/1475-7516/2020/04/038}{\emph{JCAP} {\bfseries
  04} (2020) 038}, [\href{https://arxiv.org/abs/1911.09073}{{\ttfamily
  1911.09073}}].

\bibitem{Nunes:2020hzy}
R.~C. Nunes, S.~K. Yadav, J.~F. Jesus and A.~Bernui, \emph{{Cosmological
  parameter analyses using transversal BAO data}},
  \href{https://doi.org/10.1093/mnras/staa2036}{\emph{Mon. Not. Roy. Astron.
  Soc.} {\bfseries 497} (2020) 2133--2141},
  [\href{https://arxiv.org/abs/2002.09293}{{\ttfamily 2002.09293}}].

\bibitem{Yang:2020tax}
W.~Yang, E.~Di~Valentino, O.~Mena and S.~Pan, \emph{{Dynamical Dark sectors and
  Neutrino masses and abundances}},
  \href{https://doi.org/10.1103/PhysRevD.102.023535}{\emph{Phys. Rev. D}
  {\bfseries 102} (2020) 023535},
  [\href{https://arxiv.org/abs/2003.12552}{{\ttfamily 2003.12552}}].

\bibitem{Zhang:2020mox}
M.~Zhang, J.-F. Zhang and X.~Zhang, \emph{{Impacts of dark energy on
  constraining neutrino mass after Planck 2018}},
  \href{https://doi.org/10.1088/1572-9494/abbb84}{\emph{Commun. Theor. Phys.}
  {\bfseries 72} (2020) 125402},
  [\href{https://arxiv.org/abs/2005.04647}{{\ttfamily 2005.04647}}].

\bibitem{Li:2020gtk}
H.-L. Li, J.-F. Zhang and X.~Zhang, \emph{{Constraints on neutrino mass in the
  scenario of vacuum energy interacting with cold dark matter after Planck
  2018}}, \href{https://doi.org/10.1088/1572-9494/abb7c9}{\emph{Commun. Theor.
  Phys.} {\bfseries 72} (2020) 125401},
  [\href{https://arxiv.org/abs/2005.12041}{{\ttfamily 2005.12041}}].

\bibitem{Yang:2020ope}
W.~Yang, E.~Di~Valentino, S.~Pan and O.~Mena, \emph{{Emergent Dark Energy,
  neutrinos and cosmological tensions}},
  \href{https://doi.org/10.1016/j.dark.2020.100762}{\emph{Phys. Dark Univ.}
  {\bfseries 31} (2021) 100762},
  [\href{https://arxiv.org/abs/2007.02927}{{\ttfamily 2007.02927}}].

\bibitem{Giare:2020vzo}
W.~Giar\`e, E.~Di~Valentino, A.~Melchiorri and O.~Mena, \emph{{New cosmological
  bounds on hot relics: axions and neutrinos}},
  \href{https://doi.org/10.1093/mnras/stab1442}{\emph{Mon. Not. Roy. Astron.
  Soc.} {\bfseries 505} (2021) 2703--2711},
  [\href{https://arxiv.org/abs/2011.14704}{{\ttfamily 2011.14704}}].

\bibitem{RoyChoudhury:2020dmd}
S.~Roy~Choudhury, S.~Hannestad and T.~Tram, \emph{{Updated constraints on
  massive neutrino self-interactions from cosmology in light of the $H_0$
  tension}}, \href{https://doi.org/10.1088/1475-7516/2021/03/084}{\emph{JCAP}
  {\bfseries 03} (2021) 084},
  [\href{https://arxiv.org/abs/2012.07519}{{\ttfamily 2012.07519}}].

\bibitem{Brinckmann:2020bcn}
T.~Brinckmann, J.~H. Chang and M.~LoVerde, \emph{{Self-interacting neutrinos,
  the Hubble parameter tension, and the cosmic microwave background}},
  \href{https://doi.org/10.1103/PhysRevD.104.063523}{\emph{Phys. Rev. D}
  {\bfseries 104} (2021) 063523},
  [\href{https://arxiv.org/abs/2012.11830}{{\ttfamily 2012.11830}}].

\bibitem{DiValentino:2021zxy}
E.~Di~Valentino, S.~Pan, W.~Yang and L.~A. Anchordoqui, \emph{{Touch of
  neutrinos on the vacuum metamorphosis: Is the $H_0$ solution back?}},
  \href{https://doi.org/10.1103/PhysRevD.103.123527}{\emph{Phys. Rev. D}
  {\bfseries 103} (2021) 123527},
  [\href{https://arxiv.org/abs/2102.05641}{{\ttfamily 2102.05641}}].

\bibitem{DiValentino:2021hoh}
E.~Di~Valentino, S.~Gariazzo and O.~Mena, \emph{{Most constraining cosmological
  neutrino mass bounds}},
  \href{https://doi.org/10.1103/PhysRevD.104.083504}{\emph{Phys. Rev. D}
  {\bfseries 104} (2021) 083504},
  [\href{https://arxiv.org/abs/2106.15267}{{\ttfamily 2106.15267}}].

\bibitem{Anchordoqui:2021gji}
L.~A. Anchordoqui, E.~Di~Valentino, S.~Pan and W.~Yang, \emph{{Dissecting the
  H0 and S8 tensions with Planck + BAO + supernova type Ia in multi-parameter
  cosmologies}},
  \href{https://doi.org/10.1016/j.jheap.2021.08.001}{\emph{JHEAp} {\bfseries
  32} (2021) 28--64}, [\href{https://arxiv.org/abs/2107.13932}{{\ttfamily
  2107.13932}}].

\bibitem{Feng:2021ipq}
L.~Feng, R.-Y. Guo, J.-F. Zhang and X.~Zhang, \emph{{Cosmological search for
  sterile neutrinos after Planck 2018}},
  \href{https://doi.org/10.1016/j.physletb.2022.136940}{\emph{Phys. Lett. B}
  {\bfseries 827} (2022) 136940},
  [\href{https://arxiv.org/abs/2109.06111}{{\ttfamily 2109.06111}}].

\bibitem{DiValentino:2021rjj}
E.~Di~Valentino, S.~Gariazzo, C.~Giunti, O.~Mena, S.~Pan and W.~Yang,
  \emph{{Minimal dark energy: Key to sterile neutrino and Hubble constant
  tensions?}}, \href{https://doi.org/10.1103/PhysRevD.105.103511}{\emph{Phys.
  Rev. D} {\bfseries 105} (2022) 103511},
  [\href{https://arxiv.org/abs/2110.03990}{{\ttfamily 2110.03990}}].

\bibitem{Renzi:2021xii}
F.~Renzi, N.~B. Hogg and W.~Giar\`e, \emph{{The resilience of the
  Etherington\textendash{}Hubble relation}},
  \href{https://doi.org/10.1093/mnras/stac1030}{\emph{Mon. Not. Roy. Astron.
  Soc.} {\bfseries 513} (2022) 4004--4014},
  [\href{https://arxiv.org/abs/2112.05701}{{\ttfamily 2112.05701}}].

\bibitem{Jin:2022tdf}
S.-J. {Jin}, R.-Q. {Zhu}, L.-F. {Wang}, H.-L. {Li}, J.-F. {Zhang} and
  X.~{Zhang}, \emph{{Impacts of gravitational-wave standard siren observations
  from Einstein Telescope and Cosmic Explorer on weighing neutrinos in
  interacting dark energy models}}, {\emph{arXiv e-prints} (Apr., 2022)
  arXiv:2204.04689}, [\href{https://arxiv.org/abs/2204.04689}{{\ttfamily
  2204.04689}}].

\bibitem{Kumar:2022vee}
S.~{Kumar}, R.~C. {Nunes} and P.~{Yadav}, \emph{{Updating non-standard
  neutrinos properties with Planck-CMB data and full-shape analysis of BOSS and
  eBOSS galaxies}}, {\emph{arXiv e-prints} (May, 2022) arXiv:2205.04292},
  [\href{https://arxiv.org/abs/2205.04292}{{\ttfamily 2205.04292}}].

\bibitem{Reeves:2022aoi}
A.~{Reeves}, L.~{Herold}, S.~{Vagnozzi}, B.~D. {Sherwin} and E.~G.~M.
  {Ferreira}, \emph{{Restoring cosmological concordance with early dark energy
  and massive neutrinos?}}, {\emph{arXiv e-prints} (July, 2022)
  arXiv:2207.01501}, [\href{https://arxiv.org/abs/2207.01501}{{\ttfamily
  2207.01501}}].

\bibitem{DiValentino:2022njd}
E.~{Di Valentino}, S.~{Gariazzo} and O.~{Mena}, \emph{{Model marginalized
  constraints on neutrino properties from cosmology}}, {\emph{arXiv e-prints}
  (July, 2022) arXiv:2207.05167},
  [\href{https://arxiv.org/abs/2207.05167}{{\ttfamily 2207.05167}}].

\bibitem{Wetterich:2007kr}
C.~Wetterich, \emph{{Growing neutrinos and cosmological selection}},
  \href{https://doi.org/10.1016/j.physletb.2007.08.060}{\emph{Phys. Lett. B}
  {\bfseries 655} (2007) 201--208},
  [\href{https://arxiv.org/abs/0706.4427}{{\ttfamily 0706.4427}}].

\bibitem{WaliHossain:2014usl}
M.~Wali~Hossain, R.~Myrzakulov, M.~Sami and E.~N. Saridakis, \emph{{Unification
  of inflation and dark energy \`a la quintessential inflation}},
  \href{https://doi.org/10.1142/S0218271815300141}{\emph{Int. J. Mod. Phys. D}
  {\bfseries 24} (2015) 1530014},
  [\href{https://arxiv.org/abs/1410.6100}{{\ttfamily 1410.6100}}].

\bibitem{Geng:2015haa}
C.-Q. Geng, C.-C. Lee, R.~Myrzakulov, M.~Sami and E.~N. Saridakis,
  \emph{{Observational constraints on varying neutrino-mass cosmology}},
  \href{https://doi.org/10.1088/1475-7516/2016/01/049}{\emph{JCAP} {\bfseries
  01} (2016) 049}, [\href{https://arxiv.org/abs/1504.08141}{{\ttfamily
  1504.08141}}].

\bibitem{Chacko:2020hmh}
Z.~Chacko, A.~Dev, P.~Du, V.~Poulin and Y.~Tsai, \emph{{Determining the
  Neutrino Lifetime from Cosmology}},
  \href{https://doi.org/10.1103/PhysRevD.103.043519}{\emph{Phys. Rev. D}
  {\bfseries 103} (2021) 043519},
  [\href{https://arxiv.org/abs/2002.08401}{{\ttfamily 2002.08401}}].

\bibitem{Lorenz:2021alz}
C.~S. Lorenz, L.~Funcke, M.~L\"offler and E.~Calabrese, \emph{{Reconstruction
  of the neutrino mass as a function of redshift}},
  \href{https://doi.org/10.1103/PhysRevD.104.123518}{\emph{Phys. Rev. D}
  {\bfseries 104} (2021) 123518},
  [\href{https://arxiv.org/abs/2102.13618}{{\ttfamily 2102.13618}}].

\bibitem{Green:2021xzn}
D.~{Green} and J.~{Meyers}, \emph{{Cosmological Implications of a Neutrino Mass
  Detection}}, {\emph{arXiv e-prints} (Nov., 2021) arXiv:2111.01096},
  [\href{https://arxiv.org/abs/2111.01096}{{\ttfamily 2111.01096}}].

\bibitem{Alvey:2021xmq}
J.~Alvey, M.~Escudero, N.~Sabti and T.~Schwetz, \emph{{Cosmic neutrino
  background detection in large-neutrino-mass cosmologies}},
  \href{https://doi.org/10.1103/PhysRevD.105.063501}{\emph{Phys. Rev. D}
  {\bfseries 105} (2022) 063501},
  [\href{https://arxiv.org/abs/2111.14870}{{\ttfamily 2111.14870}}].

\bibitem{Huang:2015wrx}
Q.-G. Huang, K.~Wang and S.~Wang, \emph{{Constraints on the neutrino mass and
  mass hierarchy from cosmological observations}},
  \href{https://doi.org/10.1140/epjc/s10052-016-4334-z}{\emph{Eur. Phys. J. C}
  {\bfseries 76} (2016) 489},
  [\href{https://arxiv.org/abs/1512.05899}{{\ttfamily 1512.05899}}].

\bibitem{Hannestad:2016fog}
S.~Hannestad and T.~Schwetz, \emph{{Cosmology and the neutrino mass ordering}},
  \href{https://doi.org/10.1088/1475-7516/2016/11/035}{\emph{JCAP} {\bfseries
  11} (2016) 035}, [\href{https://arxiv.org/abs/1606.04691}{{\ttfamily
  1606.04691}}].

\bibitem{Xu:2016ddc}
L.~Xu and Q.-G. Huang, \emph{{Detecting the Neutrinos Mass Hierarchy from
  Cosmological Data}},
  \href{https://doi.org/10.1007/s11433-017-9125-0}{\emph{Sci. China Phys. Mech.
  Astron.} {\bfseries 61} (2018) 039521},
  [\href{https://arxiv.org/abs/1611.05178}{{\ttfamily 1611.05178}}].

\bibitem{Gerbino:2016ehw}
M.~Gerbino, M.~Lattanzi, O.~Mena and K.~Freese, \emph{{A novel approach to
  quantifying the sensitivity of current and future cosmological datasets to
  the neutrino mass ordering through Bayesian hierarchical modeling}},
  \href{https://doi.org/10.1016/j.physletb.2017.10.052}{\emph{Phys. Lett. B}
  {\bfseries 775} (2017) 239--250},
  [\href{https://arxiv.org/abs/1611.07847}{{\ttfamily 1611.07847}}].

\bibitem{Yang:2017amu}
W.~Yang, R.~C. Nunes, S.~Pan and D.~F. Mota, \emph{{Effects of neutrino mass
  hierarchies on dynamical dark energy models}},
  \href{https://doi.org/10.1103/PhysRevD.95.103522}{\emph{Phys. Rev. D}
  {\bfseries 95} (2017) 103522},
  [\href{https://arxiv.org/abs/1703.02556}{{\ttfamily 1703.02556}}].

\bibitem{Simpson:2017qvj}
F.~Simpson, R.~Jimenez, C.~Pena-Garay and L.~Verde, \emph{{Strong Bayesian
  Evidence for the Normal Neutrino Hierarchy}},
  \href{https://doi.org/10.1088/1475-7516/2017/06/029}{\emph{JCAP} {\bfseries
  06} (2017) 029}, [\href{https://arxiv.org/abs/1703.03425}{{\ttfamily
  1703.03425}}].

\bibitem{Schwetz:2017fey}
T.~{Schwetz}, K.~{Freese}, M.~{Gerbino}, E.~{Giusarma}, S.~{Hannestad},
  M.~{Lattanzi} et~al., \emph{{Comment on ``Strong Evidence for the Normal
  Neutrino Hierarchy''}}, {\emph{arXiv e-prints} (Mar., 2017)
  arXiv:1703.04585}, [\href{https://arxiv.org/abs/1703.04585}{{\ttfamily
  1703.04585}}].

\bibitem{Long:2017dru}
A.~J. Long, M.~Raveri, W.~Hu and S.~Dodelson, \emph{{Neutrino Mass Priors for
  Cosmology from Random Matrices}},
  \href{https://doi.org/10.1103/PhysRevD.97.043510}{\emph{Phys. Rev. D}
  {\bfseries 97} (2018) 043510},
  [\href{https://arxiv.org/abs/1711.08434}{{\ttfamily 1711.08434}}].

\bibitem{Gariazzo:2018pei}
S.~Gariazzo, M.~Archidiacono, P.~F. de~Salas, O.~Mena, C.~A. Ternes and
  M.~T\'ortola, \emph{{Neutrino masses and their ordering: Global Data, Priors
  and Models}},
  \href{https://doi.org/10.1088/1475-7516/2018/03/011}{\emph{JCAP} {\bfseries
  03} (2018) 011}, [\href{https://arxiv.org/abs/1801.04946}{{\ttfamily
  1801.04946}}].

\bibitem{Heavens:2018adv}
A.~F. Heavens and E.~Sellentin, \emph{{Objective Bayesian analysis of neutrino
  masses and hierarchy}},
  \href{https://doi.org/10.1088/1475-7516/2018/04/047}{\emph{JCAP} {\bfseries
  04} (2018) 047}, [\href{https://arxiv.org/abs/1802.09450}{{\ttfamily
  1802.09450}}].

\bibitem{Mahony:2019fyb}
C.~Mahony, B.~Leistedt, H.~V. Peiris, J.~Braden, B.~Joachimi, A.~Korn et~al.,
  \emph{{Target Neutrino Mass Precision for Determining the Neutrino
  Hierarchy}}, \href{https://doi.org/10.1103/PhysRevD.101.083513}{\emph{Phys.
  Rev. D} {\bfseries 101} (2020) 083513},
  [\href{https://arxiv.org/abs/1907.04331}{{\ttfamily 1907.04331}}].

\bibitem{RoyChoudhury:2019hls}
S.~Roy~Choudhury and S.~Hannestad, \emph{{Updated results on neutrino mass and
  mass hierarchy from cosmology with Planck 2018 likelihoods}},
  \href{https://doi.org/10.1088/1475-7516/2020/07/037}{\emph{JCAP} {\bfseries
  07} (2020) 037}, [\href{https://arxiv.org/abs/1907.12598}{{\ttfamily
  1907.12598}}].

\bibitem{Hergt:2021qlh}
L.~T. Hergt, W.~J. Handley, M.~P. Hobson and A.~N. Lasenby, \emph{{Bayesian
  evidence for the tensor-to-scalar ratio $r$ and neutrino masses $m_\nu$:
  Effects of uniform vs logarithmic priors}},
  \href{https://doi.org/10.1103/PhysRevD.103.123511}{\emph{Phys. Rev. D}
  {\bfseries 103} (2021) 123511},
  [\href{https://arxiv.org/abs/2102.11511}{{\ttfamily 2102.11511}}].

\bibitem{Jimenez:2022dkn}
R.~{Jimenez}, C.~{Pena-Garay}, K.~{Short}, F.~{Simpson} and L.~{Verde},
  \emph{{Neutrino Masses and Mass Hierarchy: Evidence for the Normal
  Hierarchy}}, {\emph{arXiv e-prints} (Mar., 2022) arXiv:2203.14247},
  [\href{https://arxiv.org/abs/2203.14247}{{\ttfamily 2203.14247}}].

\bibitem{Gariazzo:2022ahe}
S.~{Gariazzo}, M.~{Gerbino}, T.~{Brinckmann}, M.~{Lattanzi}, O.~{Mena},
  T.~{Schwetz} et~al., \emph{{Neutrino mass and mass ordering: No conclusive
  evidence for normal ordering}}, {\emph{arXiv e-prints} (May, 2022)
  arXiv:2205.02195}, [\href{https://arxiv.org/abs/2205.02195}{{\ttfamily
  2205.02195}}].

\bibitem{Bond_1980}
J.~R. Bond, G.~Efstathiou and J.~Silk, \emph{{Massive Neutrinos and the Large
  Scale Structure of the Universe}},
  \href{https://doi.org/10.1103/PhysRevLett.45.1980}{\emph{Phys. Rev. Lett.}
  {\bfseries 45} (1980) 1980--1984}.

\bibitem{Eisenstein:1997jh}
D.~J. Eisenstein and W.~Hu, \emph{{Power spectra for cold dark matter and its
  variants}}, \href{https://doi.org/10.1086/306640}{\emph{Astrophys. J.}
  {\bfseries 511} (1997) 5},
  [\href{https://arxiv.org/abs/astro-ph/9710252}{{\ttfamily
  astro-ph/9710252}}].

\bibitem{Hu_1998}
W.~Hu, D.~J. Eisenstein and M.~Tegmark, \emph{{Weighing neutrinos with galaxy
  surveys}}, \href{https://doi.org/10.1103/PhysRevLett.80.5255}{\emph{Phys.
  Rev. Lett.} {\bfseries 80} (1998) 5255--5258},
  [\href{https://arxiv.org/abs/astro-ph/9712057}{{\ttfamily
  astro-ph/9712057}}].

\bibitem{Vagnozzi:2019utt}
S.~{Vagnozzi}, \emph{{Cosmological searches for the neutrino mass scale and
  mass ordering}}, {\emph{arXiv e-prints} (July, 2019) arXiv:1907.08010},
  [\href{https://arxiv.org/abs/1907.08010}{{\ttfamily 1907.08010}}].

\bibitem{Hou:2011ec}
Z.~Hou, R.~Keisler, L.~Knox, M.~Millea and C.~Reichardt, \emph{{How Massless
  Neutrinos Affect the Cosmic Microwave Background Damping Tail}},
  \href{https://doi.org/10.1103/PhysRevD.87.083008}{\emph{Phys. Rev. D}
  {\bfseries 87} (2013) 083008},
  [\href{https://arxiv.org/abs/1104.2333}{{\ttfamily 1104.2333}}].

\bibitem{Cabass:2015xfa}
G.~Cabass, M.~Gerbino, E.~Giusarma, A.~Melchiorri, L.~Pagano and L.~Salvati,
  \emph{{Constraints on the early and late integrated Sachs-Wolfe effects from
  the Planck 2015 cosmic microwave background anisotropies in the angular power
  spectra}}, \href{https://doi.org/10.1103/PhysRevD.92.063534}{\emph{Phys. Rev.
  D} {\bfseries 92} (2015) 063534},
  [\href{https://arxiv.org/abs/1507.07586}{{\ttfamily 1507.07586}}].

\bibitem{Kable:2020hcw}
J.~A. Kable, G.~E. Addison and C.~L. Bennett, \emph{{Deconstructing the Planck
  TT Power Spectrum to Constrain Deviations from \ensuremath{\Lambda}CDM}},
  \href{https://doi.org/10.3847/1538-4357/abc4e7}{\emph{Astrophys. J.}
  {\bfseries 905} (2020) 164},
  [\href{https://arxiv.org/abs/2008.01785}{{\ttfamily 2008.01785}}].

\bibitem{Vagnozzi:2021gjh}
S.~Vagnozzi, \emph{{Consistency tests of \ensuremath{\Lambda}CDM from the early
  integrated Sachs-Wolfe effect: Implications for early-time new physics and
  the Hubble tension}},
  \href{https://doi.org/10.1103/PhysRevD.104.063524}{\emph{Phys. Rev. D}
  {\bfseries 104} (2021) 063524},
  [\href{https://arxiv.org/abs/2105.10425}{{\ttfamily 2105.10425}}].

\bibitem{Ruiz-Granda:2022bcn}
M.~{Ruiz-Granda} and P.~{Vielva}, \emph{{Constraining CMB physical processes
  using Planck 2018 data}}, {\emph{arXiv e-prints} (June, 2022)
  arXiv:2206.00731}, [\href{https://arxiv.org/abs/2206.00731}{{\ttfamily
  2206.00731}}].

\bibitem{abazajian2016cmbs4}
K.~N. {Abazajian}, P.~{Adshead}, Z.~{Ahmed}, S.~W. {Allen}, D.~{Alonso}, K.~S.
  {Arnold} et~al., \emph{{CMB-S4 Science Book, First Edition}}, {\emph{arXiv
  e-prints} (Oct., 2016) arXiv:1610.02743},
  [\href{https://arxiv.org/abs/1610.02743}{{\ttfamily 1610.02743}}].

\bibitem{Oyama_2016}
Y.~Oyama, K.~Kohri and M.~Hazumi, \emph{{Constraints on the neutrino parameters
  by future cosmological 21 cm line and precise CMB polarization
  observations}},
  \href{https://doi.org/10.1088/1475-7516/2016/02/008}{\emph{JCAP} {\bfseries
  02} (2016) 008}, [\href{https://arxiv.org/abs/1510.03806}{{\ttfamily
  1510.03806}}].

\bibitem{Sprenger_2019}
T.~Sprenger, M.~Archidiacono, T.~Brinckmann, S.~Clesse and J.~Lesgourgues,
  \emph{{Cosmology in the era of Euclid and the Square Kilometre Array}},
  \href{https://doi.org/10.1088/1475-7516/2019/02/047}{\emph{JCAP} {\bfseries
  02} (2019) 047}, [\href{https://arxiv.org/abs/1801.08331}{{\ttfamily
  1801.08331}}].

\bibitem{Bernardeau_2002}
F.~Bernardeau, S.~Colombi, E.~Gaztanaga and R.~Scoccimarro, \emph{{Large scale
  structure of the universe and cosmological perturbation theory}},
  \href{https://doi.org/10.1016/S0370-1573(02)00135-7}{\emph{Phys. Rept.}
  {\bfseries 367} (2002) 1--248},
  [\href{https://arxiv.org/abs/astro-ph/0112551}{{\ttfamily
  astro-ph/0112551}}].

\bibitem{Desjacques_2018_review}
V.~Desjacques, D.~Jeong and F.~Schmidt, \emph{{Large-Scale Galaxy Bias}},
  \href{https://doi.org/10.1016/j.physrep.2017.12.002}{\emph{Phys. Rept.}
  {\bfseries 733} (2018) 1--193},
  [\href{https://arxiv.org/abs/1611.09787}{{\ttfamily 1611.09787}}].

\bibitem{LEWIS_2006}
A.~Lewis and A.~Challinor, \emph{{Weak gravitational lensing of the CMB}},
  \href{https://doi.org/10.1016/j.physrep.2006.03.002}{\emph{Phys. Rept.}
  {\bfseries 429} (2006) 1--65},
  [\href{https://arxiv.org/abs/astro-ph/0601594}{{\ttfamily
  astro-ph/0601594}}].

\bibitem{Hildebrandt_2016}
H.~Hildebrandt et~al., \emph{{KiDS-450: Cosmological parameter constraints from
  tomographic weak gravitational lensing}},
  \href{https://doi.org/10.1093/mnras/stw2805}{\emph{Mon. Not. Roy. Astron.
  Soc.} {\bfseries 465} (2017) 1454},
  [\href{https://arxiv.org/abs/1606.05338}{{\ttfamily 1606.05338}}].

\bibitem{Abbott_2021_DESy3}
{\scshape DES} collaboration, T.~M.~C. Abbott et~al., \emph{{Dark Energy Survey
  Year 3 results: Cosmological constraints from galaxy clustering and weak
  lensing}}, \href{https://doi.org/10.1103/PhysRevD.105.023520}{\emph{Phys.
  Rev. D} {\bfseries 105} (2022) 023520},
  [\href{https://arxiv.org/abs/2105.13549}{{\ttfamily 2105.13549}}].

\bibitem{Gil-Marin:2014baa}
H.~Gil-Mar\'\i{}n, L.~Verde, J.~Nore\~na, A.~J. Cuesta, L.~Samushia, W.~J.
  Percival et~al., \emph{{The power spectrum and bispectrum of SDSS DR11 BOSS
  galaxies \textendash{} II. Cosmological interpretation}},
  \href{https://doi.org/10.1093/mnras/stv1359}{\emph{Mon. Not. Roy. Astron.
  Soc.} {\bfseries 452} (2015) 1914--1921},
  [\href{https://arxiv.org/abs/1408.0027}{{\ttfamily 1408.0027}}].

\bibitem{BOSS:2016teh}
{\scshape BOSS} collaboration, J.~N. Grieb et~al., \emph{{The clustering of
  galaxies in the completed SDSS-III Baryon Oscillation Spectroscopic Survey:
  Cosmological implications of the Fourier space wedges of the final sample}},
  \href{https://doi.org/10.1093/mnras/stw3384}{\emph{Mon. Not. Roy. Astron.
  Soc.} {\bfseries 467} (2017) 2085--2112},
  [\href{https://arxiv.org/abs/1607.03143}{{\ttfamily 1607.03143}}].

\bibitem{BOSS:2016chr}
{\scshape BOSS} collaboration, A.~G. Sanchez et~al., \emph{{The clustering of
  galaxies in the completed SDSS-III Baryon Oscillation Spectroscopic Survey:
  combining correlated Gaussian posterior distributions}},
  \href{https://doi.org/10.1093/mnras/stw2495}{\emph{Mon. Not. Roy. Astron.
  Soc.} {\bfseries 464} (2017) 1493--1501},
  [\href{https://arxiv.org/abs/1607.03146}{{\ttfamily 1607.03146}}].

\bibitem{BOSS:2016off}
{\scshape BOSS} collaboration, A.~G. Sanchez et~al., \emph{{The clustering of
  galaxies in the completed SDSS-III Baryon Oscillation Spectroscopic Survey:
  cosmological implications of the configuration-space clustering wedges}},
  \href{https://doi.org/10.1093/mnras/stw2443}{\emph{Mon. Not. Roy. Astron.
  Soc.} {\bfseries 464} (2017) 1640--1658},
  [\href{https://arxiv.org/abs/1607.03147}{{\ttfamily 1607.03147}}].

\bibitem{BOSS:2016hvq}
{\scshape BOSS} collaboration, F.~Beutler et~al., \emph{{The clustering of
  galaxies in the completed SDSS-III Baryon Oscillation Spectroscopic Survey:
  baryon acoustic oscillations in the Fourier space}},
  \href{https://doi.org/10.1093/mnras/stw2373}{\emph{Mon. Not. Roy. Astron.
  Soc.} {\bfseries 464} (2017) 3409--3430},
  [\href{https://arxiv.org/abs/1607.03149}{{\ttfamily 1607.03149}}].

\bibitem{BOSS:2016psr}
{\scshape BOSS} collaboration, F.~Beutler et~al., \emph{{The clustering of
  galaxies in the completed SDSS-III Baryon Oscillation Spectroscopic Survey:
  Anisotropic galaxy clustering in Fourier-space}},
  \href{https://doi.org/10.1093/mnras/stw3298}{\emph{Mon. Not. Roy. Astron.
  Soc.} {\bfseries 466} (2017) 2242--2260},
  [\href{https://arxiv.org/abs/1607.03150}{{\ttfamily 1607.03150}}].

\bibitem{Gil-Marin:2018cgo}
H.~Gil-Mar\'\i{}n et~al., \emph{{The clustering of the SDSS-IV extended Baryon
  Oscillation Spectroscopic Survey DR14 quasar sample: structure growth rate
  measurement from the anisotropic quasar power spectrum in the redshift range
  $0.8 < z < 2.2$}}, \href{https://doi.org/10.1093/mnras/sty453}{\emph{Mon.
  Not. Roy. Astron. Soc.} {\bfseries 477} (2018) 1604--1638},
  [\href{https://arxiv.org/abs/1801.02689}{{\ttfamily 1801.02689}}].

\bibitem{Gil-Marin:2020bct}
H.~Gil-Mar\'{i}n et~al., \emph{{The Completed SDSS-IV extended Baryon
  Oscillation Spectroscopic Survey: measurement of the BAO and growth rate of
  structure of the luminous red galaxy sample from the anisotropic power
  spectrum between redshifts 0.6 and 1.0}},
  \href{https://doi.org/10.1093/mnras/staa2455}{\emph{Mon. Not. Roy. Astron.
  Soc.} {\bfseries 498} (2020) 2492--2531},
  [\href{https://arxiv.org/abs/2007.08994}{{\ttfamily 2007.08994}}].

\bibitem{Semenaite:2021aen}
A.~Semenaite et~al., \emph{{Cosmological implications of the full shape of
  anisotropic clustering measurements in BOSS and eBOSS}},
  \href{https://doi.org/10.1093/mnras/stac829}{\emph{Mon. Not. Roy. Astron.
  Soc.} {\bfseries 512} (2022) 5657--5670},
  [\href{https://arxiv.org/abs/2111.03156}{{\ttfamily 2111.03156}}].

\bibitem{Neveux:2022tuk}
R.~{Neveux}, E.~{Burtin}, A.~{de Mattia}, A.~{Semenaite}, K.~S. {Dawson},
  A.~{de la Macorra} et~al., \emph{{Combined full shape analysis of BOSS
  galaxies and eBOSS quasars using an iterative emulator}}, {\emph{arXiv
  e-prints} (Jan., 2022) arXiv:2201.04679},
  [\href{https://arxiv.org/abs/2201.04679}{{\ttfamily 2201.04679}}].

\bibitem{Baumann:2010tm}
D.~Baumann, A.~Nicolis, L.~Senatore and M.~Zaldarriaga, \emph{{Cosmological
  Non-Linearities as an Effective Fluid}},
  \href{https://doi.org/10.1088/1475-7516/2012/07/051}{\emph{JCAP} {\bfseries
  07} (2012) 051}, [\href{https://arxiv.org/abs/1004.2488}{{\ttfamily
  1004.2488}}].

\bibitem{Carrasco:2012cv}
J.~J.~M. Carrasco, M.~P. Hertzberg and L.~Senatore, \emph{{The Effective Field
  Theory of Cosmological Large Scale Structures}},
  \href{https://doi.org/10.1007/JHEP09(2012)082}{\emph{JHEP} {\bfseries 09}
  (2012) 082}, [\href{https://arxiv.org/abs/1206.2926}{{\ttfamily 1206.2926}}].

\bibitem{Pajer:2013jj}
E.~Pajer and M.~Zaldarriaga, \emph{{On the Renormalization of the Effective
  Field Theory of Large Scale Structures}},
  \href{https://doi.org/10.1088/1475-7516/2013/08/037}{\emph{JCAP} {\bfseries
  08} (2013) 037}, [\href{https://arxiv.org/abs/1301.7182}{{\ttfamily
  1301.7182}}].

\bibitem{Senatore:2014via}
L.~Senatore and M.~Zaldarriaga, \emph{{The IR-resummed Effective Field Theory
  of Large Scale Structures}},
  \href{https://doi.org/10.1088/1475-7516/2015/02/013}{\emph{JCAP} {\bfseries
  02} (2015) 013}, [\href{https://arxiv.org/abs/1404.5954}{{\ttfamily
  1404.5954}}].

\bibitem{Senatore:2014eva}
L.~Senatore, \emph{{Bias in the Effective Field Theory of Large Scale
  Structures}},
  \href{https://doi.org/10.1088/1475-7516/2015/11/007}{\emph{JCAP} {\bfseries
  11} (2015) 007}, [\href{https://arxiv.org/abs/1406.7843}{{\ttfamily
  1406.7843}}].

\bibitem{DAmico:2019fhj}
G.~D'Amico, J.~Gleyzes, N.~Kokron, K.~Markovic, L.~Senatore, P.~Zhang et~al.,
  \emph{{The Cosmological Analysis of the SDSS/BOSS data from the Effective
  Field Theory of Large-Scale Structure}},
  \href{https://doi.org/10.1088/1475-7516/2020/05/005}{\emph{JCAP} {\bfseries
  05} (2020) 005}, [\href{https://arxiv.org/abs/1909.05271}{{\ttfamily
  1909.05271}}].

\bibitem{Ivanov_2020}
M.~M. Ivanov, M.~Simonovi\'c and M.~Zaldarriaga, \emph{{Cosmological Parameters
  from the BOSS Galaxy Power Spectrum}},
  \href{https://doi.org/10.1088/1475-7516/2020/05/042}{\emph{JCAP} {\bfseries
  05} (2020) 042}, [\href{https://arxiv.org/abs/1909.05277}{{\ttfamily
  1909.05277}}].

\bibitem{Colas:2019ret}
T.~Colas, G.~D'amico, L.~Senatore, P.~Zhang and F.~Beutler, \emph{{Efficient
  Cosmological Analysis of the SDSS/BOSS data from the Effective Field Theory
  of Large-Scale Structure}},
  \href{https://doi.org/10.1088/1475-7516/2020/06/001}{\emph{JCAP} {\bfseries
  06} (2020) 001}, [\href{https://arxiv.org/abs/1909.07951}{{\ttfamily
  1909.07951}}].

\bibitem{Ivanov:2019hqk}
M.~M. Ivanov, M.~Simonovi\'c and M.~Zaldarriaga, \emph{{Cosmological Parameters
  and Neutrino Masses from the Final Planck and Full-Shape BOSS Data}},
  \href{https://doi.org/10.1103/PhysRevD.101.083504}{\emph{Phys. Rev. D}
  {\bfseries 101} (2020) 083504},
  [\href{https://arxiv.org/abs/1912.08208}{{\ttfamily 1912.08208}}].

\bibitem{Philcox:2020vvt}
O.~H.~E. Philcox, M.~M. Ivanov, M.~Simonovi\'c and M.~Zaldarriaga,
  \emph{{Combining Full-Shape and BAO Analyses of Galaxy Power Spectra: A
  1.6\textbackslash{}\% CMB-independent constraint on H$_0$}},
  \href{https://doi.org/10.1088/1475-7516/2020/05/032}{\emph{JCAP} {\bfseries
  05} (2020) 032}, [\href{https://arxiv.org/abs/2002.04035}{{\ttfamily
  2002.04035}}].

\bibitem{DAmico:2020kxu}
G.~D'Amico, L.~Senatore and P.~Zhang, \emph{{Limits on $w$CDM from the EFTofLSS
  with the PyBird code}},
  \href{https://doi.org/10.1088/1475-7516/2021/01/006}{\emph{JCAP} {\bfseries
  01} (2021) 006}, [\href{https://arxiv.org/abs/2003.07956}{{\ttfamily
  2003.07956}}].

\bibitem{Nishimichi:2020tvu}
T.~Nishimichi, G.~D'Amico, M.~M. Ivanov, L.~Senatore, M.~Simonovi\'c, M.~Takada
  et~al., \emph{{Blinded challenge for precision cosmology with large-scale
  structure: results from effective field theory for the redshift-space galaxy
  power spectrum}},
  \href{https://doi.org/10.1103/PhysRevD.102.123541}{\emph{Phys. Rev. D}
  {\bfseries 102} (2020) 123541},
  [\href{https://arxiv.org/abs/2003.08277}{{\ttfamily 2003.08277}}].

\bibitem{Chudaykin:2020aoj}
A.~Chudaykin, M.~M. Ivanov, O.~H.~E. Philcox and M.~Simonovi\'c,
  \emph{{Nonlinear perturbation theory extension of the Boltzmann code CLASS}},
  \href{https://doi.org/10.1103/PhysRevD.102.063533}{\emph{Phys. Rev. D}
  {\bfseries 102} (2020) 063533},
  [\href{https://arxiv.org/abs/2004.10607}{{\ttfamily 2004.10607}}].

\bibitem{Ivanov:2020ril}
M.~M. Ivanov, E.~McDonough, J.~C. Hill, M.~Simonovi\'c, M.~W. Toomey,
  S.~Alexander et~al., \emph{{Constraining Early Dark Energy with Large-Scale
  Structure}}, \href{https://doi.org/10.1103/PhysRevD.102.103502}{\emph{Phys.
  Rev. D} {\bfseries 102} (2020) 103502},
  [\href{https://arxiv.org/abs/2006.11235}{{\ttfamily 2006.11235}}].

\bibitem{DAmico:2020ods}
G.~D'Amico, L.~Senatore, P.~Zhang and H.~Zheng, \emph{{The Hubble Tension in
  Light of the Full-Shape Analysis of Large-Scale Structure Data}},
  \href{https://doi.org/10.1088/1475-7516/2021/05/072}{\emph{JCAP} {\bfseries
  05} (2021) 072}, [\href{https://arxiv.org/abs/2006.12420}{{\ttfamily
  2006.12420}}].

\bibitem{Philcox:2020xbv}
O.~H.~E. Philcox, B.~D. Sherwin, G.~S. Farren and E.~J. Baxter,
  \emph{{Determining the Hubble Constant without the Sound Horizon:
  Measurements from Galaxy Surveys}},
  \href{https://doi.org/10.1103/PhysRevD.103.023538}{\emph{Phys. Rev. D}
  {\bfseries 103} (2021) 023538},
  [\href{https://arxiv.org/abs/2008.08084}{{\ttfamily 2008.08084}}].

\bibitem{Wadekar:2020hax}
D.~Wadekar, M.~M. Ivanov and R.~Scoccimarro, \emph{{Cosmological constraints
  from BOSS with analytic covariance matrices}},
  \href{https://doi.org/10.1103/PhysRevD.102.123521}{\emph{Phys. Rev. D}
  {\bfseries 102} (2020) 123521},
  [\href{https://arxiv.org/abs/2009.00622}{{\ttfamily 2009.00622}}].

\bibitem{Chudaykin:2020ghx}
A.~Chudaykin, K.~Dolgikh and M.~M. Ivanov, \emph{{Constraints on the curvature
  of the Universe and dynamical dark energy from the Full-shape and BAO data}},
  \href{https://doi.org/10.1103/PhysRevD.103.023507}{\emph{Phys. Rev. D}
  {\bfseries 103} (2021) 023507},
  [\href{https://arxiv.org/abs/2009.10106}{{\ttfamily 2009.10106}}].

\bibitem{DAmico:2020tty}
G.~{D'Amico}, Y.~{Donath}, L.~{Senatore} and P.~{Zhang}, \emph{{Limits on
  Clustering and Smooth Quintessence from the EFTofLSS}}, {\emph{arXiv
  e-prints} (Dec., 2020) arXiv:2012.07554},
  [\href{https://arxiv.org/abs/2012.07554}{{\ttfamily 2012.07554}}].

\bibitem{Ivanov:2021zmi}
M.~M. Ivanov, \emph{{Cosmological constraints from the power spectrum of eBOSS
  emission line galaxies}},
  \href{https://doi.org/10.1103/PhysRevD.104.103514}{\emph{Phys. Rev. D}
  {\bfseries 104} (2021) 103514},
  [\href{https://arxiv.org/abs/2106.12580}{{\ttfamily 2106.12580}}].

\bibitem{Philcox:2021ukg}
O.~H.~E. Philcox, \emph{{Cosmology without window functions. II. Cubic
  estimators for the galaxy bispectrum}},
  \href{https://doi.org/10.1103/PhysRevD.104.123529}{\emph{Phys. Rev. D}
  {\bfseries 104} (2021) 123529},
  [\href{https://arxiv.org/abs/2107.06287}{{\ttfamily 2107.06287}}].

\bibitem{Philcox:2021hbm}
O.~H.~E. {Philcox}, J.~{Hou} and Z.~{Slepian}, \emph{{A First Detection of the
  Connected 4-Point Correlation Function of Galaxies Using the BOSS CMASS
  Sample}}, {\emph{arXiv e-prints} (Aug., 2021) arXiv:2108.01670},
  [\href{https://arxiv.org/abs/2108.01670}{{\ttfamily 2108.01670}}].

\bibitem{Ivanov:2021kcd}
M.~M. Ivanov, O.~H.~E. Philcox, T.~Nishimichi, M.~Simonovi\'c, M.~Takada and
  M.~Zaldarriaga, \emph{{Precision analysis of the redshift-space galaxy
  bispectrum}}, \href{https://doi.org/10.1103/PhysRevD.105.063512}{\emph{Phys.
  Rev. D} {\bfseries 105} (2022) 063512},
  [\href{https://arxiv.org/abs/2110.10161}{{\ttfamily 2110.10161}}].

\bibitem{Philcox:2021kcw}
O.~H.~E. Philcox and M.~M. Ivanov, \emph{{BOSS DR12 full-shape cosmology:
  \ensuremath{\Lambda}CDM constraints from the large-scale galaxy power
  spectrum and bispectrum monopole}},
  \href{https://doi.org/10.1103/PhysRevD.105.043517}{\emph{Phys. Rev. D}
  {\bfseries 105} (2022) 043517},
  [\href{https://arxiv.org/abs/2112.04515}{{\ttfamily 2112.04515}}].

\bibitem{Cabass:2022wjy}
G.~{Cabass}, M.~M. {Ivanov}, O.~H.~E. {Philcox}, M.~{Simonovi{\'c}} and
  M.~{Zaldarriaga}, \emph{{Constraints on Single-Field Inflation from the BOSS
  Galaxy Survey}}, {\emph{arXiv e-prints} (Jan., 2022) arXiv:2201.07238},
  [\href{https://arxiv.org/abs/2201.07238}{{\ttfamily 2201.07238}}].

\bibitem{DAmico:2022gki}
G.~{D'Amico}, M.~{Lewandowski}, L.~{Senatore} and P.~{Zhang}, \emph{{Limits on
  primordial non-Gaussianities from BOSS galaxy-clustering data}}, {\emph{arXiv
  e-prints} (Jan., 2022) arXiv:2201.11518},
  [\href{https://arxiv.org/abs/2201.11518}{{\ttfamily 2201.11518}}].

\bibitem{Cabass:2022ymb}
G.~{Cabass}, M.~M. {Ivanov}, O.~H.~E. {Philcox}, M.~{Simonovi{\'c}} and
  M.~{Zaldarriaga}, \emph{{Constraints on Multi-Field Inflation from the BOSS
  Galaxy Survey}}, {\emph{arXiv e-prints} (Apr., 2022) arXiv:2204.01781},
  [\href{https://arxiv.org/abs/2204.01781}{{\ttfamily 2204.01781}}].

\bibitem{Philcox:2022frc}
O.~H.~E. {Philcox}, M.~M. {Ivanov}, G.~{Cabass}, M.~{Simonovi{\'c}},
  M.~{Zaldarriaga} and T.~{Nishimichi}, \emph{{Cosmology with the
  Redshift-Space Galaxy Bispectrum Monopole at One-Loop Order}}, {\emph{arXiv
  e-prints} (June, 2022) arXiv:2206.02800},
  [\href{https://arxiv.org/abs/2206.02800}{{\ttfamily 2206.02800}}].

\bibitem{Hou:2022wfj}
J.~{Hou}, Z.~{Slepian} and R.~N. {Cahn}, \emph{{Measurement of Parity-Odd Modes
  in the Large-Scale 4-Point Correlation Function of SDSS BOSS DR12 CMASS and
  LOWZ Galaxies}}, {\emph{arXiv e-prints} (June, 2022) arXiv:2206.03625},
  [\href{https://arxiv.org/abs/2206.03625}{{\ttfamily 2206.03625}}].

\bibitem{Philcox:2022hkh}
O.~H.~E. {Philcox}, \emph{{Probing Parity-Violation with the Four-Point
  Correlation Function of BOSS Galaxies}}, {\emph{arXiv e-prints} (June, 2022)
  arXiv:2206.04227}, [\href{https://arxiv.org/abs/2206.04227}{{\ttfamily
  2206.04227}}].

\bibitem{DAmico:2022osl}
G.~{D'Amico}, Y.~{Donath}, M.~{Lewandowski}, L.~{Senatore} and P.~{Zhang},
  \emph{{The BOSS bispectrum analysis at one loop from the Effective Field
  Theory of Large-Scale Structure}}, {\emph{arXiv e-prints} (June, 2022)
  arXiv:2206.08327}, [\href{https://arxiv.org/abs/2206.08327}{{\ttfamily
  2206.08327}}].

\bibitem{Chen:2020zjt}
S.-F. Chen, Z.~Vlah, E.~Castorina and M.~White, \emph{{Redshift-Space
  Distortions in Lagrangian Perturbation Theory}},
  \href{https://doi.org/10.1088/1475-7516/2021/03/100}{\emph{JCAP} {\bfseries
  03} (2021) 100}, [\href{https://arxiv.org/abs/2012.04636}{{\ttfamily
  2012.04636}}].

\bibitem{Chen:2021wdi}
S.-F. Chen, Z.~Vlah and M.~White, \emph{{A new analysis of galaxy 2-point
  functions in the BOSS survey, including full-shape information and
  post-reconstruction BAO}},
  \href{https://doi.org/10.1088/1475-7516/2022/02/008}{\emph{JCAP} {\bfseries
  02} (2022) 008}, [\href{https://arxiv.org/abs/2110.05530}{{\ttfamily
  2110.05530}}].

\bibitem{Upadhye:2017hdl}
A.~Upadhye, \emph{{Neutrino mass and dark energy constraints from
  redshift-space distortions}},
  \href{https://doi.org/10.1088/1475-7516/2019/05/041}{\emph{JCAP} {\bfseries
  05} (2019) 041}, [\href{https://arxiv.org/abs/1707.09354}{{\ttfamily
  1707.09354}}].

\bibitem{Loureiro:2018qva}
A.~Loureiro et~al., \emph{{Cosmological measurements from angular power spectra
  analysis of BOSS DR12 tomography}},
  \href{https://doi.org/10.1093/mnras/stz191}{\emph{Mon. Not. Roy. Astron.
  Soc.} {\bfseries 485} (2019) 326--355},
  [\href{https://arxiv.org/abs/1809.07204}{{\ttfamily 1809.07204}}].

\bibitem{Boyle:2017lzt}
A.~Boyle and E.~Komatsu, \emph{{Deconstructing the neutrino mass constraint
  from galaxy redshift surveys}},
  \href{https://doi.org/10.1088/1475-7516/2018/03/035}{\emph{JCAP} {\bfseries
  03} (2018) 035}, [\href{https://arxiv.org/abs/1712.01857}{{\ttfamily
  1712.01857}}].

\bibitem{Boyle:2018rva}
A.~Boyle, \emph{{Understanding the neutrino mass constraints achievable by
  combining CMB lensing and spectroscopic galaxy surveys}},
  \href{https://doi.org/10.1088/1475-7516/2019/04/038}{\emph{JCAP} {\bfseries
  04} (2019) 038}, [\href{https://arxiv.org/abs/1811.07636}{{\ttfamily
  1811.07636}}].

\bibitem{Chudaykin_2019}
A.~Chudaykin and M.~M. Ivanov, \emph{{Measuring neutrino masses with
  large-scale structure: Euclid forecast with controlled theoretical error}},
  \href{https://doi.org/10.1088/1475-7516/2019/11/034}{\emph{JCAP} {\bfseries
  11} (2019) 034}, [\href{https://arxiv.org/abs/1907.06666}{{\ttfamily
  1907.06666}}].

\bibitem{Boyle_2020}
A.~Boyle and F.~Schmidt, \emph{{Neutrino mass constraints beyond linear order:
  cosmology dependence and systematic biases}},
  \href{https://doi.org/10.1088/1475-7516/2021/04/022}{\emph{JCAP} {\bfseries
  04} (2021) 022}, [\href{https://arxiv.org/abs/2011.10594}{{\ttfamily
  2011.10594}}].

\bibitem{Donald-McCann:2021nxc}
J.~{Donald-McCann}, F.~{Beutler}, K.~{Koyama} and M.~{Karamanis},
  \emph{{matryoshka: Halo Model Emulator for the Galaxy Power Spectrum}},
  \href{https://doi.org/10.1093/mnras/stac239}{\emph{Mon. Not. Roy. Astron.
  Soc} (Jan., 2022) }, [\href{https://arxiv.org/abs/2109.15236}{{\ttfamily
  2109.15236}}].

\bibitem{Donald-McCann:2022pac}
J.~{Donald-McCann}, K.~{Koyama} and F.~{Beutler}, \emph{{$\texttt{matryoshka}$
  II: Accelerating Effective Field Theory Analyses of the Galaxy Power
  Spectrum}}, {\emph{arXiv e-prints} (Feb., 2022) arXiv:2202.07557},
  [\href{https://arxiv.org/abs/2202.07557}{{\ttfamily 2202.07557}}].

\bibitem{Angulo:2015eqa}
R.~Angulo, M.~Fasiello, L.~Senatore and Z.~Vlah, \emph{{On the Statistics of
  Biased Tracers in the Effective Field Theory of Large Scale Structures}},
  \href{https://doi.org/10.1088/1475-7516/2015/9/029}{\emph{JCAP} {\bfseries
  09} (2015) 029}, [\href{https://arxiv.org/abs/1503.08826}{{\ttfamily
  1503.08826}}].

\bibitem{Eggemeier:2020umu}
A.~Eggemeier, R.~Scoccimarro, M.~Crocce, A.~Pezzotta and A.~G. S\'anchez,
  \emph{{Testing one-loop galaxy bias: Power spectrum}},
  \href{https://doi.org/10.1103/PhysRevD.102.103530}{\emph{Phys. Rev. D}
  {\bfseries 102} (2020) 103530},
  [\href{https://arxiv.org/abs/2006.09729}{{\ttfamily 2006.09729}}].

\bibitem{Eggemeier:2021cam}
A.~Eggemeier, R.~Scoccimarro, R.~E. Smith, M.~Crocce, A.~Pezzotta and A.~G.
  S\'anchez, \emph{{Testing one-loop galaxy bias: Joint analysis of power
  spectrum and bispectrum}},
  \href{https://doi.org/10.1103/PhysRevD.103.123550}{\emph{Phys. Rev. D}
  {\bfseries 103} (2021) 123550},
  [\href{https://arxiv.org/abs/2102.06902}{{\ttfamily 2102.06902}}].

\bibitem{pezzotta2021testing}
A.~Pezzotta, M.~Crocce, A.~Eggemeier, A.~G. S\'anchez and R.~Scoccimarro,
  \emph{{Testing one-loop galaxy bias: Cosmological constraints from the power
  spectrum}}, \href{https://doi.org/10.1103/PhysRevD.104.043531}{\emph{Phys.
  Rev. D} {\bfseries 104} (2021) 043531},
  [\href{https://arxiv.org/abs/2102.08315}{{\ttfamily 2102.08315}}].

\bibitem{Gualdi:2017iey}
D.~Gualdi, M.~Manera, B.~Joachimi and O.~Lahav, \emph{{Maximal compression of
  the redshift space galaxy power spectrum and bispectrum}},
  \href{https://doi.org/10.1093/mnras/sty261}{\emph{Mon. Not. Roy. Astron.
  Soc.} {\bfseries 476} (2018) 4045--4070},
  [\href{https://arxiv.org/abs/1709.03600}{{\ttfamily 1709.03600}}].

\bibitem{Gualdi:2019ybt}
D.~Gualdi, H.~Gil-Mar\'\i{}n, M.~Manera, B.~Joachimi and O.~Lahav,
  \emph{{Geometrical compression: a new method to enhance the BOSS galaxy
  bispectrum monopole constraints}},
  \href{https://doi.org/10.1093/mnrasl/sly242}{\emph{Mon. Not. Roy. Astron.
  Soc.} {\bfseries 484} (2019) L29--L34},
  [\href{https://arxiv.org/abs/1901.00987}{{\ttfamily 1901.00987}}].

\bibitem{Gualdi:2018pyw}
D.~Gualdi, H.~Gil-Mar\'\i{}n, R.~L. Schuhmann, M.~Manera, B.~Joachimi and
  O.~Lahav, \emph{{Enhancing BOSS bispectrum cosmological constraints with
  maximal compression}}, \href{https://doi.org/10.1093/mnras/stz051}{\emph{Mon.
  Not. Roy. Astron. Soc.} {\bfseries 484} (2019) 3713--3730},
  [\href{https://arxiv.org/abs/1806.02853}{{\ttfamily 1806.02853}}].

\bibitem{Gualdi:2019sfc}
D.~Gualdi, H.~Gil-Mar\'\i{}n, M.~Manera, B.~Joachimi and O.~Lahav,
  \emph{{GEOMAX: beyond linear compression for three-point galaxy clustering
  statistics}}, \href{https://doi.org/10.1093/mnras/staa1941}{\emph{Mon. Not.
  Roy. Astron. Soc.} {\bfseries 497} (2020) 776--792},
  [\href{https://arxiv.org/abs/1912.01011}{{\ttfamily 1912.01011}}].

\bibitem{Brieden:2021edu}
S.~Brieden, H.~Gil-Mar\'\i{}n and L.~Verde, \emph{{ShapeFit: extracting the
  power spectrum shape information in galaxy surveys beyond BAO and RSD}},
  \href{https://doi.org/10.1088/1475-7516/2021/12/054}{\emph{JCAP} {\bfseries
  12} (2021) 054}, [\href{https://arxiv.org/abs/2106.07641}{{\ttfamily
  2106.07641}}].

\bibitem{Brieden:2021cfg}
S.~Brieden, H.~Gil-Mar\'\i{}n and L.~Verde, \emph{{Model-independent versus
  model-dependent interpretation of the SDSS-III BOSS power spectrum: Bridging
  the divide}}, \href{https://doi.org/10.1103/PhysRevD.104.L121301}{\emph{Phys.
  Rev. D} {\bfseries 104} (2021) L121301},
  [\href{https://arxiv.org/abs/2106.11931}{{\ttfamily 2106.11931}}].

\bibitem{Valogiannis:2021chp}
G.~Valogiannis and C.~Dvorkin, \emph{{Towards an Optimal Estimation of
  Cosmological Parameters with the Wavelet Scattering Transform}},
  \href{https://doi.org/10.1103/PhysRevD.105.103534}{\emph{Phys. Rev. D}
  {\bfseries 105} (2022) 103534},
  [\href{https://arxiv.org/abs/2108.07821}{{\ttfamily 2108.07821}}].

\bibitem{Gualdi:2022kwz}
D.~{Gualdi} and L.~{Verde}, \emph{{Integrated trispectrum detection from BOSS
  DR12 NGC CMASS}}, {\emph{arXiv e-prints} (Jan., 2022) arXiv:2201.06932},
  [\href{https://arxiv.org/abs/2201.06932}{{\ttfamily 2201.06932}}].

\bibitem{Gil-Marin:2022hnv}
H.~Gil-Mar\'\i{}n, \emph{{How to optimally combine pre-reconstruction full
  shape and post-reconstruction BAO signals}},
  \href{https://doi.org/10.1088/1475-7516/2022/05/040}{\emph{JCAP} {\bfseries
  05} (2022) 040}, [\href{https://arxiv.org/abs/2203.05581}{{\ttfamily
  2203.05581}}].

\bibitem{Brieden:2022ieb}
S.~Brieden, H.~Gil-Mar\'\i{}n and L.~Verde, \emph{{PT challenge: validation of
  ShapeFit on large-volume, high-resolution mocks}},
  \href{https://doi.org/10.1088/1475-7516/2022/06/005}{\emph{JCAP} {\bfseries
  06} (2022) 005}, [\href{https://arxiv.org/abs/2201.08400}{{\ttfamily
  2201.08400}}].

\bibitem{Brieden:2022lsd}
S.~{Brieden}, H.~{Gil-Mar{\'\i}n} and L.~{Verde}, \emph{{Model-agnostic
  interpretation of 10 billion years of cosmic evolution traced by BOSS and
  eBOSS data}}, {\emph{arXiv e-prints} (Apr., 2022) arXiv:2204.11868},
  [\href{https://arxiv.org/abs/2204.11868}{{\ttfamily 2204.11868}}].

\bibitem{Valogiannis:2022xwu}
G.~{Valogiannis} and C.~{Dvorkin}, \emph{{Going Beyond the Galaxy Power
  Spectrum: an Analysis of BOSS Data with Wavelet Scattering Transforms}},
  {\emph{arXiv e-prints} (Apr., 2022) arXiv:2204.13717},
  [\href{https://arxiv.org/abs/2204.13717}{{\ttfamily 2204.13717}}].

\bibitem{Carbone:2016nzj}
C.~Carbone, M.~Petkova and K.~Dolag, \emph{{DEMNUni: ISW, Rees-Sciama, and
  weak-lensing in the presence of massive neutrinos}},
  \href{https://doi.org/10.1088/1475-7516/2016/07/034}{\emph{JCAP} {\bfseries
  07} (2016) 034}, [\href{https://arxiv.org/abs/1605.02024}{{\ttfamily
  1605.02024}}].

\bibitem{Ruggeri:2017dda}
R.~Ruggeri, E.~Castorina, C.~Carbone and E.~Sefusatti, \emph{{DEMNUni: Massive
  neutrinos and the bispectrum of large scale structures}},
  \href{https://doi.org/10.1088/1475-7516/2018/03/003}{\emph{JCAP} {\bfseries
  03} (2018) 003}, [\href{https://arxiv.org/abs/1712.02334}{{\ttfamily
  1712.02334}}].

\bibitem{Villaescusa-Navarro:2017mfx}
F.~Villaescusa-Navarro, A.~Banerjee, N.~Dalal, E.~Castorina, R.~Scoccimarro,
  R.~Angulo et~al., \emph{{The imprint of neutrinos on clustering in
  redshift-space}},
  \href{https://doi.org/10.3847/1538-4357/aac6bf}{\emph{Astrophys. J.}
  {\bfseries 861} (2018) 53},
  [\href{https://arxiv.org/abs/1708.01154}{{\ttfamily 1708.01154}}].

\bibitem{Zennaro:2017qnp}
M.~Zennaro, J.~Bel, J.~Dossett, C.~Carbone and L.~Guzzo, \emph{{Cosmological
  constraints from galaxy clustering in the presence of massive neutrinos}},
  \href{https://doi.org/10.1093/mnras/sty670}{\emph{Mon. Not. Roy. Astron.
  Soc.} {\bfseries 477} (2018) 491--506},
  [\href{https://arxiv.org/abs/1712.02886}{{\ttfamily 1712.02886}}].

\bibitem{Bird:2018all}
S.~Bird, Y.~Ali-Ha\"\i{}moud, Y.~Feng and J.~Liu, \emph{{An Efficient and
  Accurate Hybrid Method for Simulating Non-Linear Neutrino Structure}},
  \href{https://doi.org/10.1093/mnras/sty2376}{\emph{Mon. Not. Roy. Astron.
  Soc.} {\bfseries 481} (2018) 1486--1500},
  [\href{https://arxiv.org/abs/1803.09854}{{\ttfamily 1803.09854}}].

\bibitem{2019MNRAS.489.5938Z}
M.~{Zennaro}, R.~E. {Angulo}, G.~{Aric{\`o}}, S.~{Contreras} and
  M.~{Pellejero-Ib{\'a}{\~n}ez}, \emph{{How to add massive neutrinos to your
  {\ensuremath{\Lambda}}CDM simulation - extending cosmology rescaling
  algorithms}}, \href{https://doi.org/10.1093/mnras/stz2612}{\emph{Mon. Not.
  Roy. Astron. Soc} {\bfseries 489} (Nov., 2019) 5938--5951},
  [\href{https://arxiv.org/abs/1905.08696}{{\ttfamily 1905.08696}}].

\bibitem{2019arXiv191004255G}
E.~{Giusarma}, M.~{Reyes Hurtado}, F.~{Villaescusa-Navarro}, S.~{He}, S.~{Ho}
  and C.~{Hahn}, \emph{{Learning neutrino effects in Cosmology with
  Convolutional Neural Networks}}, {\emph{arXiv e-prints} (Oct., 2019)
  arXiv:1910.04255}, [\href{https://arxiv.org/abs/1910.04255}{{\ttfamily
  1910.04255}}].

\bibitem{Massara:2020pli}
E.~Massara, F.~Villaescusa-Navarro, S.~Ho, N.~Dalal and D.~N. Spergel,
  \emph{{Using the Marked Power Spectrum to Detect the Signature of Neutrinos
  in Large-Scale Structure}},
  \href{https://doi.org/10.1103/PhysRevLett.126.011301}{\emph{Phys. Rev. Lett.}
  {\bfseries 126} (2021) 011301},
  [\href{https://arxiv.org/abs/2001.11024}{{\ttfamily 2001.11024}}].

\bibitem{Euclid:2020rfv}
{\scshape Euclid} collaboration, M.~Knabenhans et~al., \emph{{Euclid
  preparation: IX. EuclidEmulator2 \textendash{} power spectrum emulation with
  massive neutrinos and self-consistent dark energy perturbations}},
  \href{https://doi.org/10.1093/mnras/stab1366}{\emph{Mon. Not. Roy. Astron.
  Soc.} {\bfseries 505} (2021) 2840--2869},
  [\href{https://arxiv.org/abs/2010.11288}{{\ttfamily 2010.11288}}].

\bibitem{Bose:2021mkz}
B.~Bose, B.~S. Wright, M.~Cataneo, A.~Pourtsidou, C.~Giocoli, L.~Lombriser
  et~al., \emph{{On the road to per cent accuracy \textendash{} V. The
  non-linear power spectrum beyond \ensuremath{\Lambda}CDM with massive
  neutrinos and baryonic feedback}},
  \href{https://doi.org/10.1093/mnras/stab2731}{\emph{Mon. Not. Roy. Astron.
  Soc.} {\bfseries 508} (2021) 2479--2491},
  [\href{https://arxiv.org/abs/2105.12114}{{\ttfamily 2105.12114}}].

\bibitem{Li:2018owg}
Z.~Li, J.~Liu, J.~M.~Z. Matilla and W.~R. Coulton, \emph{{Constraining neutrino
  mass with tomographic weak lensing peak counts}},
  \href{https://doi.org/10.1103/PhysRevD.99.063527}{\emph{Phys. Rev. D}
  {\bfseries 99} (2019) 063527},
  [\href{https://arxiv.org/abs/1810.01781}{{\ttfamily 1810.01781}}].

\bibitem{Coulton:2018ebd}
W.~R. Coulton, J.~Liu, M.~S. Madhavacheril, V.~B\"ohm and D.~N. Spergel,
  \emph{{Constraining Neutrino Mass with the Tomographic Weak Lensing
  Bispectrum}},
  \href{https://doi.org/10.1088/1475-7516/2019/05/043}{\emph{JCAP} {\bfseries
  05} (2019) 043}, [\href{https://arxiv.org/abs/1810.02374}{{\ttfamily
  1810.02374}}].

\bibitem{Uhlemann:2019gni}
C.~Uhlemann, O.~Friedrich, F.~Villaescusa-Navarro, A.~Banerjee and S.~Codis,
  \emph{{Fisher for complements: Extracting cosmology and neutrino mass from
  the counts-in-cells PDF}},
  \href{https://doi.org/10.1093/mnras/staa1155}{\emph{Mon. Not. Roy. Astron.
  Soc.} {\bfseries 495} (2020) 4006--4027},
  [\href{https://arxiv.org/abs/1911.11158}{{\ttfamily 1911.11158}}].

\bibitem{Kamalinejad:2020izi}
F.~{Kamalinejad} and Z.~{Slepian}, \emph{{A Non-Degenerate Neutrino Mass
  Signature in the Galaxy Bispectrum}}, {\emph{arXiv e-prints} (Nov., 2020)
  arXiv:2011.00899}, [\href{https://arxiv.org/abs/2011.00899}{{\ttfamily
  2011.00899}}].

\bibitem{Boyle:2020bqn}
A.~Boyle, C.~Uhlemann, O.~Friedrich, A.~Barthelemy, S.~Codis, F.~Bernardeau
  et~al., \emph{{Nuw CDM cosmology from the weak-lensing convergence PDF}},
  \href{https://doi.org/10.1093/mnras/stab1381}{\emph{Mon. Not. Roy. Astron.
  Soc.} {\bfseries 505} (2021) 2886--2902},
  [\href{https://arxiv.org/abs/2012.07771}{{\ttfamily 2012.07771}}].

\bibitem{ChengChengSiHao:2021hja}
S.~Cheng and B.~M\'enard, \emph{{Weak lensing scattering transform: dark energy
  and neutrino mass sensitivity}},
  \href{https://doi.org/10.1093/mnras/stab2102}{\emph{Mon. Not. Roy. Astron.
  Soc.} {\bfseries 507} (2021) 1012--1020},
  [\href{https://arxiv.org/abs/2103.09247}{{\ttfamily 2103.09247}}].

\bibitem{Aviles:2021que}
A.~Aviles, A.~Banerjee, G.~Niz and Z.~Slepian, \emph{{Clustering in massive
  neutrino cosmologies via Eulerian Perturbation Theory}},
  \href{https://doi.org/10.1088/1475-7516/2021/11/028}{\emph{JCAP} {\bfseries
  11} (2021) 028}, [\href{https://arxiv.org/abs/2106.13771}{{\ttfamily
  2106.13771}}].

\bibitem{Zhou:2021sgl}
S.~Zhou et~al., \emph{{Sensitivity tests of cosmic velocity fields to massive
  neutrinos}}, \href{https://doi.org/10.1093/mnras/stac529}{\emph{Mon. Not.
  Roy. Astron. Soc.} {\bfseries 512} (2022) 3319--3330},
  [\href{https://arxiv.org/abs/2108.12568}{{\ttfamily 2108.12568}}].

\bibitem{MoradinezhadDizgah:2021upg}
A.~Moradinezhad~Dizgah, G.~K. Keating, K.~S. Karkare, A.~Crites and S.~R.
  Choudhury, \emph{{Neutrino Properties with Ground-based Millimeter-wavelength
  Line Intensity Mapping}},
  \href{https://doi.org/10.3847/1538-4357/ac3edd}{\emph{Astrophys. J.}
  {\bfseries 926} (2022) 137},
  [\href{https://arxiv.org/abs/2110.00014}{{\ttfamily 2110.00014}}].

\bibitem{Kamalinejad:2022yyl}
F.~{Kamalinejad} and Z.~{Slepian}, \emph{{A Two-Fluid Treatment of the Effect
  of Neutrinos on the Matter Density}}, {\emph{arXiv e-prints} (Mar., 2022)
  arXiv:2203.13103}, [\href{https://arxiv.org/abs/2203.13103}{{\ttfamily
  2203.13103}}].

\bibitem{Liu:2022vtr}
W.~{Liu}, A.~{Jiang} and W.~{Fang}, \emph{{Probing massive neutrinos with the
  Minkowski functionals of large-scale structure}}, {\emph{arXiv e-prints}
  (Apr., 2022) arXiv:2204.02945},
  [\href{https://arxiv.org/abs/2204.02945}{{\ttfamily 2204.02945}}].

\bibitem{Lee:2022lbu}
J.~{Lee}, S.~{Ryu} and M.~{Baldi}, \emph{{Disentangling Modified Gravity and
  Massive Neutrinos with Intrinsic Shape Alignments of Massive Halos}},
  {\emph{arXiv e-prints} (June, 2022) arXiv:2206.03406},
  [\href{https://arxiv.org/abs/2206.03406}{{\ttfamily 2206.03406}}].

\bibitem{Ryu:2022npy}
S.~{Ryu} and J.~{Lee}, \emph{{The Splashback Mass Function in the Presence of
  Massive Neutrinos}}, {\emph{arXiv e-prints} (June, 2022) arXiv:2206.10068},
  [\href{https://arxiv.org/abs/2206.10068}{{\ttfamily 2206.10068}}].

\bibitem{Cabass:2022avo}
G.~{Cabass}, M.~M. {Ivanov}, M.~{Lewandowski}, M.~{Mirbabayi} and
  M.~{Simonovi{\'c}}, \emph{{Snowmass White Paper: Effective Field Theories in
  Cosmology}}, {\emph{arXiv e-prints} (Mar., 2022) arXiv:2203.08232},
  [\href{https://arxiv.org/abs/2203.08232}{{\ttfamily 2203.08232}}].

\bibitem{Modi:2017wds}
C.~Modi, M.~White and Z.~Vlah, \emph{{Modeling CMB lensing cross correlations
  with CLEFT}},
  \href{https://doi.org/10.1088/1475-7516/2017/08/009}{\emph{JCAP} {\bfseries
  08} (2017) 009}, [\href{https://arxiv.org/abs/1706.03173}{{\ttfamily
  1706.03173}}].

\bibitem{Giusarma:2016phn}
E.~Giusarma, M.~Gerbino, O.~Mena, S.~Vagnozzi, S.~Ho and K.~Freese,
  \emph{{Improvement of cosmological neutrino mass bounds}},
  \href{https://doi.org/10.1103/PhysRevD.94.083522}{\emph{Phys. Rev. D}
  {\bfseries 94} (2016) 083522},
  [\href{https://arxiv.org/abs/1605.04320}{{\ttfamily 1605.04320}}].

\bibitem{Vagnozzi:2017ovm}
S.~Vagnozzi, E.~Giusarma, O.~Mena, K.~Freese, M.~Gerbino, S.~Ho et~al.,
  \emph{{Unveiling $\nu$ secrets with cosmological data: neutrino masses and
  mass hierarchy}},
  \href{https://doi.org/10.1103/PhysRevD.96.123503}{\emph{Phys. Rev. D}
  {\bfseries 96} (2017) 123503},
  [\href{https://arxiv.org/abs/1701.08172}{{\ttfamily 1701.08172}}].

\bibitem{Giusarma:2018jei}
E.~Giusarma, S.~Vagnozzi, S.~Ho, S.~Ferraro, K.~Freese, R.~Kamen-Rubio et~al.,
  \emph{{Scale-dependent galaxy bias, CMB lensing-galaxy cross-correlation, and
  neutrino masses}},
  \href{https://doi.org/10.1103/PhysRevD.98.123526}{\emph{Phys. Rev. D}
  {\bfseries 98} (2018) 123526},
  [\href{https://arxiv.org/abs/1802.08694}{{\ttfamily 1802.08694}}].

\bibitem{Bird_2012}
S.~Bird, M.~Viel and M.~G. Haehnelt, \emph{{Massive Neutrinos and the
  Non-linear Matter Power Spectrum}},
  \href{https://doi.org/10.1111/j.1365-2966.2011.20222.x}{\emph{Mon. Not. Roy.
  Astron. Soc.} {\bfseries 420} (2012) 2551--2561},
  [\href{https://arxiv.org/abs/1109.4416}{{\ttfamily 1109.4416}}].

\bibitem{Desjacques_2010}
V.~Desjacques and R.~K. Sheth, \emph{{Redshift space correlations and
  scale-dependent stochastic biasing of density peaks}},
  \href{https://doi.org/10.1103/PhysRevD.81.023526}{\emph{Phys. Rev. D}
  {\bfseries 81} (2010) 023526},
  [\href{https://arxiv.org/abs/0909.4544}{{\ttfamily 0909.4544}}].

\bibitem{Schmidt_2013}
F.~Schmidt, D.~Jeong and V.~Desjacques, \emph{{Peak-Background Split,
  Renormalization, and Galaxy Clustering}},
  \href{https://doi.org/10.1103/PhysRevD.88.023515}{\emph{Phys. Rev. D}
  {\bfseries 88} (2013) 023515},
  [\href{https://arxiv.org/abs/1212.0868}{{\ttfamily 1212.0868}}].

\bibitem{Musso_2012}
M.~Musso, A.~Paranjape and R.~K. Sheth, \emph{{Scale dependent halo bias in the
  excursion set approach}},
  \href{https://doi.org/10.1111/j.1365-2966.2012.21903.x}{\emph{Mon. Not. Roy.
  Astron. Soc.} {\bfseries 427} (2012) 3145--3158},
  [\href{https://arxiv.org/abs/1205.3401}{{\ttfamily 1205.3401}}].

\bibitem{Senatore_2015}
L.~Senatore, \emph{{Bias in the Effective Field Theory of Large Scale
  Structures}},
  \href{https://doi.org/10.1088/1475-7516/2015/11/007}{\emph{JCAP} {\bfseries
  11} (2015) 007}, [\href{https://arxiv.org/abs/1406.7843}{{\ttfamily
  1406.7843}}].

\bibitem{Raccanelli:2017kht}
A.~Raccanelli, L.~Verde and F.~Villaescusa-Navarro, \emph{{Biases from neutrino
  bias: to worry or not to worry?}},
  \href{https://doi.org/10.1093/mnras/sty2162}{\emph{Mon. Not. Roy. Astron.
  Soc.} {\bfseries 483} (2019) 734--743},
  [\href{https://arxiv.org/abs/1704.07837}{{\ttfamily 1704.07837}}].

\bibitem{Chiang:2018laa}
C.-T. Chiang, M.~LoVerde and F.~Villaescusa-Navarro, \emph{{First detection of
  scale-dependent linear halo bias in $N$-body simulations with massive
  neutrinos}},
  \href{https://doi.org/10.1103/PhysRevLett.122.041302}{\emph{Phys. Rev. Lett.}
  {\bfseries 122} (2019) 041302},
  [\href{https://arxiv.org/abs/1811.12412}{{\ttfamily 1811.12412}}].

\bibitem{Valcin:2019fxe}
D.~Valcin, F.~Villaescusa-Navarro, L.~Verde and A.~Raccanelli, \emph{{BE-HaPPY:
  Bias Emulator for Halo Power Spectrum including massive neutrinos}},
  \href{https://doi.org/10.1088/1475-7516/2019/12/057}{\emph{JCAP} {\bfseries
  12} (2019) 057}, [\href{https://arxiv.org/abs/1901.06045}{{\ttfamily
  1901.06045}}].

\bibitem{Sherwin:2012mr}
B.~D. Sherwin et~al., \emph{{The Atacama Cosmology Telescope: Cross-Correlation
  of CMB Lensing and Quasars}},
  \href{https://doi.org/10.1103/PhysRevD.86.083006}{\emph{Phys. Rev. D}
  {\bfseries 86} (2012) 083006},
  [\href{https://arxiv.org/abs/1207.4543}{{\ttfamily 1207.4543}}].

\bibitem{ACT:2014ivk}
{\scshape ACT} collaboration, M.~Madhavacheril et~al., \emph{{Evidence of
  Lensing of the Cosmic Microwave Background by Dark Matter Halos}},
  \href{https://doi.org/10.1103/PhysRevLett.114.151302}{\emph{Phys. Rev. Lett.}
  {\bfseries 114} (2015) 151302},
  [\href{https://arxiv.org/abs/1411.7999}{{\ttfamily 1411.7999}}].

\bibitem{Baxter:2014frs}
E.~J. Baxter et~al., \emph{{A Measurement of Gravitational Lensing of the
  Cosmic Microwave Background by Galaxy Clusters Using Data from the South Pole
  Telescope}},
  \href{https://doi.org/10.1088/0004-637X/806/2/247}{\emph{Astrophys. J.}
  {\bfseries 806} (2015) 247},
  [\href{https://arxiv.org/abs/1412.7521}{{\ttfamily 1412.7521}}].

\bibitem{ACT:2015eyl}
{\scshape ACT} collaboration, R.~Allison et~al., \emph{{The Atacama Cosmology
  Telescope: measuring radio galaxy bias through cross-correlation with
  lensing}}, \href{https://doi.org/10.1093/mnras/stv991}{\emph{Mon. Not. Roy.
  Astron. Soc.} {\bfseries 451} (2015) 849--858},
  [\href{https://arxiv.org/abs/1502.06456}{{\ttfamily 1502.06456}}].

\bibitem{Bianchini:2015yly}
F.~Bianchini et~al., \emph{{Toward a tomographic analysis of the
  cross-correlation between Planck CMB lensing and H-ATLAS galaxies}},
  \href{https://doi.org/10.3847/0004-637X/825/1/24}{\emph{Astrophys. J.}
  {\bfseries 825} (2016) 24},
  [\href{https://arxiv.org/abs/1511.05116}{{\ttfamily 1511.05116}}].

\bibitem{DES:2017fyz}
{\scshape DES, SPT} collaboration, E.~J. Baxter et~al., \emph{{A measurement of
  CMB cluster lensing with SPT and DES year 1 data}},
  \href{https://doi.org/10.1093/mnras/sty305}{\emph{Mon. Not. Roy. Astron.
  Soc.} {\bfseries 476} (2018) 2674--2688},
  [\href{https://arxiv.org/abs/1708.01360}{{\ttfamily 1708.01360}}].

\bibitem{He:2017owu}
S.~He, S.~Alam, S.~Ferraro, Y.-C. Chen and S.~Ho, \emph{{The detection of the
  imprint of filaments on cosmic microwave background lensing}},
  \href{https://doi.org/10.1038/s41550-018-0426-z}{\emph{Nature Astron.}
  {\bfseries 2} (2018) 401--406},
  [\href{https://arxiv.org/abs/1709.02543}{{\ttfamily 1709.02543}}].

\bibitem{Raghunathan:2017qai}
S.~Raghunathan, F.~Bianchini and C.~L. Reichardt, \emph{{Imprints of
  gravitational lensing in the Planck cosmic microwave background data at the
  location of WISE\texttimes{}SCOS galaxies}},
  \href{https://doi.org/10.1103/PhysRevD.98.043506}{\emph{Phys. Rev. D}
  {\bfseries 98} (2018) 043506},
  [\href{https://arxiv.org/abs/1710.09770}{{\ttfamily 1710.09770}}].

\bibitem{Han:2018izq}
J.~Han, S.~Ferraro, E.~Giusarma and S.~Ho, \emph{{Probing Gravitational Lensing
  of the CMB with SDSS-IV Quasars}},
  \href{https://doi.org/10.1093/mnras/stz528}{\emph{Mon. Not. Roy. Astron.
  Soc.} {\bfseries 485} (2019) 1720--1726},
  [\href{https://arxiv.org/abs/1809.04196}{{\ttfamily 1809.04196}}].

\bibitem{DES:2018miq}
{\scshape DES, SPT} collaboration, Y.~Omori et~al., \emph{{Dark Energy Survey
  Year 1 Results: Tomographic cross-correlations between Dark Energy Survey
  galaxies and CMB lensing from South Pole Telescope+Planck}},
  \href{https://doi.org/10.1103/PhysRevD.100.043501}{\emph{Phys. Rev. D}
  {\bfseries 100} (2019) 043501},
  [\href{https://arxiv.org/abs/1810.02342}{{\ttfamily 1810.02342}}].

\bibitem{Singh:2018kmr}
S.~Singh, R.~Mandelbaum, U.~Seljak, S.~Rodr\'\i{}guez-Torres and A.~Slosar,
  \emph{{Cosmological constraints from galaxy\textendash{}lensing
  cross-correlations using BOSS galaxies with SDSS and CMB lensing}},
  \href{https://doi.org/10.1093/mnras/stz2922}{\emph{Mon. Not. Roy. Astron.
  Soc.} {\bfseries 491} (2020) 51--68},
  [\href{https://arxiv.org/abs/1811.06499}{{\ttfamily 1811.06499}}].

\bibitem{Marques:2019aug}
G.~A. Marques and A.~Bernui, \emph{{Tomographic analyses of the CMB lensing and
  galaxy clustering to probe the linear structure growth}},
  \href{https://doi.org/10.1088/1475-7516/2020/05/052}{\emph{JCAP} {\bfseries
  05} (2020) 052}, [\href{https://arxiv.org/abs/1908.04854}{{\ttfamily
  1908.04854}}].

\bibitem{Krolewski:2019yrv}
A.~Krolewski, S.~Ferraro, E.~F. Schlafly and M.~White, \emph{{unWISE tomography
  of Planck CMB lensing}},
  \href{https://doi.org/10.1088/1475-7516/2020/05/047}{\emph{JCAP} {\bfseries
  05} (2020) 047}, [\href{https://arxiv.org/abs/1909.07412}{{\ttfamily
  1909.07412}}].

\bibitem{Alonso:2020jcy}
D.~Alonso, E.~Bellini, C.~Hale, M.~J. Jarvis and D.~J. Schwarz,
  \emph{{Cross-correlating radio continuum surveys and CMB lensing:
  constraining redshift distributions, galaxy bias and cosmology}},
  \href{https://doi.org/10.1093/mnras/stab046}{\emph{Mon. Not. Roy. Astron.
  Soc.} {\bfseries 502} (2021) 876--887},
  [\href{https://arxiv.org/abs/2009.01817}{{\ttfamily 2009.01817}}].

\bibitem{ACT:2020izl}
{\scshape ACT} collaboration, M.~S. Madhavacheril et~al., \emph{{The Atacama
  Cosmology Telescope: Weighing Distant Clusters with the Most Ancient Light}},
  \href{https://doi.org/10.3847/2041-8213/abbccb}{\emph{Astrophys. J. Lett.}
  {\bfseries 903} (2020) L13},
  [\href{https://arxiv.org/abs/2009.07772}{{\ttfamily 2009.07772}}].

\bibitem{Hang:2020gwn}
Q.~Hang, S.~Alam, J.~A. Peacock and Y.-C. Cai, \emph{{Galaxy clustering in the
  DESI Legacy Survey and its imprint on the CMB}},
  \href{https://doi.org/10.1093/mnras/staa3738}{\emph{Mon. Not. Roy. Astron.
  Soc.} {\bfseries 501} (2021) 1481--1498},
  [\href{https://arxiv.org/abs/2010.00466}{{\ttfamily 2010.00466}}].

\bibitem{Kitanidis:2020xno}
E.~Kitanidis and M.~White, \emph{{Cross-Correlation of Planck CMB Lensing with
  DESI-Like LRGs}}, \href{https://doi.org/10.1093/mnras/staa3927}{\emph{Mon.
  Not. Roy. Astron. Soc.} {\bfseries 501} (2021) 6181--6198},
  [\href{https://arxiv.org/abs/2010.04698}{{\ttfamily 2010.04698}}].

\bibitem{Lin:2020sbb}
X.~Lin, Z.~Cai, Y.~Li, A.~Krolewski and S.~Ferraro, \emph{{Constraining the
  Halo Mass of Damped Ly$\alpha$ Absorption Systems (DLAs) at z =
  2\textendash{}3.5 Using the Quasar-CMB Lensing Cross-correlation}},
  \href{https://doi.org/10.3847/1538-4357/abc620}{\emph{Astrophys. J.}
  {\bfseries 905} (2020) 176},
  [\href{https://arxiv.org/abs/2011.01234}{{\ttfamily 2011.01234}}].

\bibitem{Krolewski:2021yqy}
A.~Krolewski, S.~Ferraro and M.~White, \emph{{Cosmological constraints from
  unWISE and Planck CMB lensing tomography}},
  \href{https://doi.org/10.1088/1475-7516/2021/12/028}{\emph{JCAP} {\bfseries
  12} (2021) 028}, [\href{https://arxiv.org/abs/2105.03421}{{\ttfamily
  2105.03421}}].

\bibitem{Dong:2021nmq}
F.~Dong, P.~Zhang, L.~Zhang, J.~Yao, Z.~Sun, C.~Park et~al., \emph{{Detection
  of a Cross-correlation between Cosmic Microwave Background Lensing and
  Low-density Points}},
  \href{https://doi.org/10.3847/1538-4357/ac2d31}{\emph{Astrophys. J.}
  {\bfseries 923} (2021) 153},
  [\href{https://arxiv.org/abs/2107.08694}{{\ttfamily 2107.08694}}].

\bibitem{Sun:2021rhp}
Z.~Sun, J.~Yao, F.~Dong, X.~Yang, L.~Zhang and P.~Zhang,
  \emph{{Cross-correlation of Planck CMB lensing with DESI galaxy groups}},
  \href{https://doi.org/10.1093/mnras/stac138}{\emph{Mon. Not. Roy. Astron.
  Soc.} {\bfseries 511} (2022) 3548--3560},
  [\href{https://arxiv.org/abs/2109.07387}{{\ttfamily 2109.07387}}].

\bibitem{White:2021yvw}
M.~White et~al., \emph{{Cosmological constraints from the tomographic
  cross-correlation of DESI Luminous Red Galaxies and Planck CMB lensing}},
  \href{https://doi.org/10.1088/1475-7516/2022/02/007}{\emph{JCAP} {\bfseries
  02} (2022) 007}, [\href{https://arxiv.org/abs/2111.09898}{{\ttfamily
  2111.09898}}].

\bibitem{Chen:2022itx}
T.~Chen and M.~Remazeilles, \emph{{Impact of thermal SZ effect on
  cross-correlations between Planck CMB lensing and SDSS galaxy density
  fields}}, \href{https://doi.org/10.1093/mnras/stac1436}{\emph{Mon. Not. Roy.
  Astron. Soc.} {\bfseries 514} (2022) 596--606},
  [\href{https://arxiv.org/abs/2203.04809}{{\ttfamily 2203.04809}}].

\bibitem{Kusiak:2022xkt}
A.~{Kusiak}, B.~{Bolliet}, A.~{Krolewski} and J.~C. {Hill}, \emph{{Constraining
  the galaxy-halo connection of infrared-selected unWISE galaxies with galaxy
  clustering and galaxy-CMB lensing power spectra}}, {\emph{arXiv e-prints}
  (Mar., 2022) arXiv:2203.12583},
  [\href{https://arxiv.org/abs/2203.12583}{{\ttfamily 2203.12583}}].

\bibitem{DES:2022ign}
C.~{Chang}, Y.~{Omori}, E.~J. {Baxter}, C.~{Doux}, A.~{Choi}, S.~{Pandey}
  et~al., \emph{{Joint analysis of DES Year 3 data and CMB lensing from SPT and
  Planck II: Cross-correlation measurements and cosmological constraints}},
  {\emph{arXiv e-prints} (Mar., 2022) arXiv:2203.12440},
  [\href{https://arxiv.org/abs/2203.12440}{{\ttfamily 2203.12440}}].

\bibitem{Chen:2022jzq}
S.-F. {Chen}, M.~{White}, J.~{DeRose} and N.~{Kokron}, \emph{{Cosmological
  Analysis of Three-Dimensional BOSS Galaxy Clustering and Planck CMB Lensing
  Cross Correlations via Lagrangian Perturbation Theory}}, {\emph{arXiv
  e-prints} (Apr., 2022) arXiv:2204.10392},
  [\href{https://arxiv.org/abs/2204.10392}{{\ttfamily 2204.10392}}].

\bibitem{Amendola:2016saw}
L.~Amendola et~al., \emph{{Cosmology and fundamental physics with the Euclid
  satellite}}, \href{https://doi.org/10.1007/s41114-017-0010-3}{\emph{Living
  Rev. Rel.} {\bfseries 21} (2018) 2},
  [\href{https://arxiv.org/abs/1606.00180}{{\ttfamily 1606.00180}}].

\bibitem{2016arXiv161100036D}
{DESI Collaboration}, \emph{{The DESI Experiment Part I: Science,Targeting, and
  Survey Design}}, {\emph{arXiv e-prints} (Oct., 2016) arXiv:1611.00036},
  [\href{https://arxiv.org/abs/1611.00036}{{\ttfamily 1611.00036}}].

\bibitem{Vagnozzi:2018pwo}
S.~Vagnozzi, T.~Brinckmann, M.~Archidiacono, K.~Freese, M.~Gerbino,
  J.~Lesgourgues et~al., \emph{{Bias due to neutrinos must not uncorrect'd
  go}}, \href{https://doi.org/10.1088/1475-7516/2018/09/001}{\emph{JCAP}
  {\bfseries 09} (2018) 001},
  [\href{https://arxiv.org/abs/1807.04672}{{\ttfamily 1807.04672}}].

\bibitem{Castorina:2013wga}
E.~Castorina, E.~Sefusatti, R.~K. Sheth, F.~Villaescusa-Navarro and M.~Viel,
  \emph{{Cosmology with massive neutrinos II: on the universality of the halo
  mass function and bias}},
  \href{https://doi.org/10.1088/1475-7516/2014/02/049}{\emph{JCAP} {\bfseries
  02} (2014) 049}, [\href{https://arxiv.org/abs/1311.1212}{{\ttfamily
  1311.1212}}].

\bibitem{Fidler:2018dcy}
C.~Fidler, N.~Sujata and M.~Archidiacono, \emph{{Relativistic bias in neutrino
  cosmologies}},
  \href{https://doi.org/10.1088/1475-7516/2019/06/035}{\emph{JCAP} {\bfseries
  06} (2019) 035}, [\href{https://arxiv.org/abs/1812.09266}{{\ttfamily
  1812.09266}}].

\bibitem{Gerbino:2016sgw}
M.~Gerbino, K.~Freese, S.~Vagnozzi, M.~Lattanzi, O.~Mena, E.~Giusarma et~al.,
  \emph{{Impact of neutrino properties on the estimation of inflationary
  parameters from current and future observations}},
  \href{https://doi.org/10.1103/PhysRevD.95.043512}{\emph{Phys. Rev. D}
  {\bfseries 95} (2017) 043512},
  [\href{https://arxiv.org/abs/1610.08830}{{\ttfamily 1610.08830}}].

\bibitem{Archidiacono:2016lnv}
M.~Archidiacono, T.~Brinckmann, J.~Lesgourgues and V.~Poulin, \emph{{Physical
  effects involved in the measurements of neutrino masses with future
  cosmological data}},
  \href{https://doi.org/10.1088/1475-7516/2017/02/052}{\emph{JCAP} {\bfseries
  02} (2017) 052}, [\href{https://arxiv.org/abs/1610.09852}{{\ttfamily
  1610.09852}}].

\bibitem{Archidiacono:2020dvx}
M.~Archidiacono, S.~Hannestad and J.~Lesgourgues, \emph{{What will it take to
  measure individual neutrino mass states using cosmology?}},
  \href{https://doi.org/10.1088/1475-7516/2020/09/021}{\emph{JCAP} {\bfseries
  09} (2020) 021}, [\href{https://arxiv.org/abs/2003.03354}{{\ttfamily
  2003.03354}}].

\bibitem{Brandbyge:2008rv}
J.~Brandbyge, S.~Hannestad, T.~Haugb\o{}lle and B.~Thomsen, \emph{{The Effect
  of Thermal Neutrino Motion on the Non-linear Cosmological Matter Power
  Spectrum}}, \href{https://doi.org/10.1088/1475-7516/2008/08/020}{\emph{JCAP}
  {\bfseries 08} (2008) 020},
  [\href{https://arxiv.org/abs/0802.3700}{{\ttfamily 0802.3700}}].

\bibitem{Brandbyge:2008js}
J.~Brandbyge and S.~Hannestad, \emph{{Grid Based Linear Neutrino Perturbations
  in Cosmological N-body Simulations}},
  \href{https://doi.org/10.1088/1475-7516/2009/05/002}{\emph{JCAP} {\bfseries
  05} (2009) 002}, [\href{https://arxiv.org/abs/0812.3149}{{\ttfamily
  0812.3149}}].

\bibitem{Brandbyge:2009ce}
J.~Brandbyge and S.~Hannestad, \emph{{Resolving Cosmic Neutrino Structure: A
  Hybrid Neutrino N-body Scheme}},
  \href{https://doi.org/10.1088/1475-7516/2010/01/021}{\emph{JCAP} {\bfseries
  01} (2010) 021}, [\href{https://arxiv.org/abs/0908.1969}{{\ttfamily
  0908.1969}}].

\bibitem{Viel:2010bn}
M.~Viel, M.~G. Haehnelt and V.~Springel, \emph{{The effect of neutrinos on the
  matter distribution as probed by the Intergalactic Medium}},
  \href{https://doi.org/10.1088/1475-7516/2010/06/015}{\emph{JCAP} {\bfseries
  06} (2010) 015}, [\href{https://arxiv.org/abs/1003.2422}{{\ttfamily
  1003.2422}}].

\bibitem{Ali-Haimoud:2012fzp}
Y.~Ali-Haimoud and S.~Bird, \emph{{An efficient implementation of massive
  neutrinos in non-linear structure formation simulations}},
  \href{https://doi.org/10.1093/mnras/sts286}{\emph{Mon. Not. Roy. Astron.
  Soc.} {\bfseries 428} (2012) 3375--3389},
  [\href{https://arxiv.org/abs/1209.0461}{{\ttfamily 1209.0461}}].

\bibitem{Castorina:2015bma}
E.~Castorina, C.~Carbone, J.~Bel, E.~Sefusatti and K.~Dolag, \emph{{DEMNUni:
  The clustering of large-scale structures in the presence of massive
  neutrinos}}, \href{https://doi.org/10.1088/1475-7516/2015/07/043}{\emph{JCAP}
  {\bfseries 07} (2015) 043},
  [\href{https://arxiv.org/abs/1505.07148}{{\ttfamily 1505.07148}}].

\bibitem{Banerjee:2016zaa}
A.~Banerjee and N.~Dalal, \emph{{Simulating nonlinear cosmological structure
  formation with massive neutrinos}},
  \href{https://doi.org/10.1088/1475-7516/2016/11/015}{\emph{JCAP} {\bfseries
  11} (2016) 015}, [\href{https://arxiv.org/abs/1606.06167}{{\ttfamily
  1606.06167}}].

\bibitem{Liu:2017now}
J.~Liu, S.~Bird, J.~M.~Z. Matilla, J.~C. Hill, Z.~Haiman, M.~S. Madhavacheril
  et~al., \emph{{MassiveNuS: Cosmological Massive Neutrino Simulations}},
  \href{https://doi.org/10.1088/1475-7516/2018/03/049}{\emph{JCAP} {\bfseries
  03} (2018) 049}, [\href{https://arxiv.org/abs/1711.10524}{{\ttfamily
  1711.10524}}].

\bibitem{Banerjee:2018bxy}
A.~Banerjee, D.~Powell, T.~Abel and F.~Villaescusa-Navarro, \emph{{Reducing
  Noise in Cosmological N-body Simulations with Neutrinos}},
  \href{https://doi.org/10.1088/1475-7516/2018/09/028}{\emph{JCAP} {\bfseries
  09} (2018) 028}, [\href{https://arxiv.org/abs/1801.03906}{{\ttfamily
  1801.03906}}].

\bibitem{Partmann:2020qzb}
C.~Partmann, C.~Fidler, C.~Rampf and O.~Hahn, \emph{{Fast simulations of cosmic
  large-scale structure with massive neutrinos}},
  \href{https://doi.org/10.1088/1475-7516/2020/09/018}{\emph{JCAP} {\bfseries
  09} (2020) 018}, [\href{https://arxiv.org/abs/2003.07387}{{\ttfamily
  2003.07387}}].

\bibitem{Wong:2008ws}
Y.~Y.~Y. Wong, \emph{{Higher order corrections to the large scale matter power
  spectrum in the presence of massive neutrinos}},
  \href{https://doi.org/10.1088/1475-7516/2008/10/035}{\emph{JCAP} {\bfseries
  10} (2008) 035}, [\href{https://arxiv.org/abs/0809.0693}{{\ttfamily
  0809.0693}}].

\bibitem{Lesgourgues:2009am}
J.~Lesgourgues, S.~Matarrese, M.~Pietroni and A.~Riotto, \emph{{Non-linear
  Power Spectrum including Massive Neutrinos: the Time-RG Flow Approach}},
  \href{https://doi.org/10.1088/1475-7516/2009/06/017}{\emph{JCAP} {\bfseries
  06} (2009) 017}, [\href{https://arxiv.org/abs/0901.4550}{{\ttfamily
  0901.4550}}].

\bibitem{Blas:2014hya}
D.~Blas, M.~Garny, T.~Konstandin and J.~Lesgourgues, \emph{{Structure formation
  with massive neutrinos: going beyond linear theory}},
  \href{https://doi.org/10.1088/1475-7516/2014/11/039}{\emph{JCAP} {\bfseries
  11} (2014) 039}, [\href{https://arxiv.org/abs/1408.2995}{{\ttfamily
  1408.2995}}].

\bibitem{Fuhrer:2014zka}
F.~F\"uhrer and Y.~Y.~Y. Wong, \emph{{Higher-order massive neutrino
  perturbations in large-scale structure}},
  \href{https://doi.org/10.1088/1475-7516/2015/03/046}{\emph{JCAP} {\bfseries
  03} (2015) 046}, [\href{https://arxiv.org/abs/1412.2764}{{\ttfamily
  1412.2764}}].

\bibitem{Hannestad:2020rzl}
S.~Hannestad, A.~Upadhye and Y.~Y.~Y. Wong, \emph{{Spoon or slide? The
  non-linear matter power spectrum in the presence of massive neutrinos}},
  \href{https://doi.org/10.1088/1475-7516/2020/11/062}{\emph{JCAP} {\bfseries
  11} (2020) 062}, [\href{https://arxiv.org/abs/2006.04995}{{\ttfamily
  2006.04995}}].

\bibitem{Viel:2005qj}
M.~Viel, J.~Lesgourgues, M.~G. Haehnelt, S.~Matarrese and A.~Riotto,
  \emph{{Constraining warm dark matter candidates including sterile neutrinos
  and light gravitinos with WMAP and the Lyman-alpha forest}},
  \href{https://doi.org/10.1103/PhysRevD.71.063534}{\emph{Phys. Rev. D}
  {\bfseries 71} (2005) 063534},
  [\href{https://arxiv.org/abs/astro-ph/0501562}{{\ttfamily
  astro-ph/0501562}}].

\bibitem{DePorzio:2020wcz}
N.~DePorzio, W.~L. Xu, J.~B. Mu\~noz and C.~Dvorkin, \emph{{Finding eV-scale
  light relics with cosmological observables}},
  \href{https://doi.org/10.1103/PhysRevD.103.023504}{\emph{Phys. Rev. D}
  {\bfseries 103} (2021) 023504},
  [\href{https://arxiv.org/abs/2006.09380}{{\ttfamily 2006.09380}}].

\bibitem{Xu:2021rwg}
W.~L. Xu, J.~B. Mu\~noz and C.~Dvorkin, \emph{{Cosmological constraints on
  light but massive relics}},
  \href{https://doi.org/10.1103/PhysRevD.105.095029}{\emph{Phys. Rev. D}
  {\bfseries 105} (2022) 095029},
  [\href{https://arxiv.org/abs/2107.09664}{{\ttfamily 2107.09664}}].

\bibitem{Vagnozzi:2020rcz}
S.~Vagnozzi, E.~Di~Valentino, S.~Gariazzo, A.~Melchiorri, O.~Mena and J.~Silk,
  \emph{{The galaxy power spectrum take on spatial curvature and cosmic
  concordance}}, \href{https://doi.org/10.1016/j.dark.2021.100851}{\emph{Phys.
  Dark Univ.} {\bfseries 33} (2021) 100851},
  [\href{https://arxiv.org/abs/2010.02230}{{\ttfamily 2010.02230}}].

\bibitem{Bond:1997wr}
J.~R. Bond, G.~Efstathiou and M.~Tegmark, \emph{{Forecasting cosmic parameter
  errors from microwave background anisotropy experiments}},
  \href{https://doi.org/10.1093/mnras/291.1.L33}{\emph{Mon. Not. Roy. Astron.
  Soc.} {\bfseries 291} (1997) L33--L41},
  [\href{https://arxiv.org/abs/astro-ph/9702100}{{\ttfamily
  astro-ph/9702100}}].

\bibitem{Zaldarriaga:1997ch}
M.~Zaldarriaga, D.~N. Spergel and U.~Seljak, \emph{{Microwave background
  constraints on cosmological parameters}},
  \href{https://doi.org/10.1086/304692}{\emph{Astrophys. J.} {\bfseries 488}
  (1997) 1--13}, [\href{https://arxiv.org/abs/astro-ph/9702157}{{\ttfamily
  astro-ph/9702157}}].

\bibitem{Efstathiou:1998xx}
G.~Efstathiou and J.~R. Bond, \emph{{Cosmic confusion: Degeneracies among
  cosmological parameters derived from measurements of microwave background
  anisotropies}},
  \href{https://doi.org/10.1046/j.1365-8711.1999.02274.x}{\emph{Mon. Not. Roy.
  Astron. Soc.} {\bfseries 304} (1999) 75--97},
  [\href{https://arxiv.org/abs/astro-ph/9807103}{{\ttfamily
  astro-ph/9807103}}].

\bibitem{Efstathiou:2020wem}
G.~Efstathiou and S.~Gratton, \emph{{The evidence for a spatially flat
  Universe}}, \href{https://doi.org/10.1093/mnrasl/slaa093}{\emph{Mon. Not.
  Roy. Astron. Soc.} {\bfseries 496} (2020) L91--L95},
  [\href{https://arxiv.org/abs/2002.06892}{{\ttfamily 2002.06892}}].

\bibitem{Vagnozzi:2020dfn}
S.~Vagnozzi, A.~Loeb and M.~Moresco, \emph{{Eppur \`e piatto? The Cosmic
  Chronometers Take on Spatial Curvature and Cosmic Concordance}},
  \href{https://doi.org/10.3847/1538-4357/abd4df}{\emph{Astrophys. J.}
  {\bfseries 908} (2021) 84},
  [\href{https://arxiv.org/abs/2011.11645}{{\ttfamily 2011.11645}}].

\bibitem{Dhawan:2021mel}
S.~Dhawan, J.~Alsing and S.~Vagnozzi, \emph{{Non-parametric spatial curvature
  inference using late-Universe cosmological probes}},
  \href{https://doi.org/10.1093/mnrasl/slab058}{\emph{Mon. Not. Roy. Astron.
  Soc.} {\bfseries 506} (2021) L1--L5},
  [\href{https://arxiv.org/abs/2104.02485}{{\ttfamily 2104.02485}}].

\bibitem{Khadka:2020vlh}
N.~Khadka and B.~Ratra, \emph{{Using quasar X-ray and UV flux measurements to
  constrain cosmological model parameters}},
  \href{https://doi.org/10.1093/mnras/staa1855}{\emph{Mon. Not. Roy. Astron.
  Soc.} {\bfseries 497} (2020) 263--278},
  [\href{https://arxiv.org/abs/2004.09979}{{\ttfamily 2004.09979}}].

\bibitem{Khadka:2020hvb}
N.~Khadka and B.~Ratra, \emph{{Constraints on cosmological parameters from
  gamma-ray burst peak photon energy and bolometric fluence measurements and
  other data}}, \href{https://doi.org/10.1093/mnras/staa2779}{\emph{Mon. Not.
  Roy. Astron. Soc.} {\bfseries 499} (2020) 391--403},
  [\href{https://arxiv.org/abs/2007.13907}{{\ttfamily 2007.13907}}].

\bibitem{Khadka:2020tlm}
N.~Khadka and B.~Ratra, \emph{{Determining the range of validity of quasar
  X-ray and UV flux measurements for constraining cosmological model
  parameters}}, \href{https://doi.org/10.1093/mnras/stab486}{\emph{Mon. Not.
  Roy. Astron. Soc.} {\bfseries 502} (2021) 6140--6156},
  [\href{https://arxiv.org/abs/2012.09291}{{\ttfamily 2012.09291}}].

\bibitem{Khadka:2021vqa}
N.~Khadka, O.~Luongo, M.~Muccino and B.~Ratra, \emph{{Do gamma-ray burst
  measurements provide a useful test of cosmological models?}},
  \href{https://doi.org/10.1088/1475-7516/2021/09/042}{\emph{JCAP} {\bfseries
  09} (2021) 042}, [\href{https://arxiv.org/abs/2105.12692}{{\ttfamily
  2105.12692}}].

\bibitem{Khadka:2021ukv}
N.~Khadka, Z.~Yu, M.~Zaja\v{c}ek, M.~L. Martinez-Aldama, B.~Czerny and
  B.~Ratra, \emph{{Standardizing reverberation-measured Mg II time-lag quasars,
  by using the radius\textendash{}luminosity relation, and constraining
  cosmological model parameters}},
  \href{https://doi.org/10.1093/mnras/stab2807}{\emph{Mon. Not. Roy. Astron.
  Soc.} {\bfseries 508} (2021) 4722--4737},
  [\href{https://arxiv.org/abs/2106.11136}{{\ttfamily 2106.11136}}].

\bibitem{Cao:2021ldv}
S.~Cao, J.~Ryan and B.~Ratra, \emph{{Using Pantheon and DES supernova, baryon
  acoustic oscillation, and Hubble parameter data to constrain the Hubble
  constant, dark energy dynamics, and spatial curvature}},
  \href{https://doi.org/10.1093/mnras/stab942}{\emph{Mon. Not. Roy. Astron.
  Soc.} {\bfseries 504} (2021) 300--310},
  [\href{https://arxiv.org/abs/2101.08817}{{\ttfamily 2101.08817}}].

\bibitem{Cao:2021cix}
S.~Cao, J.~Ryan and B.~Ratra, \emph{{Cosmological constraints from H\,ii
  starburst galaxy, quasar angular size, and other measurements}},
  \href{https://doi.org/10.1093/mnras/stab3304}{\emph{Mon. Not. Roy. Astron.
  Soc.} {\bfseries 509} (2022) 4745--4757},
  [\href{https://arxiv.org/abs/2109.01987}{{\ttfamily 2109.01987}}].

\bibitem{Cao:2022wlg}
S.~Cao, M.~Dainotti and B.~Ratra, \emph{{Standardizing Platinum
  Dainotti-correlated gamma-ray bursts, and using them with standardized
  Amati-correlated gamma-ray bursts to constrain cosmological model
  parameters}}, \href{https://doi.org/10.1093/mnras/stac517}{\emph{Mon. Not.
  Roy. Astron. Soc.} {\bfseries 512} (2022) 439--454},
  [\href{https://arxiv.org/abs/2201.05245}{{\ttfamily 2201.05245}}].

\bibitem{Kaiser_1984}
N.~Kaiser, \emph{{On the Spatial correlations of Abell clusters}},
  \href{https://doi.org/10.1086/184341}{\emph{Astrophys. J. Lett.} {\bfseries
  284} (1984) L9--L12}.

\bibitem{Sheth:1999mn}
R.~K. Sheth and G.~Tormen, \emph{{Large scale bias and the peak background
  split}}, \href{https://doi.org/10.1046/j.1365-8711.1999.02692.x}{\emph{Mon.
  Not. Roy. Astron. Soc.} {\bfseries 308} (1999) 119},
  [\href{https://arxiv.org/abs/astro-ph/9901122}{{\ttfamily
  astro-ph/9901122}}].

\bibitem{Seljak:2000jg}
U.~Seljak, \emph{{Redshift space bias and beta from the halo model}},
  \href{https://doi.org/10.1046/j.1365-8711.2001.04508.x}{\emph{Mon. Not. Roy.
  Astron. Soc.} {\bfseries 325} (2001) 1359},
  [\href{https://arxiv.org/abs/astro-ph/0009016}{{\ttfamily
  astro-ph/0009016}}].

\bibitem{Schulz:2005kj}
A.~E. Schulz and M.~J. White, \emph{{Scale-dependent bias and the halo model}},
  \href{https://doi.org/10.1016/j.astropartphys.2005.11.007}{\emph{Astropart.
  Phys.} {\bfseries 25} (2006) 172--177},
  [\href{https://arxiv.org/abs/astro-ph/0510100}{{\ttfamily
  astro-ph/0510100}}].

\bibitem{Matsubara:2011ck}
T.~Matsubara, \emph{{Nonlinear Perturbation Theory Integrated with Nonlocal
  Bias, Redshift-space Distortions, and Primordial Non-Gaussianity}},
  \href{https://doi.org/10.1103/PhysRevD.83.083518}{\emph{Phys. Rev. D}
  {\bfseries 83} (2011) 083518},
  [\href{https://arxiv.org/abs/1102.4619}{{\ttfamily 1102.4619}}].

\bibitem{Saito:2014qha}
S.~Saito, T.~Baldauf, Z.~Vlah, U.~Seljak, T.~Okumura and P.~McDonald,
  \emph{{Understanding higher-order nonlocal halo bias at large scales by
  combining the power spectrum with the bispectrum}},
  \href{https://doi.org/10.1103/PhysRevD.90.123522}{\emph{Phys. Rev. D}
  {\bfseries 90} (2014) 123522},
  [\href{https://arxiv.org/abs/1405.1447}{{\ttfamily 1405.1447}}].

\bibitem{Desjacques_2018_1loop}
V.~Desjacques, D.~Jeong and F.~Schmidt, \emph{{The Galaxy Power Spectrum and
  Bispectrum in Redshift Space}},
  \href{https://doi.org/10.1088/1475-7516/2018/12/035}{\emph{JCAP} {\bfseries
  12} (2018) 035}, [\href{https://arxiv.org/abs/1806.04015}{{\ttfamily
  1806.04015}}].

\bibitem{Okumura:2012xh}
T.~Okumura, U.~Seljak and V.~Desjacques, \emph{{Distribution function approach
  to redshift space distortions, Part III: halos and galaxies}},
  \href{https://doi.org/10.1088/1475-7516/2012/11/014}{\emph{JCAP} {\bfseries
  11} (2012) 014}, [\href{https://arxiv.org/abs/1206.4070}{{\ttfamily
  1206.4070}}].

\bibitem{Modi_2020}
C.~Modi, S.-F. Chen and M.~White, \emph{{Simulations and symmetries}},
  \href{https://doi.org/10.1093/mnras/staa251}{\emph{Mon. Not. Roy. Astron.
  Soc.} {\bfseries 492} (2020) 5754--5763},
  [\href{https://arxiv.org/abs/1910.07097}{{\ttfamily 1910.07097}}].

\bibitem{Hayashi:2007uk}
E.~Hayashi and S.~D.~M. White, \emph{{Understanding the shape of the halo-mass
  and galaxy-mass cross-correlation functions}},
  \href{https://doi.org/10.1111/j.1365-2966.2008.13371.x}{\emph{Mon. Not. Roy.
  Astron. Soc.} {\bfseries 388} (2008) 2},
  [\href{https://arxiv.org/abs/0709.3933}{{\ttfamily 0709.3933}}].

\bibitem{Sheth:1998xe}
R.~K. Sheth and G.~Lemson, \emph{{Biasing and the distribution of dark matter
  haloes}}, \href{https://doi.org/10.1046/j.1365-8711.1999.02378.x}{\emph{Mon.
  Not. Roy. Astron. Soc.} {\bfseries 304} (1999) 767},
  [\href{https://arxiv.org/abs/astro-ph/9808138}{{\ttfamily
  astro-ph/9808138}}].

\bibitem{reid2015sdssiii}
B.~Reid et~al., \emph{{SDSS-III Baryon Oscillation Spectroscopic Survey Data
  Release 12: galaxy target selection and large scale structure catalogues}},
  \href{https://doi.org/10.1093/mnras/stv2382}{\emph{Mon. Not. Roy. Astron.
  Soc.} {\bfseries 455} (2016) 1553--1573},
  [\href{https://arxiv.org/abs/1509.06529}{{\ttfamily 1509.06529}}].

\bibitem{Baldauf_2013}
T.~Baldauf, U.~Seljak, R.~E. Smith, N.~Hamaus and V.~Desjacques, \emph{{Halo
  stochasticity from exclusion and nonlinear clustering}},
  \href{https://doi.org/10.1103/PhysRevD.88.083507}{\emph{Phys. Rev. D}
  {\bfseries 88} (2013) 083507},
  [\href{https://arxiv.org/abs/1305.2917}{{\ttfamily 1305.2917}}].

\bibitem{Modi:2016dah}
C.~Modi, E.~Castorina and U.~Seljak, \emph{{Halo bias in Lagrangian Space:
  Estimators and theoretical predictions}},
  \href{https://doi.org/10.1093/mnras/stx2148}{\emph{Mon. Not. Roy. Astron.
  Soc.} {\bfseries 472} (2017) 3959--3970},
  [\href{https://arxiv.org/abs/1612.01621}{{\ttfamily 1612.01621}}].

\bibitem{Matsubara_1999}
T.~Matsubara, \emph{{Stochasticity of bias and nonlocality of galaxy formation:
  Linear scales}}, \href{https://doi.org/10.1086/307931}{\emph{Astrophys. J.}
  {\bfseries 525} (1999) 543--553},
  [\href{https://arxiv.org/abs/astro-ph/9906029}{{\ttfamily
  astro-ph/9906029}}].

\bibitem{Kaiser_1987}
N.~Kaiser, \emph{{Clustering in real space and in redshift space}}, {\emph{Mon.
  Not. Roy. Astron. Soc.} {\bfseries 227} (1987) 1--27}.

\bibitem{Lahav:1991wc}
O.~Lahav, P.~B. Lilje, J.~R. Primack and M.~J. Rees, \emph{{Dynamical effects
  of the cosmological constant}}, {\emph{Mon. Not. Roy. Astron. Soc.}
  {\bfseries 251} (1991) 128--136}.

\bibitem{Linder_2007}
E.~V. Linder and R.~N. Cahn, \emph{{Parameterized Beyond-Einstein Growth}},
  \href{https://doi.org/10.1016/j.astropartphys.2007.09.003}{\emph{Astropart.
  Phys.} {\bfseries 28} (2007) 481--488},
  [\href{https://arxiv.org/abs/astro-ph/0701317}{{\ttfamily
  astro-ph/0701317}}].

\bibitem{Nesseris_2008}
S.~Nesseris and L.~Perivolaropoulos, \emph{{Testing $\Lambda$CDM with the
  Growth Function $\delta(a)$: Current Constraints}},
  \href{https://doi.org/10.1103/PhysRevD.77.023504}{\emph{Phys. Rev. D}
  {\bfseries 77} (2008) 023504},
  [\href{https://arxiv.org/abs/0710.1092}{{\ttfamily 0710.1092}}].

\bibitem{LoVerde:2014pxa}
M.~LoVerde, \emph{{Halo bias in mixed dark matter cosmologies}},
  \href{https://doi.org/10.1103/PhysRevD.90.083530}{\emph{Phys. Rev. D}
  {\bfseries 90} (2014) 083530},
  [\href{https://arxiv.org/abs/1405.4855}{{\ttfamily 1405.4855}}].

\bibitem{LoVerde:2014rxa}
M.~LoVerde, \emph{{Spherical collapse in $\nu \Lambda$CDM}},
  \href{https://doi.org/10.1103/PhysRevD.90.083518}{\emph{Phys. Rev. D}
  {\bfseries 90} (2014) 083518},
  [\href{https://arxiv.org/abs/1405.4858}{{\ttfamily 1405.4858}}].

\bibitem{LoVerde:2016ahu}
M.~LoVerde, \emph{{Neutrino mass without cosmic variance}},
  \href{https://doi.org/10.1103/PhysRevD.93.103526}{\emph{Phys. Rev. D}
  {\bfseries 93} (2016) 103526},
  [\href{https://arxiv.org/abs/1602.08108}{{\ttfamily 1602.08108}}].

\bibitem{Munoz:2018ajr}
J.~B. Mu\~noz and C.~Dvorkin, \emph{{Efficient Computation of Galaxy Bias with
  Neutrinos and Other Relics}},
  \href{https://doi.org/10.1103/PhysRevD.98.043503}{\emph{Phys. Rev. D}
  {\bfseries 98} (2018) 043503},
  [\href{https://arxiv.org/abs/1805.11623}{{\ttfamily 1805.11623}}].

\bibitem{Xu:2020fyg}
W.~L. Xu, N.~DePorzio, J.~B. Mu\~noz and C.~Dvorkin, \emph{{Accurately Weighing
  Neutrinos with Cosmological Surveys}},
  \href{https://doi.org/10.1103/PhysRevD.103.023503}{\emph{Phys. Rev. D}
  {\bfseries 103} (2021) 023503},
  [\href{https://arxiv.org/abs/2006.09395}{{\ttfamily 2006.09395}}].

\bibitem{Jackson:1971sky}
J.~C. Jackson, \emph{{Fingers of God: A critique of Rees' theory of primoridal
  gravitational radiation}},
  \href{https://doi.org/10.1093/mnras/156.1.1P}{\emph{Mon. Not. Roy. Astron.
  Soc.} {\bfseries 156} (1972) 1P--5P},
  [\href{https://arxiv.org/abs/0810.3908}{{\ttfamily 0810.3908}}].

\bibitem{Bull:2014rha}
P.~Bull, P.~G. Ferreira, P.~Patel and M.~G. Santos, \emph{{Late-time cosmology
  with 21cm intensity mapping experiments}},
  \href{https://doi.org/10.1088/0004-637X/803/1/21}{\emph{Astrophys. J.}
  {\bfseries 803} (2015) 21},
  [\href{https://arxiv.org/abs/1405.1452}{{\ttfamily 1405.1452}}].

\bibitem{Ivanov:2021fbu}
M.~M. Ivanov, O.~H.~E. Philcox, M.~Simonovi\'c, M.~Zaldarriaga, T.~Nischimichi
  and M.~Takada, \emph{{Cosmological constraints without nonlinear
  redshift-space distortions}},
  \href{https://doi.org/10.1103/PhysRevD.105.043531}{\emph{Phys. Rev. D}
  {\bfseries 105} (2022) 043531},
  [\href{https://arxiv.org/abs/2110.00006}{{\ttfamily 2110.00006}}].

\bibitem{2021arXiv211000016D}
G.~{D'Amico}, L.~{Senatore}, P.~{Zhang} and T.~{Nishimichi}, \emph{{Taming
  redshift-space distortion effects in the EFTofLSS and its application to
  data}}, {\emph{arXiv e-prints} (Sept., 2021) arXiv:2110.00016},
  [\href{https://arxiv.org/abs/2110.00016}{{\ttfamily 2110.00016}}].

\bibitem{Okumura:2015fga}
T.~Okumura, N.~Hand, U.~Seljak, Z.~Vlah and V.~Desjacques, \emph{{Galaxy power
  spectrum in redshift space: combining perturbation theory with the halo
  model}}, \href{https://doi.org/10.1103/PhysRevD.92.103516}{\emph{Phys. Rev.
  D} {\bfseries 92} (2015) 103516},
  [\href{https://arxiv.org/abs/1506.05814}{{\ttfamily 1506.05814}}].

\bibitem{Pullen_2016}
A.~R. Pullen, S.~Alam, S.~He and S.~Ho, \emph{{Constraining Gravity at the
  Largest Scales through CMB Lensing and Galaxy Velocities}},
  \href{https://doi.org/10.1093/mnras/stw1249}{\emph{Mon. Not. Roy. Astron.
  Soc.} {\bfseries 460} (2016) 4098--4108},
  [\href{https://arxiv.org/abs/1511.04457}{{\ttfamily 1511.04457}}].

\bibitem{Bianchini_2015}
{\scshape Herschel ATLAS} collaboration, F.~Bianchini et~al.,
  \emph{{Cross-correlation between the CMB lensing potential measured by Planck
  and high-z sub-mm galaxies detected by the Herschel-ATLAS survey}},
  \href{https://doi.org/10.1088/0004-637X/802/1/64}{\emph{Astrophys. J.}
  {\bfseries 802} (2015) 64},
  [\href{https://arxiv.org/abs/1410.4502}{{\ttfamily 1410.4502}}].

\bibitem{LoVerde_2008}
M.~LoVerde and N.~Afshordi, \emph{{Extended Limber Approximation}},
  \href{https://doi.org/10.1103/PhysRevD.78.123506}{\emph{Phys. Rev. D}
  {\bfseries 78} (2008) 123506},
  [\href{https://arxiv.org/abs/0809.5112}{{\ttfamily 0809.5112}}].

\bibitem{Raccanelli:2013gja}
A.~Raccanelli, D.~Bertacca, R.~Maartens, C.~Clarkson and O.~Dor\'e,
  \emph{{Lensing and time-delay contributions to galaxy correlations}},
  \href{https://doi.org/10.1007/s10714-016-2076-8}{\emph{Gen. Rel. Grav.}
  {\bfseries 48} (2016) 84}, [\href{https://arxiv.org/abs/1311.6813}{{\ttfamily
  1311.6813}}].

\bibitem{MoradinezhadDizgah:2016pqy}
A.~Moradinezhad~Dizgah and R.~Durrer, \emph{{Lensing corrections to the
  $E_g(z)$ statistics from large scale structure}},
  \href{https://doi.org/10.1088/1475-7516/2016/09/035}{\emph{JCAP} {\bfseries
  09} (2016) 035}, [\href{https://arxiv.org/abs/1604.08914}{{\ttfamily
  1604.08914}}].

\bibitem{Yoo:2010ni}
J.~Yoo, \emph{{General Relativistic Description of the Observed Galaxy Power
  Spectrum: Do We Understand What We Measure?}},
  \href{https://doi.org/10.1103/PhysRevD.82.083508}{\emph{Phys. Rev. D}
  {\bfseries 82} (2010) 083508},
  [\href{https://arxiv.org/abs/1009.3021}{{\ttfamily 1009.3021}}].

\bibitem{Maartens:2012rh}
R.~Maartens, G.-B. Zhao, D.~Bacon, K.~Koyama and A.~Raccanelli,
  \emph{{Relativistic corrections and non-Gaussianity in radio continuum
  surveys}}, \href{https://doi.org/10.1088/1475-7516/2013/02/044}{\emph{JCAP}
  {\bfseries 02} (2013) 044},
  [\href{https://arxiv.org/abs/1206.0732}{{\ttfamily 1206.0732}}].

\bibitem{Raccanelli:2015vla}
A.~Raccanelli, F.~Montanari, D.~Bertacca, O.~Dor\'e and R.~Durrer,
  \emph{{Cosmological Measurements with General Relativistic Galaxy
  Correlations}},
  \href{https://doi.org/10.1088/1475-7516/2016/05/009}{\emph{JCAP} {\bfseries
  05} (2016) 009}, [\href{https://arxiv.org/abs/1505.06179}{{\ttfamily
  1505.06179}}].

\bibitem{Grimm:2020ays}
N.~Grimm, F.~Scaccabarozzi, J.~Yoo, S.~G. Biern and J.-O. Gong, \emph{{Galaxy
  Power Spectrum in General Relativity}},
  \href{https://doi.org/10.1088/1475-7516/2020/11/064}{\emph{JCAP} {\bfseries
  11} (2020) 064}, [\href{https://arxiv.org/abs/2005.06484}{{\ttfamily
  2005.06484}}].

\bibitem{Castorina:2021xzs}
E.~Castorina and E.~di~Dio, \emph{{The observed galaxy power spectrum in
  General Relativity}},
  \href{https://doi.org/10.1088/1475-7516/2022/01/061}{\emph{JCAP} {\bfseries
  01} (2022) 061}, [\href{https://arxiv.org/abs/2106.08857}{{\ttfamily
  2106.08857}}].

\bibitem{Lesgourgues:2004ps}
J.~Lesgourgues, S.~Pastor and L.~Perotto, \emph{{Probing neutrino masses with
  future galaxy redshift surveys}},
  \href{https://doi.org/10.1103/PhysRevD.70.045016}{\emph{Phys. Rev. D}
  {\bfseries 70} (2004) 045016},
  [\href{https://arxiv.org/abs/hep-ph/0403296}{{\ttfamily hep-ph/0403296}}].

\bibitem{DeBernardis:2009di}
F.~De~Bernardis, T.~D. Kitching, A.~Heavens and A.~Melchiorri,
  \emph{{Determining the Neutrino Mass Hierarchy with Cosmology}},
  \href{https://doi.org/10.1103/PhysRevD.80.123509}{\emph{Phys. Rev. D}
  {\bfseries 80} (2009) 123509},
  [\href{https://arxiv.org/abs/0907.1917}{{\ttfamily 0907.1917}}].

\bibitem{Jimenez:2010ev}
R.~Jimenez, T.~Kitching, C.~Pena-Garay and L.~Verde, \emph{{Can we measure the
  neutrino mass hierarchy in the sky?}},
  \href{https://doi.org/10.1088/1475-7516/2010/05/035}{\emph{JCAP} {\bfseries
  05} (2010) 035}, [\href{https://arxiv.org/abs/1003.5918}{{\ttfamily
  1003.5918}}].

\bibitem{Hall:2012kg}
A.~C. Hall and A.~Challinor, \emph{{Probing the neutrino mass hierarchy with
  CMB weak lensing}},
  \href{https://doi.org/10.1111/j.1365-2966.2012.21493.x}{\emph{Mon. Not. Roy.
  Astron. Soc.} {\bfseries 425} (2012) 1170--1184},
  [\href{https://arxiv.org/abs/1205.6172}{{\ttfamily 1205.6172}}].

\bibitem{Jimenez:2016ckl}
R.~Jimenez, C.~P.~n. Garay and L.~Verde, \emph{{Neutrino footprint in Large
  Scale Structure}},
  \href{https://doi.org/10.1016/j.dark.2016.11.004}{\emph{Phys. Dark Univ.}
  {\bfseries 15} (2017) 31--34},
  [\href{https://arxiv.org/abs/1602.08430}{{\ttfamily 1602.08430}}].

\bibitem{Gil-Marin:2015sqa}
H.~Gil-Mar\'\i{}n et~al., \emph{{The clustering of galaxies in the SDSS-III
  Baryon Oscillation Spectroscopic Survey: RSD measurement from the
  LOS-dependent power spectrum of DR12 BOSS galaxies}},
  \href{https://doi.org/10.1093/mnras/stw1096}{\emph{Mon. Not. Roy. Astron.
  Soc.} {\bfseries 460} (2016) 4188--4209},
  [\href{https://arxiv.org/abs/1509.06386}{{\ttfamily 1509.06386}}].

\bibitem{BOSS:2012coo}
{\scshape BOSS} collaboration, A.~J. Ross et~al., \emph{{The clustering of
  galaxies in the SDSS-III Baryon Oscillation Spectroscopic Survey: Analysis of
  potential systematics}},
  \href{https://doi.org/10.1111/j.1365-2966.2012.21235.x}{\emph{Mon. Not. Roy.
  Astron. Soc.} {\bfseries 424} (2012) 564},
  [\href{https://arxiv.org/abs/1203.6499}{{\ttfamily 1203.6499}}].

\bibitem{planck2018likelihoods}
{\scshape Planck} collaboration, N.~Aghanim et~al., \emph{{Planck 2018 results.
  V. CMB power spectra and likelihoods}},
  \href{https://doi.org/10.1051/0004-6361/201936386}{\emph{Astron. Astrophys.}
  {\bfseries 641} (2020) A5},
  [\href{https://arxiv.org/abs/1907.12875}{{\ttfamily 1907.12875}}].

\bibitem{Aiola_2020}
{\scshape ACT} collaboration, S.~Aiola et~al., \emph{{The Atacama Cosmology
  Telescope: DR4 Maps and Cosmological Parameters}},
  \href{https://doi.org/10.1088/1475-7516/2020/12/047}{\emph{JCAP} {\bfseries
  12} (2020) 047}, [\href{https://arxiv.org/abs/2007.07288}{{\ttfamily
  2007.07288}}].

\bibitem{Choi_2020}
{\scshape ACT} collaboration, S.~K. Choi et~al., \emph{{The Atacama Cosmology
  Telescope: a measurement of the Cosmic Microwave Background power spectra at
  98 and 150 GHz}},
  \href{https://doi.org/10.1088/1475-7516/2020/12/045}{\emph{JCAP} {\bfseries
  12} (2020) 045}, [\href{https://arxiv.org/abs/2007.07289}{{\ttfamily
  2007.07289}}].

\bibitem{Ross_2015}
A.~J. Ross, L.~Samushia, C.~Howlett, W.~J. Percival, A.~Burden and M.~Manera,
  \emph{{The clustering of the SDSS DR7 main Galaxy sample \textendash{} I. A 4
  per cent distance measure at $z = 0.15$}},
  \href{https://doi.org/10.1093/mnras/stv154}{\emph{Mon. Not. Roy. Astron.
  Soc.} {\bfseries 449} (2015) 835--847},
  [\href{https://arxiv.org/abs/1409.3242}{{\ttfamily 1409.3242}}].

\bibitem{Beutler_2011}
F.~Beutler, C.~Blake, M.~Colless, D.~H. Jones, L.~Staveley-Smith, L.~Campbell
  et~al., \emph{{The 6dF Galaxy Survey: Baryon Acoustic Oscillations and the
  Local Hubble Constant}},
  \href{https://doi.org/10.1111/j.1365-2966.2011.19250.x}{\emph{Mon. Not. Roy.
  Astron. Soc.} {\bfseries 416} (2011) 3017--3032},
  [\href{https://arxiv.org/abs/1106.3366}{{\ttfamily 1106.3366}}].

\bibitem{Alam_2017}
{\scshape BOSS} collaboration, S.~Alam et~al., \emph{{The clustering of
  galaxies in the completed SDSS-III Baryon Oscillation Spectroscopic Survey:
  cosmological analysis of the DR12 galaxy sample}},
  \href{https://doi.org/10.1093/mnras/stx721}{\emph{Mon. Not. Roy. Astron.
  Soc.} {\bfseries 470} (2017) 2617--2652},
  [\href{https://arxiv.org/abs/1607.03155}{{\ttfamily 1607.03155}}].

\bibitem{Alcock:1979mp}
C.~Alcock and B.~Paczynski, \emph{{An evolution free test for non-zero
  cosmological constant}},
  \href{https://doi.org/10.1038/281358a0}{\emph{Nature} {\bfseries 281} (1979)
  358--359}.

\bibitem{SDSS:2006lmn}
{\scshape SDSS} collaboration, M.~Tegmark et~al., \emph{{Cosmological
  Constraints from the SDSS Luminous Red Galaxies}},
  \href{https://doi.org/10.1103/PhysRevD.74.123507}{\emph{Phys. Rev. D}
  {\bfseries 74} (2006) 123507},
  [\href{https://arxiv.org/abs/astro-ph/0608632}{{\ttfamily
  astro-ph/0608632}}].

\bibitem{Ferramacho:2008ap}
L.~D. Ferramacho, A.~Blanchard and Y.~Zolnierowski, \emph{{Constraints on
  C.D.M. cosmology from galaxy power spectrum, CMB and SNIa evolution}},
  \href{https://doi.org/10.1051/0004-6361/200810693}{\emph{Astron. Astrophys.}
  {\bfseries 499} (2009) 21},
  [\href{https://arxiv.org/abs/0807.4608}{{\ttfamily 0807.4608}}].

\bibitem{Reid:2009xm}
B.~A. Reid et~al., \emph{{Cosmological Constraints from the Clustering of the
  Sloan Digital Sky Survey DR7 Luminous Red Galaxies}},
  \href{https://doi.org/10.1111/j.1365-2966.2010.16276.x}{\emph{Mon. Not. Roy.
  Astron. Soc.} {\bfseries 404} (2010) 60--85},
  [\href{https://arxiv.org/abs/0907.1659}{{\ttfamily 0907.1659}}].

\bibitem{Parkinson_2012}
D.~Parkinson et~al., \emph{{The WiggleZ Dark Energy Survey: Final data release
  and cosmological results}},
  \href{https://doi.org/10.1103/PhysRevD.86.103518}{\emph{Phys. Rev. D}
  {\bfseries 86} (2012) 103518},
  [\href{https://arxiv.org/abs/1210.2130}{{\ttfamily 1210.2130}}].

\bibitem{Schmittfull:2017ffw}
M.~Schmittfull and U.~Seljak, \emph{{Parameter constraints from
  cross-correlation of CMB lensing with galaxy clustering}},
  \href{https://doi.org/10.1103/PhysRevD.97.123540}{\emph{Phys. Rev. D}
  {\bfseries 97} (2018) 123540},
  [\href{https://arxiv.org/abs/1710.09465}{{\ttfamily 1710.09465}}].

\bibitem{Philcox:2019xzt}
O.~H.~E. Philcox and D.~J. Eisenstein, \emph{{Estimating Covariance Matrices
  for Two- and Three-Point Correlation Function Moments in Arbitrary Survey
  Geometries}}, \href{https://doi.org/10.1093/mnras/stz2896}{\emph{Mon. Not.
  Roy. Astron. Soc.} {\bfseries 490} (2019) 5931--5951},
  [\href{https://arxiv.org/abs/1910.04764}{{\ttfamily 1910.04764}}].

\bibitem{Philcox:2019hdi}
O.~H.~E. Philcox and D.~J. Eisenstein, \emph{{Computing the Small-Scale Galaxy
  Power Spectrum and Bispectrum in Configuration-Space}},
  \href{https://doi.org/10.1093/mnras/stz3335}{\emph{Mon. Not. Roy. Astron.
  Soc.} {\bfseries 492} (2020) 1214--1242},
  [\href{https://arxiv.org/abs/1912.01010}{{\ttfamily 1912.01010}}].

\bibitem{Kodwani:2018uaf}
D.~Kodwani, D.~Alonso and P.~Ferreira, \emph{{The effect on cosmological
  parameter estimation of a parameter dependent covariance matrix}},
  \href{https://doi.org/10.21105/astro.1811.11584}{\emph{Open J. Astrophys.}
  {\bfseries 2} (2019) 3}, [\href{https://arxiv.org/abs/1811.11584}{{\ttfamily
  1811.11584}}].

\bibitem{Harnois-Deraps:2019rsd}
J.~Harnois-Deraps, B.~Giblin and B.~Joachimi, \emph{{Cosmic Shear Covariance
  Matrix in $w$CDM: Cosmology Matters}},
  \href{https://doi.org/10.1051/0004-6361/201935912}{\emph{Astron. Astrophys.}
  {\bfseries 631} (2019) A160},
  [\href{https://arxiv.org/abs/1905.06454}{{\ttfamily 1905.06454}}].

\bibitem{Lewis:1999bs}
A.~Lewis, A.~Challinor and A.~Lasenby, \emph{{Efficient computation of CMB
  anisotropies in closed FRW models}},
  \href{https://doi.org/10.1086/309179}{\emph{Astrophys. J.} {\bfseries 538}
  (2000) 473--476}, [\href{https://arxiv.org/abs/astro-ph/9911177}{{\ttfamily
  astro-ph/9911177}}].

\bibitem{Lewis_2002_cosmomc}
A.~Lewis and S.~Bridle, \emph{{Cosmological parameters from CMB and other data:
  A Monte Carlo approach}},
  \href{https://doi.org/10.1103/PhysRevD.66.103511}{\emph{Phys. Rev. D}
  {\bfseries 66} (2002) 103511},
  [\href{https://arxiv.org/abs/astro-ph/0205436}{{\ttfamily
  astro-ph/0205436}}].

\bibitem{Gelman:1992zz}
A.~Gelman and D.~B. Rubin, \emph{{Inference from Iterative Simulation Using
  Multiple Sequences}},
  \href{https://doi.org/10.1214/ss/1177011136}{\emph{Statist. Sci.} {\bfseries
  7} (1992) 457--472}.

\bibitem{Lewis_2019_getdist}
A.~{Lewis}, \emph{{GetDist: a Python package for analysing Monte Carlo
  samples}}, {\emph{arXiv e-prints} (Oct., 2019) arXiv:1910.13970},
  [\href{https://arxiv.org/abs/1910.13970}{{\ttfamily 1910.13970}}].

\bibitem{Eisenstein_2007}
D.~J. Eisenstein, H.-j. Seo, E.~Sirko and D.~Spergel, \emph{{Improving
  Cosmological Distance Measurements by Reconstruction of the Baryon Acoustic
  Peak}}, \href{https://doi.org/10.1086/518712}{\emph{Astrophys. J.} {\bfseries
  664} (2007) 675--679},
  [\href{https://arxiv.org/abs/astro-ph/0604362}{{\ttfamily
  astro-ph/0604362}}].

\bibitem{Padmanabhan_2012}
N.~Padmanabhan, X.~Xu, D.~J. Eisenstein, R.~Scalzo, A.~J. Cuesta, K.~T. Mehta
  et~al., \emph{{A 2 per cent distance to $z$=0.35 by reconstructing baryon
  acoustic oscillations - I. Methods and application to the Sloan Digital Sky
  Survey}}, \href{https://doi.org/10.1111/j.1365-2966.2012.21888.x}{\emph{Mon.
  Not. Roy. Astron. Soc.} {\bfseries 427} (2012) 2132--2145},
  [\href{https://arxiv.org/abs/1202.0090}{{\ttfamily 1202.0090}}].

\bibitem{Levy:2020emr}
B.~L\'evy, R.~Mohayaee and S.~von Hausegger, \emph{{A fast semidiscrete optimal
  transport algorithm for a unique reconstruction of the early Universe}},
  \href{https://doi.org/10.1093/mnras/stab1676}{\emph{Mon. Not. Roy. Astron.
  Soc.} {\bfseries 506} (2021) 1165--1185},
  [\href{https://arxiv.org/abs/2012.09074}{{\ttfamily 2012.09074}}].

\bibitem{vonHausegger:2021luu}
S.~von Hausegger, B.~L\'evy and R.~Mohayaee, \emph{{Accurate Baryon Acoustic
  Oscillations Reconstruction via Semidiscrete Optimal Transport}},
  \href{https://doi.org/10.1103/PhysRevLett.128.201302}{\emph{Phys. Rev. Lett.}
  {\bfseries 128} (2022) 201302},
  [\href{https://arxiv.org/abs/2110.08868}{{\ttfamily 2110.08868}}].

\bibitem{Brinckmann:2018owf}
T.~Brinckmann, D.~C. Hooper, M.~Archidiacono, J.~Lesgourgues and T.~Sprenger,
  \emph{{The promising future of a robust cosmological neutrino mass
  measurement}},
  \href{https://doi.org/10.1088/1475-7516/2019/01/059}{\emph{JCAP} {\bfseries
  01} (2019) 059}, [\href{https://arxiv.org/abs/1808.05955}{{\ttfamily
  1808.05955}}].

\bibitem{Hamann:2010pw}
J.~Hamann, S.~Hannestad, J.~Lesgourgues, C.~Rampf and Y.~Y.~Y. Wong,
  \emph{{Cosmological parameters from large scale structure - geometric versus
  shape information}},
  \href{https://doi.org/10.1088/1475-7516/2010/07/022}{\emph{JCAP} {\bfseries
  07} (2010) 022}, [\href{https://arxiv.org/abs/1003.3999}{{\ttfamily
  1003.3999}}].

\bibitem{Nunes:2022bhn}
R.~C. Nunes, S.~Vagnozzi, S.~Kumar, E.~Di~Valentino and O.~Mena, \emph{{New
  tests of dark sector interactions from the full-shape galaxy power
  spectrum}}, \href{https://doi.org/10.1103/PhysRevD.105.123506}{\emph{Phys.
  Rev. D} {\bfseries 105} (2022) 123506},
  [\href{https://arxiv.org/abs/2203.08093}{{\ttfamily 2203.08093}}].

\bibitem{Singh_2016}
S.~Singh, R.~Mandelbaum and J.~R. Brownstein, \emph{{Cross-correlating Planck
  CMB lensing with SDSS: Lensing-lensing and galaxy-lensing
  cross-correlations}}, \href{https://doi.org/10.1093/mnras/stw2482}{\emph{Mon.
  Not. Roy. Astron. Soc.} {\bfseries 464} (2017) 2120--2138},
  [\href{https://arxiv.org/abs/1606.08841}{{\ttfamily 1606.08841}}].

\bibitem{Doux:2017tsv}
C.~Doux, M.~Penna-Lima, S.~D.~P. Vitenti, J.~Tr\'eguer, E.~Aubourg and
  K.~Ganga, \emph{{Cosmological constraints from a joint analysis of cosmic
  microwave background and spectroscopic tracers of the large-scale
  structure}}, \href{https://doi.org/10.1093/mnras/sty2160}{\emph{Mon. Not.
  Roy. Astron. Soc.} {\bfseries 480} (2018) 5386--5411},
  [\href{https://arxiv.org/abs/1706.04583}{{\ttfamily 1706.04583}}].

\bibitem{Bermejo-Climent:2021jxf}
J.~R. Bermejo-Climent, M.~Ballardini, F.~Finelli, D.~Paoletti, R.~Maartens,
  J.~A. Rubi\~no Mart\'\i{}n et~al., \emph{{Cosmological parameter forecasts by
  a joint 2D tomographic approach to CMB and galaxy clustering}},
  \href{https://doi.org/10.1103/PhysRevD.103.103502}{\emph{Phys. Rev. D}
  {\bfseries 103} (2021) 103502},
  [\href{https://arxiv.org/abs/2106.05267}{{\ttfamily 2106.05267}}].

\bibitem{Ballardini:2021frp}
M.~Ballardini and R.~Maartens, \emph{{Constraining the neutrino mass using a
  multitracer combination of two galaxy surveys and cosmic microwave background
  lensing}}, \href{https://doi.org/10.1093/mnras/stab3480}{\emph{Mon. Not. Roy.
  Astron. Soc.} {\bfseries 510} (2022) 4295--4301},
  [\href{https://arxiv.org/abs/2109.03763}{{\ttfamily 2109.03763}}].

\bibitem{DES:2015eqk}
{\scshape DES} collaboration, T.~Giannantonio et~al., \emph{{CMB lensing
  tomography with the DES Science Verification galaxies}},
  \href{https://doi.org/10.1093/mnras/stv2678}{\emph{Mon. Not. Roy. Astron.
  Soc.} {\bfseries 456} (2016) 3213--3244},
  [\href{https://arxiv.org/abs/1507.05551}{{\ttfamily 1507.05551}}].

\bibitem{Kuntz:2015wza}
A.~Kuntz, \emph{{Cross-correlation of CFHTLenS galaxy catalogue and Planck CMB
  lensing using the halo model prescription}},
  \href{https://doi.org/10.1051/0004-6361/201526940}{\emph{Astron. Astrophys.}
  {\bfseries 584} (2015) A53},
  [\href{https://arxiv.org/abs/1510.00398}{{\ttfamily 1510.00398}}].

\bibitem{saraf2021crosscorrelation}
C.~{Shekhar Saraf}, P.~{Bielewicz} and M.~{Chodorowski},
  \emph{{Cross-correlation between $Planck$ CMB lensing potential and galaxy
  catalogues from HELP}}, {\emph{arXiv e-prints} (June, 2021)
  arXiv:2106.02551}, [\href{https://arxiv.org/abs/2106.02551}{{\ttfamily
  2106.02551}}].

\bibitem{Bianchini:2018mwv}
F.~Bianchini and C.~L. Reichardt, \emph{{Constraining gravity at large scales
  with the 2MASS Photometric Redshift catalogue and Planck lensing}},
  \href{https://doi.org/10.3847/1538-4357/aacafd}{\emph{Astrophys. J.}
  {\bfseries 862} (2018) 81},
  [\href{https://arxiv.org/abs/1801.03736}{{\ttfamily 1801.03736}}].

\bibitem{Hirata:2004rp}
C.~M. Hirata, N.~Padmanabhan, U.~Seljak, D.~Schlegel and J.~Brinkmann,
  \emph{{Cross-correlation of CMB with large-scale structure: Weak
  gravitational lensing}},
  \href{https://doi.org/10.1103/PhysRevD.70.103501}{\emph{Phys. Rev. D}
  {\bfseries 70} (2004) 103501},
  [\href{https://arxiv.org/abs/astro-ph/0406004}{{\ttfamily
  astro-ph/0406004}}].

\bibitem{Smith:2007rg}
K.~M. Smith, O.~Zahn and O.~Dore, \emph{{Detection of Gravitational Lensing in
  the Cosmic Microwave Background}},
  \href{https://doi.org/10.1103/PhysRevD.76.043510}{\emph{Phys. Rev. D}
  {\bfseries 76} (2007) 043510},
  [\href{https://arxiv.org/abs/0705.3980}{{\ttfamily 0705.3980}}].

\bibitem{Das:2011ak}
S.~Das et~al., \emph{{Detection of the Power Spectrum of Cosmic Microwave
  Background Lensing by the Atacama Cosmology Telescope}},
  \href{https://doi.org/10.1103/PhysRevLett.107.021301}{\emph{Phys. Rev. Lett.}
  {\bfseries 107} (2011) 021301},
  [\href{https://arxiv.org/abs/1103.2124}{{\ttfamily 1103.2124}}].

\bibitem{vanEngelen:2013rla}
A.~van Engelen, S.~Bhattacharya, N.~Sehgal, G.~P. Holder, O.~Zahn and D.~Nagai,
  \emph{{CMB Lensing Power Spectrum Biases from Galaxies and Clusters using
  High-angular Resolution Temperature Maps}},
  \href{https://doi.org/10.1088/0004-637X/786/1/13}{\emph{Astrophys. J.}
  {\bfseries 786} (2014) 13},
  [\href{https://arxiv.org/abs/1310.7023}{{\ttfamily 1310.7023}}].

\bibitem{Liu:2015xfa}
J.~Liu and J.~C. Hill, \emph{{Cross-correlation of Planck CMB Lensing and
  CFHTLenS Galaxy Weak Lensing Maps}},
  \href{https://doi.org/10.1103/PhysRevD.92.063517}{\emph{Phys. Rev. D}
  {\bfseries 92} (2015) 063517},
  [\href{https://arxiv.org/abs/1504.05598}{{\ttfamily 1504.05598}}].

\bibitem{Ferraro:2017fac}
S.~Ferraro and J.~C. Hill, \emph{{Bias to CMB Lensing Reconstruction from
  Temperature Anisotropies due to Large-Scale Galaxy Motions}},
  \href{https://doi.org/10.1103/PhysRevD.97.023512}{\emph{Phys. Rev. D}
  {\bfseries 97} (2018) 023512},
  [\href{https://arxiv.org/abs/1705.06751}{{\ttfamily 1705.06751}}].

\bibitem{Madhavacheril:2018bxi}
M.~S. Madhavacheril and J.~C. Hill, \emph{{Mitigating Foreground Biases in CMB
  Lensing Reconstruction Using Cleaned Gradients}},
  \href{https://doi.org/10.1103/PhysRevD.98.023534}{\emph{Phys. Rev. D}
  {\bfseries 98} (2018) 023534},
  [\href{https://arxiv.org/abs/1802.08230}{{\ttfamily 1802.08230}}].

\bibitem{Darwish_2020}
O.~Darwish, M.~S. Madhavacheril, B.~D. Sherwin, S.~Aiola, N.~Battaglia, J.~A.
  Beall et~al., \emph{The atacama cosmology telescope: a cmb lensing mass map
  over 2100 square degrees of sky and its cross-correlation with boss-cmass
  galaxies}, \href{https://doi.org/10.1093/mnras/staa3438}{\emph{Monthly
  Notices of the Royal Astronomical Society} {\bfseries 500} (Nov, 2020)
  2250–2263}.

\bibitem{SimonsObservatory:2018koc}
{\scshape Simons Observatory} collaboration, P.~Ade et~al., \emph{{The Simons
  Observatory: Science goals and forecasts}},
  \href{https://doi.org/10.1088/1475-7516/2019/02/056}{\emph{JCAP} {\bfseries
  02} (2019) 056}, [\href{https://arxiv.org/abs/1808.07445}{{\ttfamily
  1808.07445}}].

\bibitem{SimonsObservatory:2019qwx}
{\scshape Simons Observatory} collaboration, M.~H. Abitbol et~al., \emph{{The
  Simons Observatory: Astro2020 Decadal Project Whitepaper}}, {\emph{Bull. Am.
  Astron. Soc.} {\bfseries 51} (2019) 147},
  [\href{https://arxiv.org/abs/1907.08284}{{\ttfamily 1907.08284}}].

\bibitem{Fang:2021ici}
X.~Fang, T.~Eifler, E.~Schaan, H.-J. Huang, E.~Krause and S.~Ferraro,
  \emph{{Cosmology from clustering, cosmic shear, CMB lensing, and cross
  correlations: combining Rubin observatory and Simons Observatory}},
  \href{https://doi.org/10.1093/mnras/stab3410}{\emph{Mon. Not. Roy. Astron.
  Soc.} {\bfseries 509} (2021) 5721--5736},
  [\href{https://arxiv.org/abs/2108.00658}{{\ttfamily 2108.00658}}].

\bibitem{Yu:2021vce}
B.~Yu, S.~Ferraro, Z.~R. Knight, L.~Knox and B.~D. Sherwin, \emph{{The physical
  origin of dark energy constraints from rubin observatory and CMB-S4 lensing
  tomography}}, \href{https://doi.org/10.1093/mnras/stac1054}{\emph{Mon. Not.
  Roy. Astron. Soc.} {\bfseries 513} (2022) 1887--1894},
  [\href{https://arxiv.org/abs/2108.02801}{{\ttfamily 2108.02801}}].

\bibitem{Handley:2020hdp}
W.~Handley and P.~Lemos, \emph{{Quantifying the global parameter tensions
  between ACT, SPT and Planck}},
  \href{https://doi.org/10.1103/PhysRevD.103.063529}{\emph{Phys. Rev. D}
  {\bfseries 103} (2021) 063529},
  [\href{https://arxiv.org/abs/2007.08496}{{\ttfamily 2007.08496}}].

\bibitem{DiValentino:2021imh}
E.~Di~Valentino and A.~Melchiorri, \emph{{Neutrino Mass Bounds in the Era of
  Tension Cosmology}},
  \href{https://doi.org/10.3847/2041-8213/ac6ef5}{\emph{Astrophys. J. Lett.}
  {\bfseries 931} (2022) L18},
  [\href{https://arxiv.org/abs/2112.02993}{{\ttfamily 2112.02993}}].

\bibitem{Sharma:2022ifr}
R.~K. {Sharma}, K.~{Lal Pandey} and S.~{Das}, \emph{{Multi-parameter Dynamical
  Dark Energy Equation of State and Present Cosmological Tensions}},
  {\emph{arXiv e-prints} (Feb., 2022) arXiv:2202.01749},
  [\href{https://arxiv.org/abs/2202.01749}{{\ttfamily 2202.01749}}].

\bibitem{Chudaykin:2022rnl}
A.~{Chudaykin}, D.~{Gorbunov} and N.~{Nedelko}, \emph{{Exploring $\Lambda$CDM
  extensions with SPT-3G and Planck data: 4$\sigma$ evidence for neutrino
  masses, full resolution of the Hubble crisis by dark energy with phantom
  crossing, and all that}}, {\emph{arXiv e-prints} (Mar., 2022)
  arXiv:2203.03666}, [\href{https://arxiv.org/abs/2203.03666}{{\ttfamily
  2203.03666}}].

\bibitem{Scoccimarro_1999}
R.~Scoccimarro, M.~Zaldarriaga and L.~Hui, \emph{{Power spectrum correlations
  induced by nonlinear clustering}},
  \href{https://doi.org/10.1086/308059}{\emph{Astrophys. J.} {\bfseries 527}
  (1999) 1}, [\href{https://arxiv.org/abs/astro-ph/9901099}{{\ttfamily
  astro-ph/9901099}}].

\bibitem{Takada_2013}
M.~Takada and W.~Hu, \emph{Power spectrum super-sample covariance},
  \href{https://doi.org/10.1103/physrevd.87.123504}{\emph{Physical Review D}
  {\bfseries 87} (Jun, 2013) }.

\bibitem{Mohammed_2016}
I.~Mohammed, U.~Seljak and Z.~Vlah, \emph{Perturbative approach to covariance
  matrix of the matter power spectrum},
  \href{https://doi.org/10.1093/mnras/stw3196}{\emph{Monthly Notices of the
  Royal Astronomical Society} {\bfseries 466} (Dec, 2016) 780–797}.

\bibitem{Cooray_2004}
A.~R. Cooray, \emph{{Non-linear galaxy power spectrum and cosmological
  parameters}},
  \href{https://doi.org/10.1111/j.1365-2966.2004.07358.x}{\emph{Mon. Not. Roy.
  Astron. Soc.} {\bfseries 348} (2004) 250--260},
  [\href{https://arxiv.org/abs/astro-ph/0311515}{{\ttfamily
  astro-ph/0311515}}].

\bibitem{Assassi:2014fva}
V.~Assassi, D.~Baumann, D.~Green and M.~Zaldarriaga, \emph{{Renormalized Halo
  Bias}}, \href{https://doi.org/10.1088/1475-7516/2014/08/056}{\emph{JCAP}
  {\bfseries 08} (2014) 056},
  [\href{https://arxiv.org/abs/1402.5916}{{\ttfamily 1402.5916}}].

\bibitem{Biagetti:2014pha}
M.~Biagetti, V.~Desjacques, A.~Kehagias and A.~Riotto, \emph{{Nonlocal halo
  bias with and without massive neutrinos}},
  \href{https://doi.org/10.1103/PhysRevD.90.045022}{\emph{Phys. Rev. D}
  {\bfseries 90} (2014) 045022},
  [\href{https://arxiv.org/abs/1405.1435}{{\ttfamily 1405.1435}}].

\bibitem{LSST:2008ijt}
{\scshape LSST} collaboration, v.~Ivezi\'c et~al., \emph{{LSST: from Science
  Drivers to Reference Design and Anticipated Data Products}},
  \href{https://doi.org/10.3847/1538-4357/ab042c}{\emph{Astrophys. J.}
  {\bfseries 873} (2019) 111},
  [\href{https://arxiv.org/abs/0805.2366}{{\ttfamily 0805.2366}}].

\end{thebibliography}\endgroup
\bibliographystyle{JHEP}

\end{document}